\documentclass{aa}  
\usepackage[table,xcdraw,dvipsnames]{xcolor}
\definecolor{LightBlue}{RGB}{173, 216, 230}
\usepackage{lscape}
\usepackage{setspace}
\usepackage{chngpage}
\usepackage[english]{varioref}
\usepackage{pythonhighlight}
\usepackage{booktabs}
\usepackage{hyperref} 
\usepackage{graphicx}
\usepackage{txfonts}
\usepackage{caption}
\usepackage{xcolor}
\usepackage{ifthen}
\newboolean{acceptcomments}
\setboolean{acceptcomments}{true}  % Cambia in true per accettare i commenti, false per nasconderli
\newcommand{\comment}[2]{%
  \ifthenelse{\boolean{acceptcomments}}%
    {\textcolor{#1}{#2}}  % Se true, mostra il commento con colore
    {}  % Se false, non mostra nulla
}

\begin{document}

   \title{The VMC Survey -- LIV. Anomalous Cepheids in the Magellanic Clouds}

   \subtitle{Period-Luminosity relations in the near-infrared bands.}

   \author{T. Sicignano
          \inst{1,2,3,4}
\and 
          V. Ripepi \inst{3}
\and 
          M. Rejkuba \inst{1}
\and 
          M. Romaniello  \inst{1}
\and 
          M. Marconi \inst{3}
\and 
          R. Molinaro \inst{3}
\and 
          A. Bhardwaj \inst{5}
\and 
          \\
          G. De Somma \inst{3,4}
\and 
          M.-R. L. Cioni \inst{6}
\and 
          F. Cusano \inst{7}
\and 
          G. Clementini \inst{7}
\and
          R. de Grijs \inst{8,9,10}
\and 
          V. D. Ivanov \inst{1} 
\and
          J. Storm \inst{6} 
\and
          \\
          M. A. T. Groenewegen \inst{11}         
    }      
   \institute{European Southern Observatory, Karl-Schwarzschild-Strasse 2, 85748 Garching bei München, Germany \\ \email{teresa.sicignano@inaf.it}
   \and Scuola Superiore Meridionale, Largo San Marcellino10 I-80138 Napoli, Italy \and INAF-Osservatorio Astronomico di Capodimonte, Salita Moiariello 16, 80131, Napoli, Italy \and Istituto Nazionale di Fisica Nucleare, Sez. di Napoli, Monte S. Angelo, Via Cinthia Edificio 6 I-80126 Napoli, Italy \and Inter-University Center for Astronomy and Astrophysics (IUCAA), Post Bag 4, Ganeshkhind, Pune 411 007, India \and Leibniz-Institut für Astrophysik Potsdam (AIP), An der Sternwarte 16, 14482 Potsdam, Germany \and INAF-Osservatorio di Astrofisica e Scienza dello Spazio, Via Piero Gobetti, 93/3, I-40129 Bologna, Italy \and School of Mathematical and Physical Sciences, Macquarie University, Balaclava Road, Sydney NSW 2109, Australia \and Astrophysics and Space Technologies Research Centre, Macquarie University, Balaclava Road, Sydney, NSW 2109, Australia \and International Space Science Institute--Beijing, 1 Nanertiao, Zhongguancun, Hai Dian District, Beijing 100190, China  \and 
   Koninklijke Sterrenwacht van België, Ringlaan 3, 1180, Brussels, Belgium
   }

   \date{Received ; accepted }

% \abstract{}{}{}{}{} 
% 5 {} token are mandatory
 
  \abstract
   { Anomalous Cepheids (ACs) are less studied metal-poor pulsating stars ([Fe/H] < –1.5) compared to Classical Cepheids (CCs) and RR Lyrae stars. They follow distinct Period-Luminosity (PL) and Period-Wesenheit (PW) relations and pulsate in either the fundamental (F) or first overtone (1O) mode. Our goal is to assess the precision and accuracy of AC-based distances and evaluate their potential for establishing an independent distance scale.
 %   Anomalous Cepheids (ACs) are less studied pulsating variable stars compared to the well-known Classical Cepheids (CCs) and RR Lyrae stars. These metal-poor stars ([Fe/H] < 1.5) follow distinct Period-Luminosity (PL) and Period-Wesenheit (PW) relations, which can be used for distance measurements, and can pulsate both in fundamental (F) and first overtone mode (1O).
   }
  { We derive new PL and PW relations for F-mode, 1O-mode, and, for the first time, combined F+1O ACs in the Magellanic Clouds. We study their wavelength dependence and apply the relations to estimate distances to Local Group stellar systems hosting ACs, while also confirming AC classifications.

 % Our goal is to evaluate the precision and accuracy of distances obtained via PL and PW relations of ACs, thus assessing if they could be used to establish a cosmic distance scale independently from CCs. To this aim we derive new precise PL and PW relations for F-mode, 1O-mode, and, for the first time, the combined F+1O mode ACs in Magellanic Clouds; investigate their wavelength dependence; and apply them to calculate the distances of different stellar systems in the Local Group hosting ACs, as well as confirm the classification of these variables.
  }
   { Our analysis is based on near-infrared time-series photometry in the $Y$, $J$, and $K_s$ bands for $\sim$200 ACs in the Magellanic Clouds from the VISTA survey of the Magellanic Clouds system (VMC, 2009–2018). VMC data are complemented with optical photometry from Gaia DR3 and OGLE-IV, which also provide periods and pulsation modes.
   Custom light-curve templates were used to derive precise intensity-averaged magnitudes for 118 ACs in the Large Magellanic Cloud (LMC) and 75 in the Small Magellanic Cloud. These data were used to derive multi-band PL and PW relations, calibrated using the geometric LMC distance from eclipsing binaries.
    %We analyse near-infrared (NIR) time-series photometry in the $Y,\, J,$ and $K_s$ bands for about 200 ACs in the Magellanic Clouds acquired during 2009--2018 in the context of the VISTA survey of the Magellanic Clouds system (VMC), an ESO public survey. The VMC NIR photometry was complemented with optical data from the $Gaia$ DR3 and the Optical Gravitational Lensing Experiment IV survey, which also provided identification, periods, and pulsation mode for the investigated ACs. 
   %Custom templates generated from our best light curves were used to derive precise intensity-averaged mean magnitudes for 118 and 75 ACs in the Large (LMC) and the Small Magellanic Cloud (SMC), respectively. 
   %Based on these data, we calculated accurate multi-band PL and PW relations in both MCs. We calibrated the zero points of these relations via geometric distances to the Large Magellanic Cloud (LMC) and $Gaia$ parallaxes.
   }
   { We find that PL relation slopes increase and dispersions decrease with wavelength. Using Gaia parallaxes, we determine the LMC distance modulus and the LMC–SMC relative distance. We also confirm the AC nature of several new candidates in Galactic Globular Clusters and derive a distance modulus for the Draco dSph galaxy of $19.425\pm0.048$ mag. A 0.1 mag discrepancy with RR Lyrae-based distances may reflect metallicity effects.
    Future spectroscopic surveys and Gaia DR4 will help refine the AC distance scale and quantify metallicity impacts.
   %Optical and NIR mean magnitudes were used to derive multi-band PL and PW relations, which were calibrated with the geometric distance modulus to the LMC based on eclipsing binaries.
  % We investigated the dependence of PL relations on wavelength, finding that slopes increase and dispersion decreases going from optical to NIR bands. We calculated the LMC distance modulus, through calibrated AC PW relations in the Milky Way using Gaia parallaxes, the LMC--SMC relative distance modulus and confirmed the AC nature of a few new pulsators in Galactic Globular Clusters. We derive a distance modulus for Draco dwarf spheroidal galaxy of $19.425\pm0.048$ mag in agreement with recent literature determinations, but a discrepancy of 0.1 mag with RR Lyrae based distance hints at possible metallicity effects on the AC PL and PW relations. 
   %Future spectroscopic surveys and Gaia DR4 will refine the AC distance scale and assess metallicity effects on PLRs and PWRs.
   }
    {}
 
   \keywords{ variable stars --
                extra-galactic distance scale --
                period-luminosity
               }

   \maketitle
%
%-------------------------------------------------------------------

\section{Introduction}
Anomalous Cepheids (ACs) are radially pulsating stars of intermediate-old age ($>$ 1 Gyr) that cross the classical Instability Strip (IS) during their central helium burning phase \citep[e.g.][and references therein]{2004caputo}.
This class of pulsating stars was called \textit{Anomalous} for the first time by \citet{zinn1976} because the Cepheids found in the dwarf spheroidal (dSph) satellites of the Milky Way (MW) obeyed a different Period-Luminosity \citep[PL,][]{leavitt1912period} relation than the "normal" Classical Cepheids (CCs) and Type II Cepheids (T2Cs), as well as infrared PL relation of RR Lyrae stars (RRLs).

Like RRLs stars and CCs, the ACs can pulsate in the fundamental (F) or first overtone (1O) mode. F mode ACs are characterized by high-amplitude asymmetric light curves with periods ranging between 0.5---2.5 days. Whereas, 1O mode ACs show more sinusoidal light curves with smaller amplitudes and periods in the 0.4--1 day range. In both cases, the AC light curves resemble those of the RRL variables for periods $<$ 1 day and those of CCs for periods $>$ 1 day.
%, so it is challenging to correctly classify the different classes without precise distances. 
At a fixed period, the ACs are brighter than both RRLs by $\sim$1.5--2 mag and BL Herculis (the T2C subclass with the shortest periods) stars by $\sim$0.7--1 mag, while they are fainter than CCs by $\sim$0.5-1 mag.  
%A clear insight into the different evolutionary status between ACs, RRLs, T2Cs, and CCs is presented in \citet{Monelli2022}.
%In particular, they were found in all the dwarf galaxies of the Local Group \citep{Monelli2022} but only two in the Galactic Globular Clusters \citep[NGC5466 and M92][]{ZinnDahn1976, Ngeow2022}. 
The differences in brightness among different classes of pulsators is a direct consequence of their different evolutionary stages.
In particular, the ACs are giant stars in their central helium-burning phase having ignited helium in a partially electron degenerate helium core at the tip of the red giant branch \citep[TRGB, e.g.][]{2013Cassisi}. Their location in the Color-Magnitude diagram and their periods suggest that they are stars of intermediate-age with masses in the range $\sim 1.3$--$2.3 M_{\odot}$ \citep[see also][]{fiorentino2012,monelli2022Univ}.
%A clear insight into the different evolutionary status between ACs, RRLs, T2Cs, and CCs is derived from the inspection of Figure~\ref{instrip} which has been taken from \citet{Monelli2022}. The three panels show the theoretical colour-magnitude diagram incorporating the ZAHB and the HB for the typical masses of RRLs and ACs, respectively, and the evolutionary tracks for T2Cs and CCs. The figure also displays the IS for RRLs and CCs. The metallicity decreases from the leftmost to the rightmost panel. 
According to these authors, ACs can be found in stellar systems where the turnover of the horizontal branch (HB)\footnote{The HB turnover refers to an increase in the effective temperature after its minimum value occurred for brighter luminosities.} enters the IS. Stellar evolution models show that this occurs only for values of [Fe/H] more metal-poor than $\sim$ $-$1.3--1.5 dex. The metallicity threshold is critical because lower metallicity stars experience different convective efficiency and mass-loss rates, influencing their evolution through the IS \citep[][]{1997bono}. This occurrence, suggesting that ACs are metal-poor variables, was confirmed observationally by \citet{2024ripepi}.    

Two channels of formation for ACs have been devised so far in the literature: either the ACs are single intermediate-age stars \citep{norris1975cepheid} or they form via the evolution of binary systems with mass transfer \citep{renzini1977,wheeler1979,gautschy2017}. The latter scenario is required to explain the presence of ACs in purely old systems, such as old dSph or Galactic Globular Clusters (GGC).

As intermediate-mass stars, the presence of ACs in stellar systems with extended intermediate-age populations, such as the Magellanic Clouds (MCs), Fornax, and Carina, is not surprising.
The presence of ACs in the MCs, particularly in the Large Magellanic Cloud (LMC), which serves as an anchor of the distance ladder at an average distance of $\sim$ 50 kpc \citep[e.g.][hereinafter P19]{2022ApJ...934L...7R,2019Natur.567..200P}, and in stellar systems at greater distances up to $\sim$ 150 kpc \citep[][]{monelli2022Univ}, establishes them as potential standard candles within the Local Group (LG). In particular, ACs in the LMC are of interest because they can be considered at the same distance with an excellent degree of confidence. Therefore, the slopes of the PL and period-Wesenheit\footnote{The Wesenheit magnitudes are constructed to be reddening free by definition \citep[][]{1982ApJ...253..575M}.} (PW) relations can be easily determined and the zero-point can be calibrated very accurately thanks to the percent-level precise geometric distance to this galaxy (P19).

Similarly, the more famous CCs, ACs PL and PW relations in the NIR bands are particularly effective for distance estimates, due to their reduced scatter, steeper slopes, and lower  (none by definition in the case of the PW relations) sensitivity to reddening \citep[see e.g.][]{1991PASP..103..933M}.  
In this framework, in our previous work, we took advantage of data from the European Southern Observatory (ESO) “Visible and Infrared Survey Telescope for Astronomy” (VISTA) public survey of the Magellanic Clouds system \citep[VMC;][]{2011A&A...527A.116C} to obtain PL and PW relations in the NIR bands for the LMC ACs \citep[][hereinafter R14]{2014ripepianomalous}, which were used to estimate the distances of several objects in the LG.
%The VMC survey has imaged the whole Magellanic system\footnote{Mainly composed by the LMC and the Small Magellanic Cloud (SMC) together with the Bridge connecting them and the Stream \citep[][]{1974IAUS...60..617M}, in the Near-Infrared (NIR) bands and obtained deep time-series photometry reaching even the faintest pulsators in the Clouds. The detailed description of the VMC survey can be found in \citet{2011A&A...527A.116C}. Our interest is focused on time-series $YJKs$ observations from the VMC survey, since they are particularly useful in providing calibration the PL and PW relations of primary stellar standard candles in the MCs.
This paper aims to extend and complete the work by R14, which was based on observations in the $K_s$ band of 48 ACs located within only 11 VISTA tiles\footnote{A tile in the context of VMC refers to a 1.5 sq.deg region of the sky observed with the VISTA telescope. } on the LMC. We use the seventh and last Data Release (DR7, based on 110 tiles) of the VMC survey (Cioni et al., submitted), to determine new and accurate multi-wavelength PL and PW relations for almost 200 ACs both in the LMC and the SMC. We adopt the same methodology of \citet[][hereinafter S24]{2024A&A...685A..41S} where we carried out a similar study of T2Cs.

The paper is structured as follows: Sections 2 and 3 present the characteristics of the VMC survey and the data analysis on ACs. Section 4 reports the fitted PL and PW relations. A comprehensive investigation and application of the inferred AC PL and PW relations are presented in Sections 5 and 6. Finally, Section 7 is dedicated to summary, conclusions, and future perspectives.

\section{Anomalous Cepheids in the VMC survey}

\begin{figure}
\centering
\sidecaption
    \includegraphics[trim=0.8cm 5.5cm 1.45cm 13cm, clip,width=0.9\hsize]{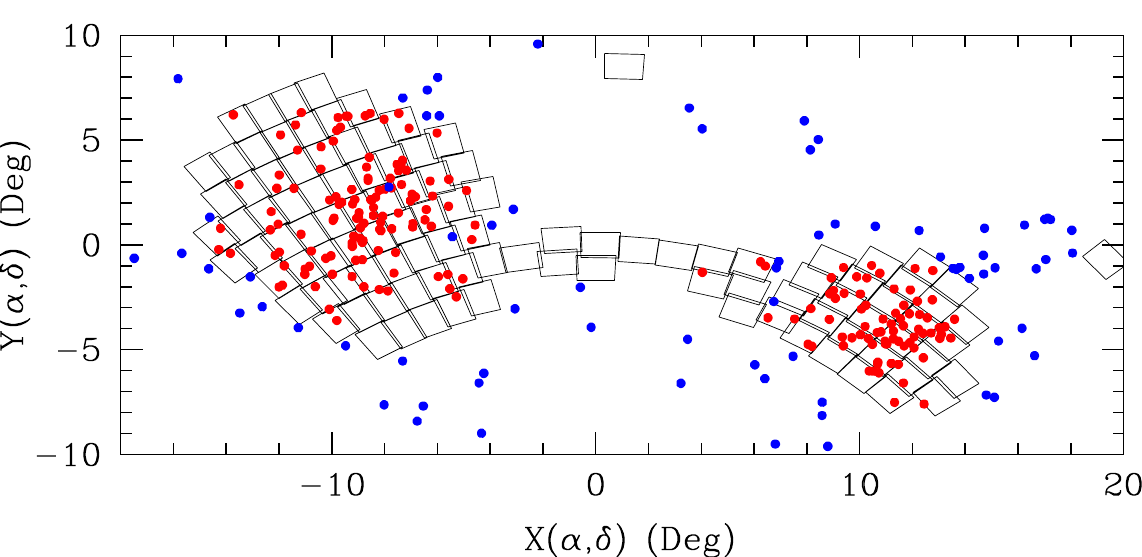}
    \caption{The distribution of the ACs data set in MCs: red points are stars present in the VMC, while blue are ACs from OGLE or Gaia with no VMC counterpart. The projection is a zenithal equidistant projection. Centre is at RA=55 deg DEC= $-$73 deg. The empty black boxes show the footprint of the VMC survey. } 
    \label{stars} 
	\end{figure}
    
The VMC survey acquired time series in the $Y$, $J$, and $K_s$ bands across the entire survey region \citep[refer to][for more information]{2011A&A...527A.116C}. As documented in numerous studies within this series \citep[R14,][]{Ripepi2012, Moretti2014,ripepi2015vmc,ripepi2016vmc, Ripepi2017vmc,ripepi2022vmc}, the VMC survey includes accurate and well-sampled $K_s$-band light curves for all categories of Cepheid variables (classical, anomalous, and type II). 
%This is achieved through 13 observations scheduled with a tailored interval \citep[][]{2011A&A...527A.116C}. Of these, eleven epochs achieve a limiting magnitude in $K_s \sim $ 19.3 mag with a signal-to-noise ratio (S/N) of approximately 5, while two additional, shallower epochs were acquired using half the exposure time. In the \textit{Y} and \textit{J} bands, four epochs are scheduled, with two being less deep. Since observations were repeated if they did not meet the constraints, the light curves typically encompass more epochs \citep[see e.g.][and references therein]{ripepi2016vmc}

%The pipeline of the VISTA Data Flow System (VDFS) \citep{2004SPIE.5493..411I}, located at the Cambridge Astronomical Survey Unit (CASU), is responsible for processing the raw images from the VMC survey. Version 1.5 of the VDFS pipeline was used to reduce the data discussed in this paper. The photometric measurements are calibrated on the VISTA photometric system, which is extensively detailed by \citet{2018MNRAS.474.5459G}. The reduced data are part of the quoted VMC DR7 and are stored in the VISTA Science Archive \citep[VSA][]{2012A&A...548A.119C} as well as at the ESO archive\footnote{https://archive.eso.org/cms.html}.

The OGLE IV survey \citep[Optical Gravitational Lensing Experiment][]{2018AcA....68...89S} and the $Gaia$ mission \citep{2016A&A...595A...1G, 2023A&A...674A...1G, 2016A&A...595A.133C, 2019A&A...622A..60C, 2019A&A...625A..14R, Ripepi2023} published catalogues of ACs in both the LMC and SMC, including accurate periods, the pulsation mode and the epochs of maximum light for each variable star. OGLE IV identified 270 ACs, whereas the $Gaia$ Data Release (DR3) identified 6 additional ACs not present in the OGLE sample. We matched the coordinates of the VMC sources downloaded from the Vista Science Archive (\href{https://vsa.roe.ac.uk/}{VSA}) database, with a tolerance of 1", and including all objects with at least two observation epochs in a single band. For 192 OGLE IV and 6 $Gaia$ ACs, we obtained the $Y,J$ and $K_s$ time-series photometry (refer to Sect.\ref{opt data}). Not all the ACs have VMC observations in all the filters: 195, 197, and 198 stars have photometry in the $Y$, $J$, and $K_s$ bands, respectively. These samples represent 70.6\%, 71.4\% and 71.7\% of the known ACs.

Fig.~\ref{stars} shows the spatial distribution of ACs within the MCs compared with the VMC footprint. Some of the known ACs lie outside the VMC observed region, while a few OGLE IV survey stars lack a VMC counterpart within 1 arcsec. At the current state of knowledge, there is no reason to believe that the missing ACs have properties different from the analyzed ones. Upon initial examination, five ACs with counterparts were omitted (further explained in the Appendix~\ref{samplean}). The final selection includes 193 ACs: 75 (48 F, 27 1O) are part of the SMC, and 118 (84 F, 34 1O) are part of the LMC.

\begin{figure}
\sidecaption
\centering
    \includegraphics[width=\hsize]{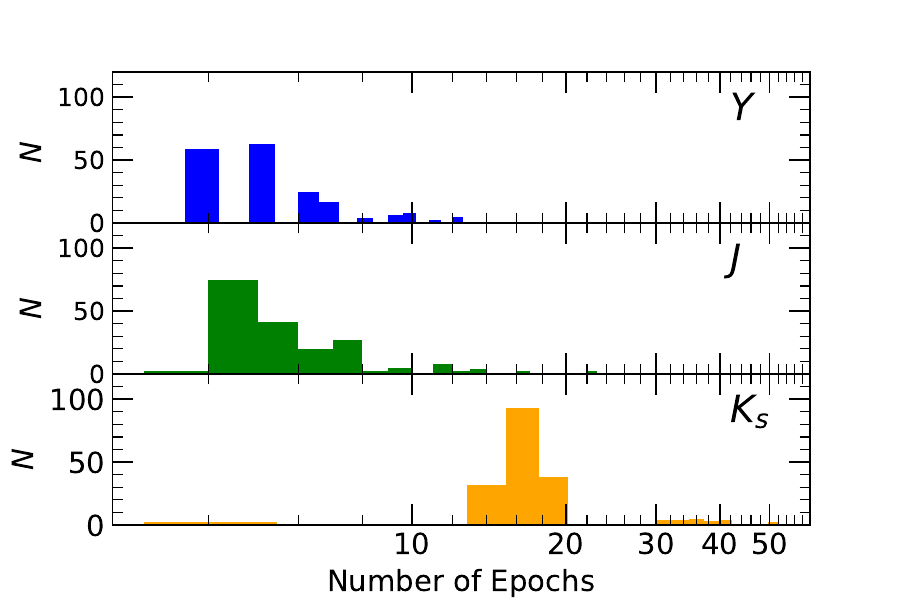}
    \caption{ Number of epochs in the VMC $Y,\,J,\,K_s$ bands for our ACs sample.}
\label{epochs}
\end{figure}

In this study, the ACs' light curves typically have 5-6 epochs in $Y$ and $J$, and 14-15 in $K_s$, as illustrated in Fig.~\ref{epochs}. In some cases, the number of epochs is larger, particularly when a variable star is positioned on the overlapping area of two neighbouring VMC tiles, effectively doubling the number of epochs. Table~\ref{tabdati} presents an example of a typical time-series photometry in the $J$ band. The complete table with $Y, J, K$ time-series photometry for ACs can be accessed at the Centre de Données astronomiques de Strasbourg (\href{https://cds.u-strasbg.fr/}{CDS}).
 \begin{table}\centering\footnotesize\caption{Example of 5 epochs of the time-series photometry for the star OGLE-SMC-ACEP-062 in the \rm{J}-band. The machine-readable version of the full table is published online at the CDS.} 
 \label{tabdati}
 \begin{center}
\begin{tabular}{l c c c } 
\\
    \hline
    ID & HJD & $J$& $\sigma_J$ \\ 
    & days & mag & mag \\[0.5ex] 
    \hline
   OGLE-SMC-ACEP-062 & 55493.609696  &  17.2679   &   0.0125   \\
   OGLE-SMC-ACEP-062 &55493.648796   &  17.2606   &   0.0118   \\
   OGLE-SMC-ACEP-062 &55495.563371  &  17.4333  &    0.0179   \\
   OGLE-SMC-ACEP-062 &55539.652021   &  17.4451  &    0.0169   \\
   OGLE-SMC-ACEP-062 &55778.764537  &   17.3937   &   0.0159   \\
  
   \hline
\end{tabular}
\end{center}
\end{table}

\section{Data analysis}

\begin{figure}
\sidecaption
    \vbox{
    \includegraphics[width=\hsize]{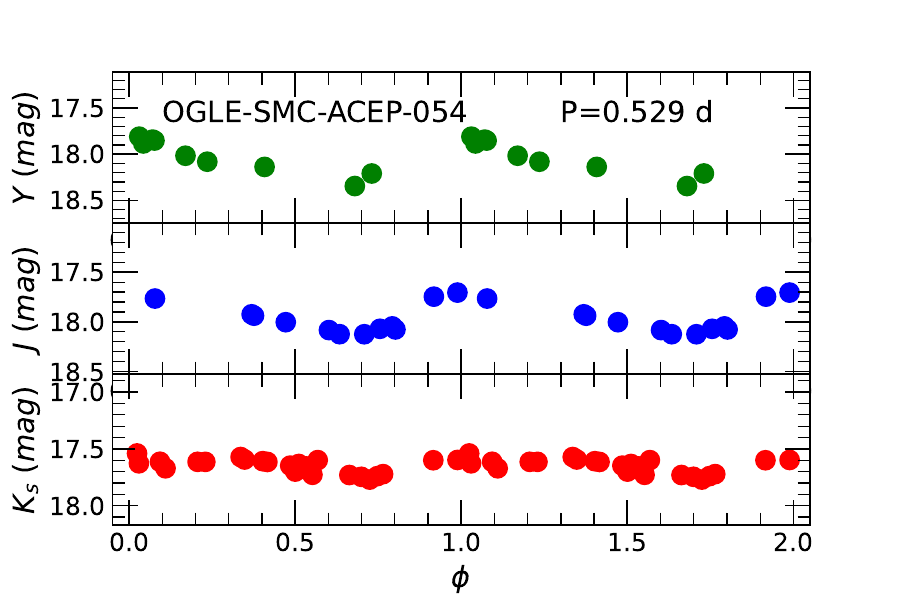}
    \includegraphics[width=\hsize]{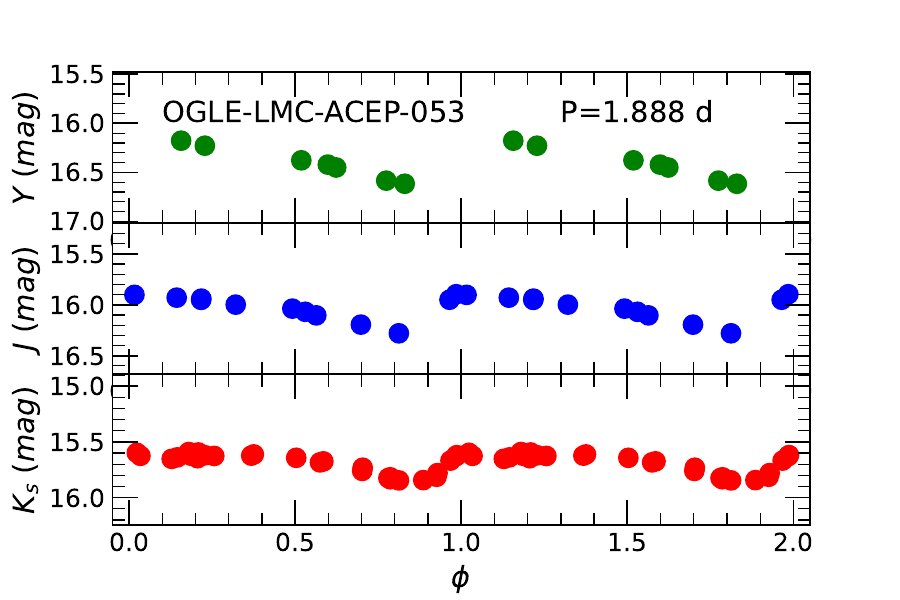}
    }           
    \caption{Examples of light curves (used to create templates) for different AC types. From the top: one 1O pulsator and one F pulsator in  $Y$ (green), $J$ (blue) and $K_s$ (red) bands, respectively. Note that the photometric errors are smaller than the dot size and the light curves are doubled to shape them better.}	
    \label{lightcurves} 
	\end{figure}

To phase the time series photometry for each target we adopted the period and epoch of peak brightness from the literature (OGLE 
IV and $Gaia$\,DR3, as explained above). Figure~\ref{lightcurves} illustrates samples of light curves for the F and 1O mode ACs. The approach for determining the $YJK_s$ intensity-averaged magnitudes follows the methodology described in \citet[][]{2016ApJS..224...21R,2017MNRAS.472..808R,2022A&A...659A.167R} and S24. Specifically, we adhered strictly to the approach detailed in S24, which is briefly summarized next.

\subsection{Template derivation and fitting to the light curves}
The intensity-averaged mean magnitudes for pulsating stars are calculated using the template-fitting method. This procedure is crucial because the majority of the sample has a number of epochs $\leq 10$, especially in $Y$ and $J$ bands.

%The first step to creating templates was to choose the best light curves among our sample stars. To this aim, we selected a subsample of stars with an adequate number of epochs ($>$ 10) to ensure a good probability to have well-sampled light curves. As shown in Fig.~\ref{epochs}, this first selection reduced significantly the number of objects to be considered in $Y$ and $J$, while several tens of stars are available in $K_s$ (see Fig.~\ref{epochs}). 

%Then, we proceeded to a visual inspection of the light curves, retaining only those showing the least scatter and taking care to conserve the entire period range spanned by the data, as well as the variety of light curve shapes representing the four different T2C types we are investigating in this work (BLHer, WVir, pWvir, RVTau).  

The templates are derived from light curves that have more than 10 epochs.
As outlined in S24, the fitting of light curves to create the templates is executed in three phases: (i) the well-sampled and selected by visual inspection light curves are fitted by employing a spline function; (ii) the spline function is rescaled to have mean magnitude equal to zero and peak-to-peak amplitude equal to 1; iii) A ten terms truncated Fourier series is fitted to the spline curve and the coefficients (amplitudes and phases) are provided in Appendix~\ref{apptemplates}.

%The first step was carried out with a {\tt Python} code specifically written to this aim, which adopts the {\tt splrep} function to fit smooth spline curves to the data. The best smoothing factors were found utilizing visual inspection of each light curve. The fitting spline curves have been subsequently transformed into templates by subtracting their intensity-averaged mean magnitude and re-scaling the amplitude by dividing for its peak-to-peak amplitude. A template constructed in this way consists of a light curve with zero mean value and peak-to-peak amplitude equal to 1. 

\begin{figure}
\sidecaption
\centering
    \includegraphics[trim=0cm 2.7cm 0cm 0cm, clip,width=2.555\hsize]{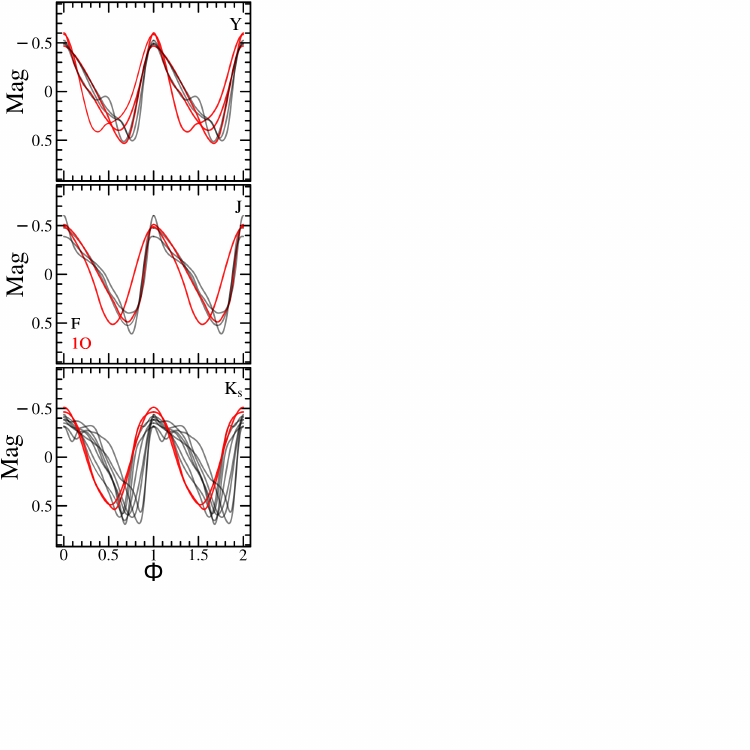}
    \caption{ Templates created in every band: $Y$ (top), $J$ (middle) and $K_s$ (bottom).}
\label{figtemplates}
\end{figure}

%The second step consisted in fitting these normalised templates with a truncated Fourier series with 15 terms to ensure a perfect correspondence of the analytical function to the spline curve. The fitting equation is:
%\begin{equation}
%    m(\phi)= A_0 + \sum^N_1 A_i cos(2\pi\phi + \Phi_i) 
%\end{equation}
%\noindent
%where N is the number of harmonics, $A_0$ the average magnitude (which in our case is zero) $\phi$ is the phase (intended as the adimensional substitute of the time; it is, therefore, an observational, independent variable) while $A_i$ and $\Phi_i$ are the unknown coefficients of the series, i.e. the amplitude and the phase of each term.  
%Ultimately, each template is converted into a set of Fourier coefficients provided in Appendix~\ref{apptemplates}. 
The final template sample, illustrated in Fig.~\ref{figtemplates}, comprises 6, 5 and 11 models for the $Y$, $J$ and $K_s$ bands, respectively. These templates are further categorized by pulsation mode (F, 1O) for each band. Fewer templates are present in the $Y$ and $J$ filters compared to the $K_s$ filters because these bands have fewer observational epochs (c.f.\, Fig.~\ref{epochs}).

As in S24, the procedure for fitting each light curve with a template requires determining three parameters: (1) a magnitude shift $\delta M$; (2) a scaling factor, $a$, that increases or reduces the template amplitude to match that of the observed light curve; and (3) a phase shift, $\delta \phi$, which takes into account possible differences in phase between templates and observed light curves.
These three unknowns for each template are retrieved by minimizing the following $\chi^2$ function:
\begin{equation} \label{chi2}
    \chi^2 = \sum_i^{N_{pts}} \frac{[m_i - (a\times M_t(\phi_i + \delta\phi)+ \delta M)]^2}{\sigma_i^2}
\end{equation}

\noindent

Where $N_{pts}$ is the number of epochs, $m_i$, $\phi_i$ and $\sigma_i$ are the observed magnitudes, the corresponding phases and uncertainties on the magnitudes, respectively; $M_t(\phi_i)$ is the template. 
%The fitted parameters of Eq.~\ref{chi2} are the magnitude shift, $\delta M$, the scaling factor, $a$, and the phase shift, $\delta \phi$. 

%The fitting routine has an outlier rejection procedure, as the VMC light curves can often show one or more bad measurements, due to various reasons (e.g. the star being hit by a  cosmic ray or falling on a bad detector column etc.). 
Outliers are detected by analyzing the distribution of the residuals from the fit and identifying points outside the interval $-3.5\times DMAD $ to +$3.5\times DMAD $, where \textit{DMAD} is Double-Median Absolute Deviation \footnote{{\it DMAD} is calculated treating separately the values smaller and larger than the median of the considered distribution.}.

Similar to our previous studies, we assess the uncertainties using a Monte Carlo method. The errors in our fitted parameters are determined by the robust standard deviation (1.4826 $\cdot$ MAD) derived from the distributions generated by the bootstrap simulations.

\begin{figure*}\
\centering
    \vbox{
    \hbox{    
    \includegraphics[width=0.30\textwidth]{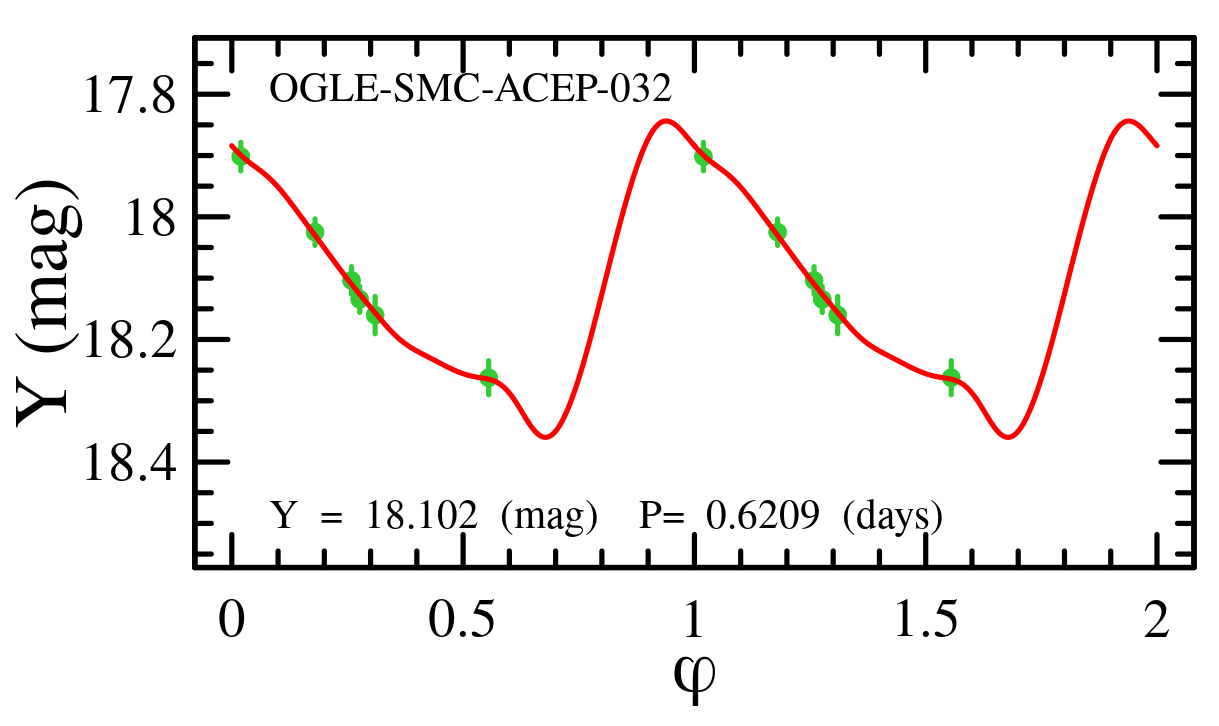}   
    \includegraphics[width=0.30\textwidth]{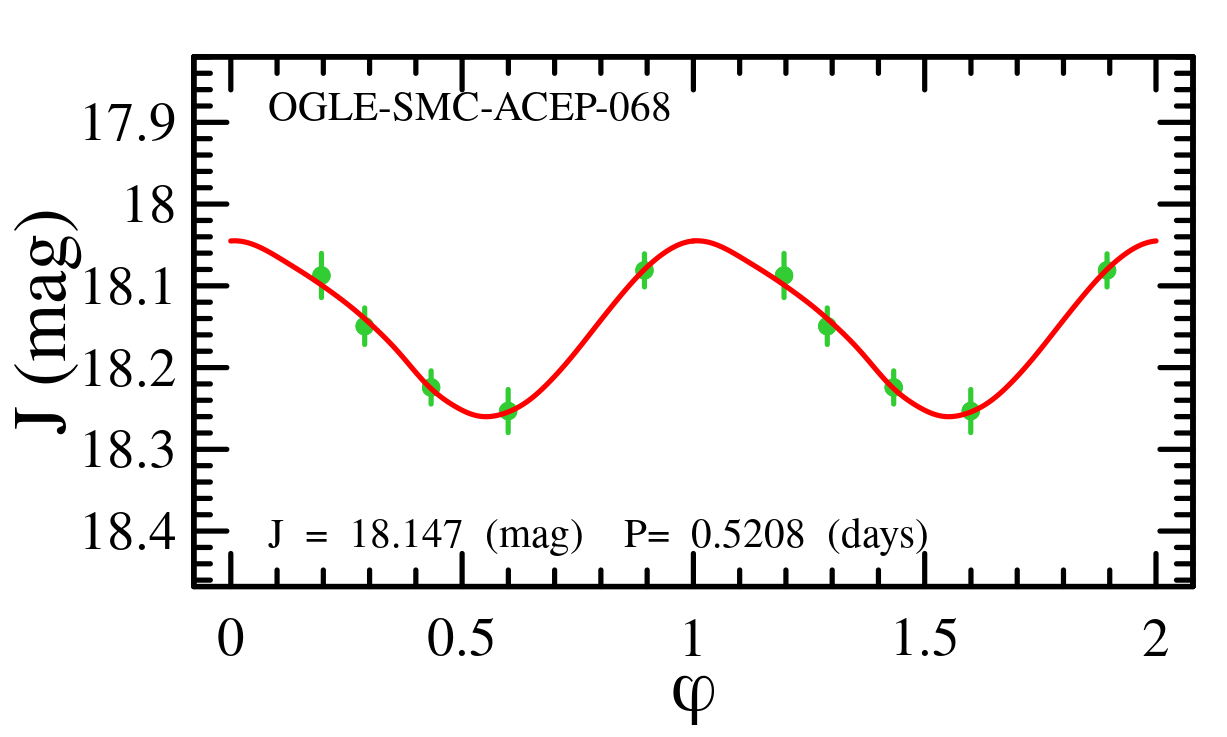}
    \includegraphics[width=0.30\textwidth]{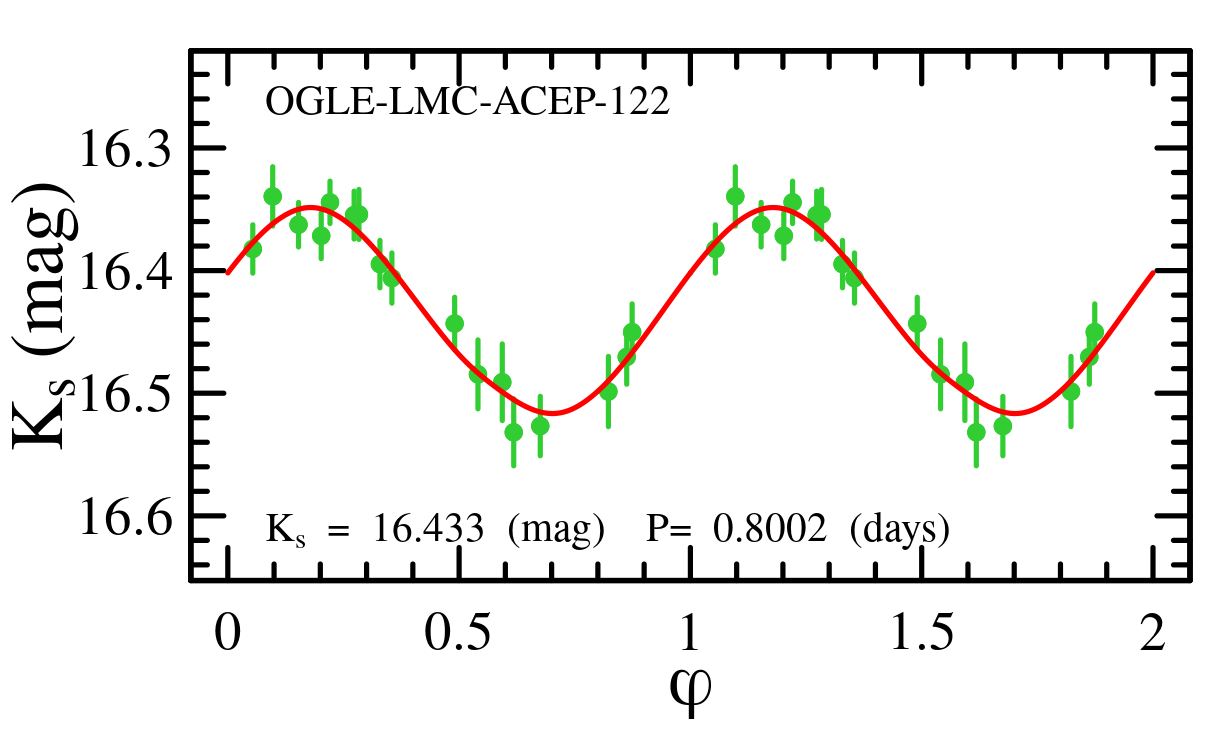}
    }
    \hbox{
    \includegraphics[width=0.30\textwidth]{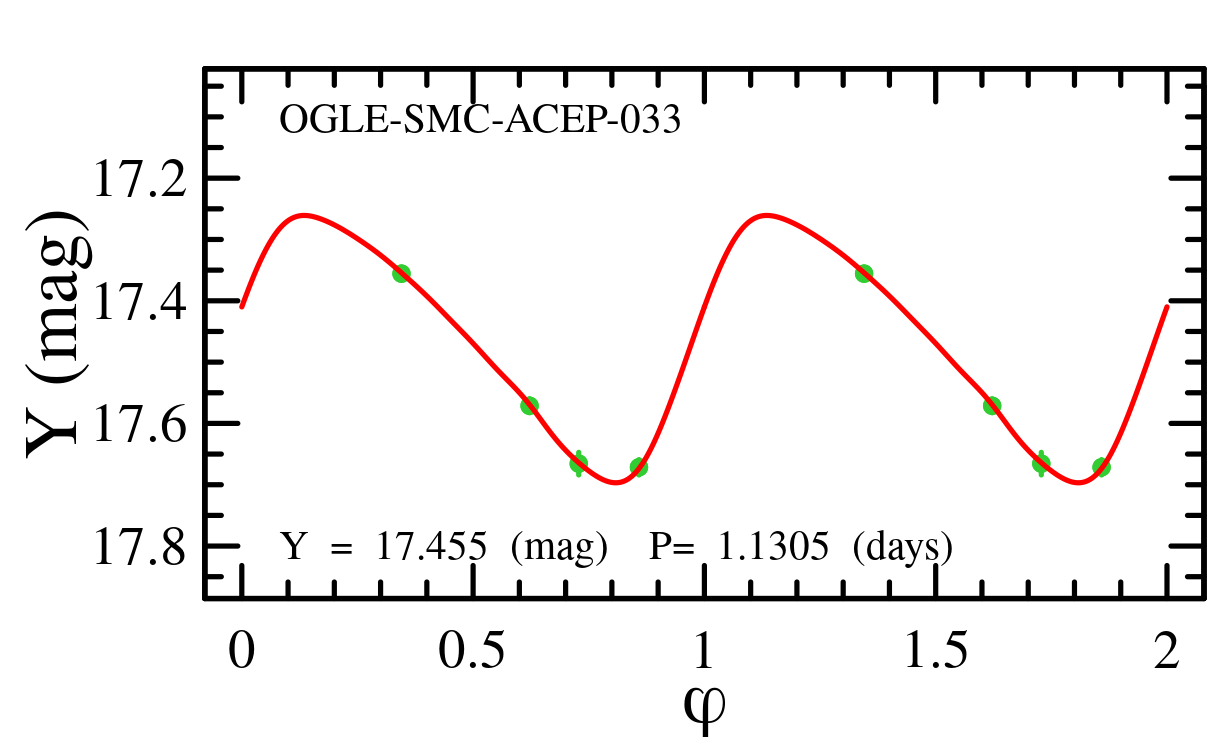}
    \includegraphics[width=0.30\textwidth]{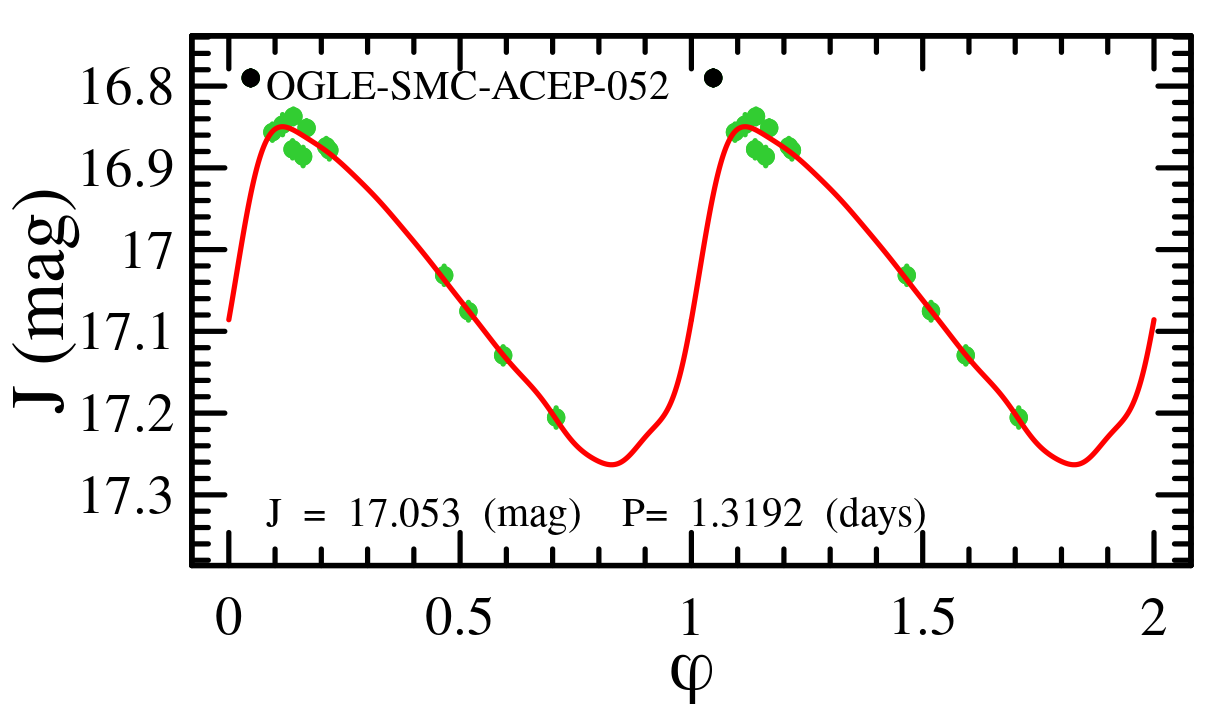}
    \includegraphics[width=0.30\textwidth]{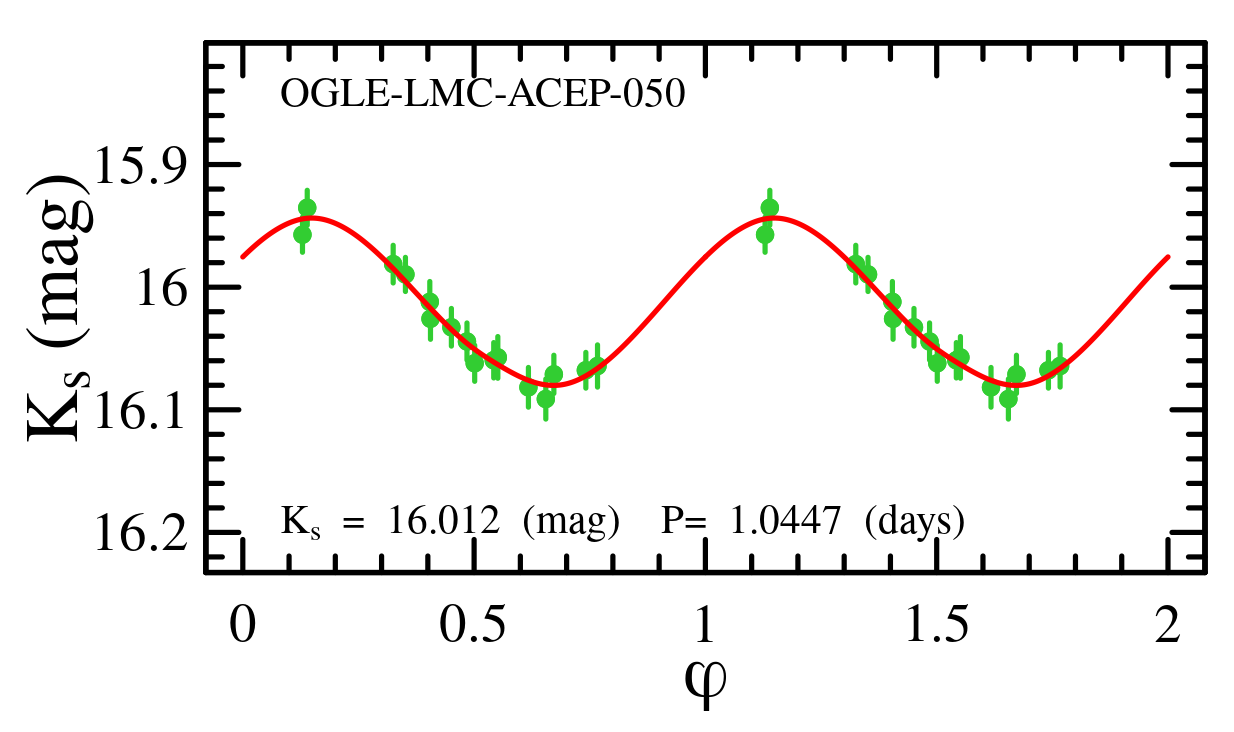}
}
    }
    \caption{Examples of template-fitted AC light curves in the $Y,\,J,\,K_s$ bands (left to right columns). Notes: Green-filled circles are the observations; the red solid line is the best template for the light curve; and black-filled circles are the outliers not used in the fit.  }	
	\label{templatefitex}
    \end{figure*}

The resulting accurate intensity-averaged magnitudes (and amplitudes) in all the bands for the 193 ACs are shown in Table~\ref{dativmc}, while examples of fitted light curves are shown in Fig.~\ref{templatefitex}. 

\begin{sidewaystable}
\footnotesize\setlength{\tabcolsep}{3pt}
\caption{VMC photometric parameters for all the 198 LMC and SMC ACs analysed in this paper. Columns: (1) Identification from OGLE\, IV or $Gaia$; (2)--(3) Ra and Dec at Epoch=J2016.0 if Source=Gaia and Epoch=J2000.0 if Source=OGLE;  (4)  Mode; (5) Period; (6)--(7) Intensity-averaged magnitude in $Y$ and relative uncertainty; (8)--(9) Peak-to-peak amplitude in $Y$ and relative uncertainty; (10) to (13) as for columns (6) to (9) but for the $J$ band; (14) to (17) As for column (6) to (9) but for the $K_\mathrm{s}$ band; (18) $E(V-I)$ values adopted in this work; (19) flag indicating the origin of the reddening value; (20) flag indicating the source (OGLE IV or $Gaia$ DR3) of period, identification and epoch of maximum. The full table is available electronically.}
\label{dativmc}
\begin{tabular}{lcclccccccccccccccccc}
\hline  
\noalign{\smallskip}   
       ID     &  Ra & Dec & Mode & $P$ & $\langle Y \rangle$ & $\sigma_{\langle Y \rangle}$& A($Y$) &  $\sigma_{{\rm A}(Y)}$  & $\langle J \rangle$ & $\sigma_{\langle J \rangle}$&   A($J$) &  $\sigma_{{\rm A}(J)}$  & $\langle K_\mathrm{s} \rangle$ & $\sigma_{\langle K_\mathrm{s} \rangle}$&   A($K_\mathrm{s}$) &  $\sigma_{{\rm A}(K_\mathrm{s})}$  & $E_{(V-I)}$ & $f_{E(V-I)}$  & Source \\
   & deg & deg &  & days & mag &mag & mag & mag  & mag & mag & mag & mag &   mag & mag &  mag & mag &     \\                     
(1)  & (2)  & (3) & (4) & (5) & (6) & (7) &(8) & (9) & (10) & (11) & (12)  & (13) & (14) & (15) &(16)  & (17) & (18) & (19) & (20)  \\                     
\noalign{\smallskip}
\hline  
\noalign{\smallskip}
  OGLE-SMC-ACEP-068 & 15.94072 & $-$73.88605 & 1O & 0.521 & 18.178 & 0.002 & 0.3604 & 0.008 & 18.147 & 0.004 & 0.215 & 0.017 & 17.914 & 0.028 & 0.120 & 0.054 & 0.062 & 1 & OGLE\\
  OGLE-SMC-ACEP-054 & 13.41347 & $-$72.32300 & 1O & 0.529 & 18.111 & 0.0280 & 0.473 & 0.062 & 17.910 & 0.006 & 0.409 & 0.015 & 17.644 & 0.007 & 0.196 & 0.023 & 0.044& 1 & OGLE\\
  DR3\_5278444889019947136 & 94.71859 & $-$70.86541 & F & 0.537 & 17.389 & 0.058 & 0.640 & 0.139 & 17.260 & 0.010 & 0.229 & 0.033 & 16.958 & 0.015 & 0.237 & 0.041 & 0.096& 1 & Gaia\\
  OGLE-LMC-ACEP-124 & 86.06129 & $-$67.24496 & 1O & 0.539 & 17.436 & 0.000 & 0.310 & 0.000 & 17.270 & 0.005 & 0.271 & 0.016 & 17.034 & 0.008 & 0.174 & 0.039 & 0.061& 1 & OGLE\\
  OGLE-SMC-ACEP-051 & 12.82191 & $-$72.94866 & 1O & 0.544 & 17.762 & 0.009 & 0.552 & 0.025 & 17.534 & 0.038 & 0.407 & 0.045 & 17.436 & 0.009 & 0.220 & 0.064 & 0.034& 1 & OGLE\\
\noalign{\smallskip}
\hline  
\noalign{\smallskip}
\end{tabular}
\end{sidewaystable}

\subsection{Reddening estimates for the target stars}
We corrected interstellar extinction using the reddening maps published by \citet{2021ApJS..252...23S}. These maps have a spatial resolution from 27" $\times$ 27" in the peripheries to 1\farcs7 $\times$ 1\farcs7 in the centers of the MCs. Since these maps do not cover the entire extent of the LMC and the SMC, they were complemented by \citet[][SFD]{1998ApJ...500..525S} maps with a lower spatial resolution (6\farcs1). Table~\ref{dativmc} reports not only the value of $E(V-I)$ but also a flag "EVI", where 0 or 1 means the use of SFD or \citet{2021ApJS..252...23S} maps, respectively.

The extinction correction requires assumption on the reddening law and $R_V$ value and we note that there are different reddening laws discussed in the literature \citep[e.g.][]{1989ApJ...345..245C,fitzpatrick1999,gordon2003,Wang_2023}. To ensure consistency and comparison with our previous works \citep[][S24]{Ripepi2012,2014ripepianomalous,ripepi2015vmc,ripepi2016vmc,ripepi2022vmc} we decided to adopt the extinction coefficients from \citet{1989ApJ...345..245C} and $R_V = A(V)/(A(B) - A(V)) = 3.23$\footnote{According to \citet[][in their Section 4.1]{Breuval2024}, a change in $R_V$ would affect the results by only $\sim$ 0.002 mag.} For the $Gaia$ bands we used the coefficients published by \citet{2018MNRAS.479L.102C}. based on \citet{2013ApJ...764...84I}. For the Gaia bands we used the coefficients published by \citet{2018MNRAS.479L.102C}. The impact of different reddening laws on the PWJK relations is discussed in the Appendix \ref{diffext}.

The resulting coefficients to be multiplied by $E(V-I)$ to obtain dereddened magnitudes in different bands are: 
\begin{itemize}
    \item in the $G_{BP}$ band 2.678;
   \item in the $V$ band 2.563;
   \item in the $G$ band 2.175;
   \item in the $G_{BP}$ band 1.615;
   \item in the $I$ band 1.564;
   \item in the $Y$ band 1.00;
   \item in the $J$ band 0.744;
   \item in the $K_s$band 0.308.
\end{itemize}

\subsection{Complementary optical data}
\label{opt data}

The VMC NIR data was complemented by optical photometry from the literature. This allowed us to examine changes in the PL relationships across different wavelengths and to build Wesenheit magnitudes by combining optical and infrared bands. Specifically, $V$ and $I$ bands were obtained from the OGLE IV survey, whereas the $G$, $G_{BP}$, and $G_{RP}$ bands were provided by the $Gaia$DR3 \citep[][]{2023A&A...674A..17R}. All optical data are presented in Table~\ref{datiottico}.

As in Table~\ref{dativmc}, the flag 'SOURCE', in Table~\ref{datiottico}, indicates whether the object was identified by OGLE IV or $Gaia$. The flag 'SOS' concerns the derivation of the average magnitudes in the $Gaia$ bands\footnote{SOS stands for Specific Object Studies \citep[see e.g.][]{2023A&A...674A..17R}}: the value 0 indicates that the standard technique was adopted \citep[see e.g.][]{2018A&A...616A...4E} without any specific assumption for pulsating variables; the value 1 means that the averaged-intensity technique through modelling the light curves was adopted \citep[e.g.][]{2016A&A...595A.133C}. 
The 193 ACs with OGLE IV identification have at least one magnitude collected by $Gaia$: 42 and 152 with flag 'SOS' = 0 and 1, respectively. 
The flag `VI', in Table~\ref{datiottico}, indicates the origin of the $V\, I$ magnitudes. In the OGLE IV catalog all the stars have the average $I$ band magnitudes, while not all have the $V$ measurement. Moreover, the $V,\, I$ data is completely missing for the 6 stars originating from the $Gaia$ catalog and not present in OGLE IV. The missing values for the $V,\, I$ bands were recovered through the photometric transformations between Johnson and $Gaia$ filters provided by \citet[][hereinafter P22]{2022A&A...664A.109P}. %\citet{2024trentin} has tested the accuracy of these transformations. The calculated $V$ magnitudes are accurate within 0.01 mag, while the $I$ magnitudes are too bright by 0.03 mag. As a consequence, we added to the transformed $I$ magnitudes an offset of 0.03 mag. 
The stars with photometry from the OGLE IV survey have the flag 'VI' equal to OGLE,OGLE. When the $V$ is estimated from $Gaia$ due to missing OGLE data, 'VI' flag is 'P22;OGLE'. For the stars identified only by $Gaia$ the 'VI' flag is equal to 'P22;P22'. 
The $V\,I$ magnitudes estimated from the $Gaia$ bands have lower quality if the flag 'SOS' is 0. 
%In any case, the effect of the adoption of the standard averaged magnitudes is expected to be contained in a few percent errors \citep[see discussions on this point in ][]{2022MNRAS.512..563R,2023A&A...674A...1G}.\\

\begin{sidewaystable}
\footnotesize\setlength{\tabcolsep}{3pt} \caption{Optical photometric parameters for all the 198 LMC and SMC ACs analyzed in this paper. Columns: (1) Identification from OGLE\,IV or $Gaia$\,DR3; (2)--(3) Ra and Dec at Epoch=J2016.0 if Source=Gaia and Epoch=J2000.0 if Source=OGLE;  (4) Mode; (5) Period; (6) period error; (7) magnitude in $V$ band from OGLE\,IV; (8) Magnitude in $I$ band from OGLE\,IV; (9) epoch; (10) amplitude in the $I$ band; (11)--(12) Magnitude in $G$ and relative uncertainty; (13)--(14) as for columns (11) and (12) but for the $G_{BP}$; (15)--(16) As for columns (11) and (12) but for the $G_{RP}$; (17) flag indicating the source used for identification; (18) flag indicating how the $Gaia$ magnitudes have been calculated; (19) flag indicating what is the origin of the $V\,I$ magnitude. The full table is available electronically.}
\label{datiottico}
\begin{tabular}{lcccccccccccccccccc}
\hline  
\noalign{\smallskip}   
       ID     &  Ra & Dec & Mode & $P$ &  $\sigma_{{\rm P}}$ & $\langle V \rangle$ &  $\langle I \rangle$ &$epoch_I$  & A(I) & $\langle G \rangle$ & $\sigma_{\langle G \rangle}$& $\langle G_{BP} \rangle$ & $\sigma_{\langle G_{BP} \rangle}$ & $\langle G_{RP} \rangle$ & $\sigma_{\langle G_{RP} \rangle}$ & Source & SOS & VI  \\
   & deg & deg &  & days & days &mag & mag & days &  mag & mag & mag & mag & mag &  mag & mag &   &  &     \\                     
(1)  & (2)  & (3) & (4) & (5) & (6) & (7) &(8) & (9) & (10) & (11) & (12)  & (13) & (14) & (15) &(16)  & (17) &(18) & (19)  \\                     
\noalign{\smallskip}
\hline  
\noalign{\smallskip}  
  OGLE-LMC-ACEP-001 & 69.32904 & $-$69.81986 & F & 0.850 & 0.000 & 18.648 & 18.038 & 6000.634 & 0.673 & 18.614 & 0.014 & 18.862 & 0.053 & 18.197 & 0.037 & OGLE & 0 & OGLE,OGLE\\
  OGLE-LMC-ACEP-054 & 82.77800 & $-$68.37494 & F & 0.980 & 0.000 & 99.999 & 17.892 & 6000.146 & 0.729 & 18.696 & 0.014 & 19.111 & 0.052 & 18.017 & 0.047 & OGLE & 0 & P22,OGLE\\
  OGLE-SMC-ACEP-072 & 16.67117 & $-$74.69983 & F & 1.234 & 0.000 & 18.068 & 17.525 & 6000.777 & 0.441 & 18.155 & 0.017 & 99.999 & 99.999 & 99.999 & 99.999 & OGLE & 0 & OGLE,OGLE\\
  DR3\_5278444889019947136 & 94.71856 & $-$70.86540 & F & 0.537 & 0.000 & 99.999 & 99.999 & 1694.610 & 99.999 & 18.094 & 0.020 & 18.342 & 0.033 & 17.740 & 0.017 & Gaia & 1 & P22,P22\\
  OGLE-LMC-ACEP-002 & 70.48429 & $-$66.86642 & F & 0.977 & 0.000 & 18.24 & 17.611 & 6000.317 & 0.514 & 18.121 & 0.003 & 18.317 & 0.054 & 17.712 & 0.010 & OGLE & 1 & OGLE,OGLE\\
\noalign{\smallskip}
\hline  
\noalign{\smallskip}
\end{tabular}
\end{sidewaystable}

\section{Period--Luminosity and Period--Wesenheit relations}

The multi-band photometry allows us to calculate PL and PW relations for the ACs, separately for the LMC and SMC. The list of PW relations considered in this study is the following: 
\begin{itemize}
    \item  PWVI = $V$ $-$ 2.54 ($V - I$) ;
    \item  PWV$K_s$ = $V$ $-$ 1.13 ($V- K_s$) ;
    \item  PWJ$K_s$ = $K_s$ $-$ 0.69 ($J- K_s$); 
  \item PWY$K_s$ = $K_s$ $-$ 0.42 ($Y- K_s$);
  \item PWG = $G$ $-$ 1.90 ($G_{BP} - G_{RP}$);
\end{itemize}
where the coefficient of the PW relation is calculated as $R_{\lambda_1}/R_{\lambda_1}-R_{\lambda_2}$, apart for the coefficients of the PW in the $Gaia$ bands that is $1.90$ according to \citet{2019A&A...625A..14R}.
\begin{figure} 
    \vbox{
    \includegraphics[width=0.5\textwidth]{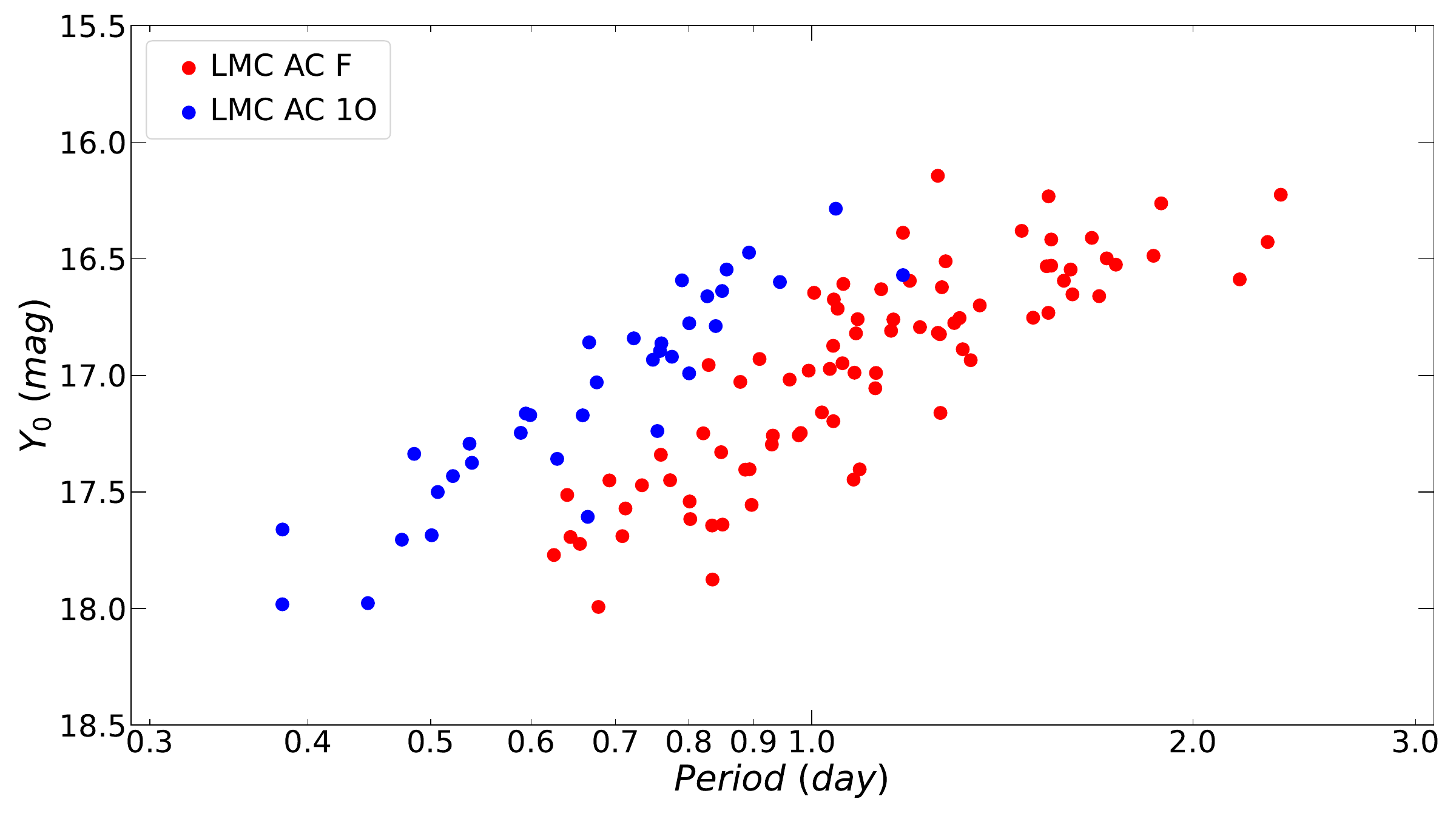}
    \includegraphics[width=0.5\textwidth]{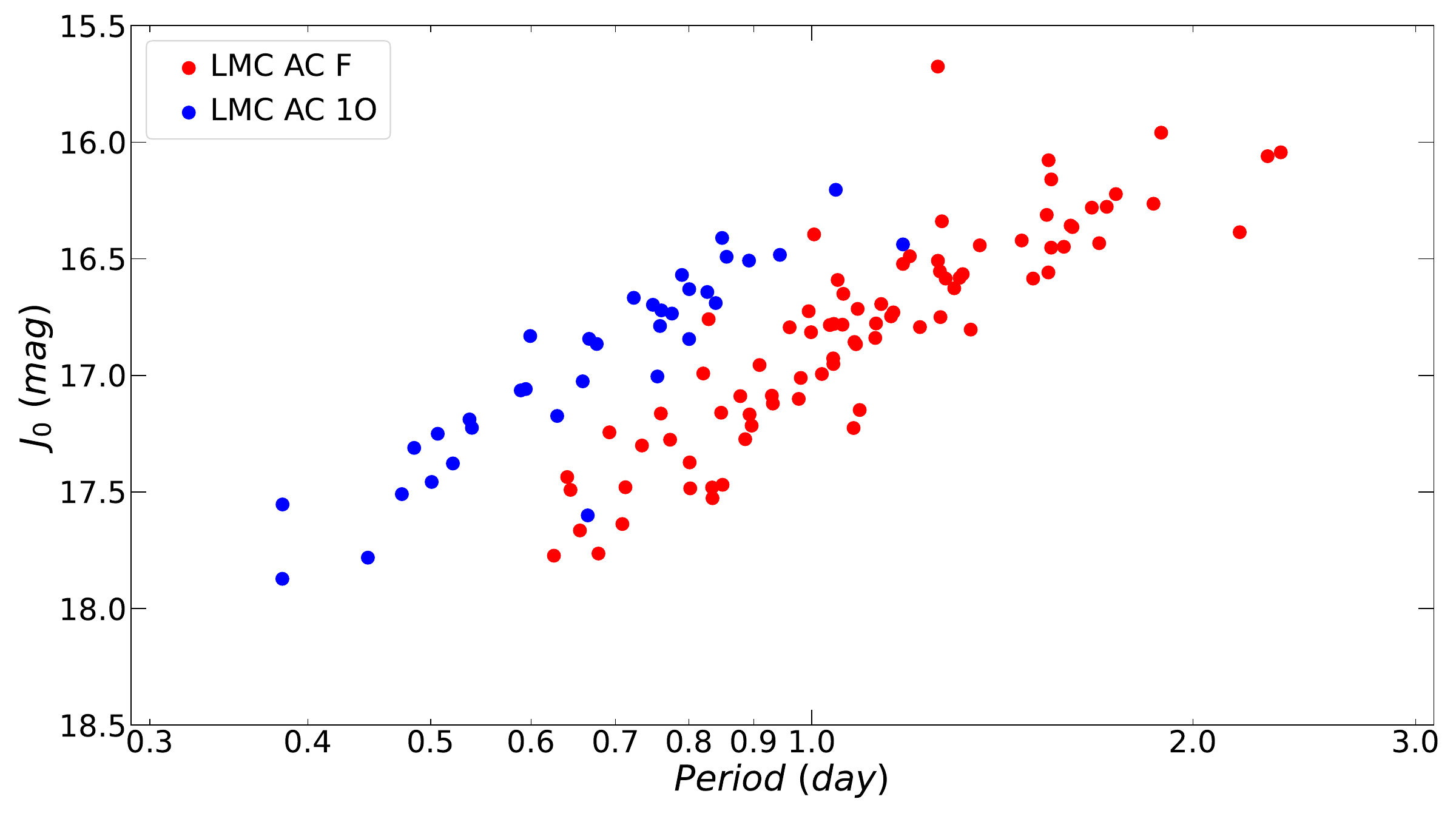}
    \includegraphics[width=0.5\textwidth]{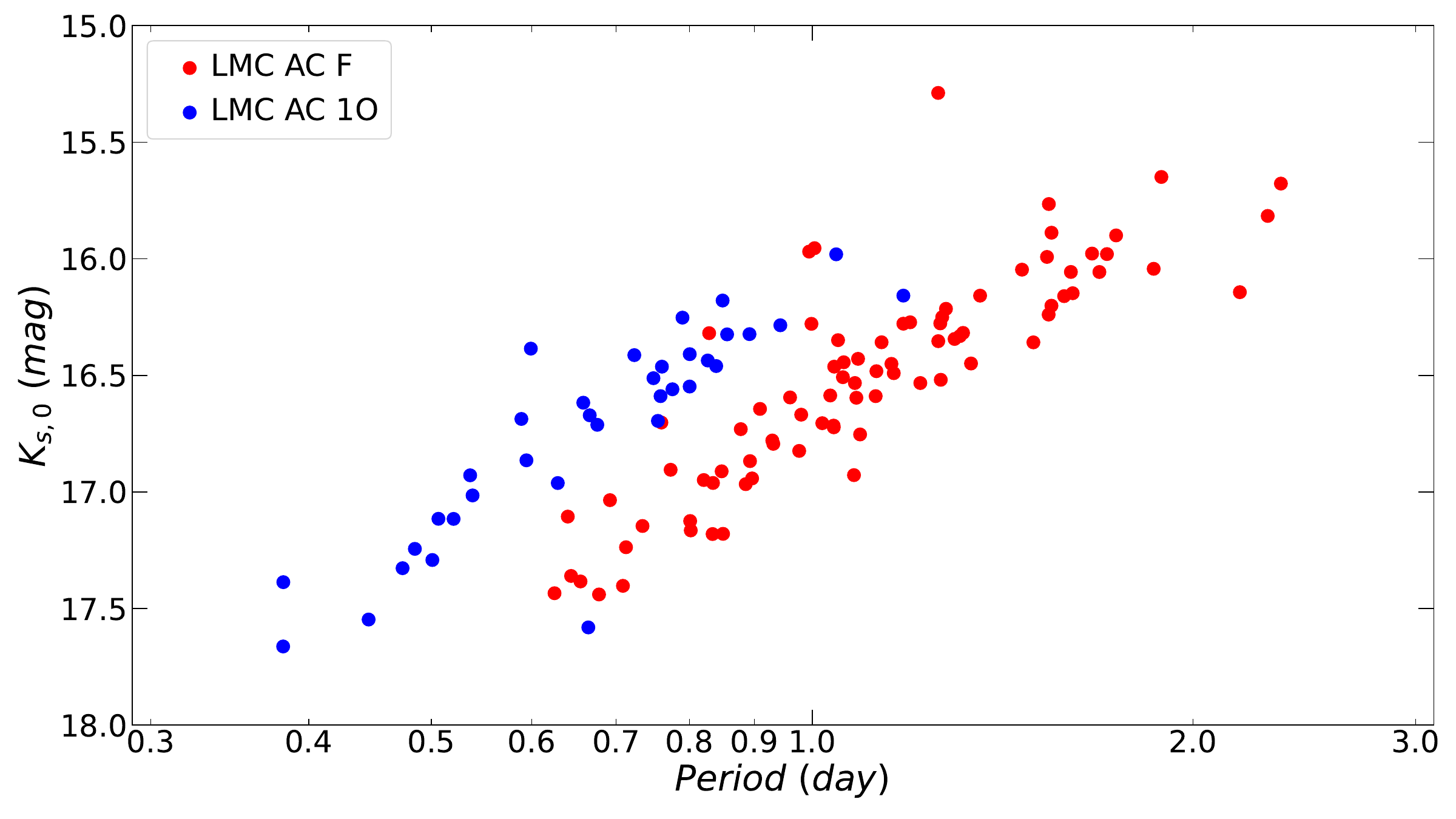}
    }
    \caption{Period--luminosity relations in the NIR bands for the ACs in the LMC. Red and blue filled circles represent F and 1O mode ACs, respectively. We note that the error bars are smaller than the size of the dots.}	
	\label{pllmc} 
    \end{figure}

\subsection{PL and PW relations derivation}
\label{ltsfit}
The observed quantities, namely the dereddened intensity averaged magnitudes and the periods, were fitted with the following linear relations to determine the coefficients of the PL/PW relations: 

\begin{eqnarray}
m_{\lambda_0} & = &\alpha +\beta\cdot \log P ~~~~~~~~~~~~~~~~~~~~~~~~~~~~~~~~~~~~~PL \\
w(\lambda_1,\lambda_2) &= &\alpha + \beta \cdot \log P ~~~~~~~~~~~~~~~~~~~~~~~~~~~~~~~~~~~~~PW 
\end{eqnarray}

To carry out these linear fits in one dimension, the {\tt Python} code LTS \citep[Least Trimmed Squares,][]{2013MNRAS.432.1709C} was adopted. 
This method makes the fit converge to the correct solution even in the presence of numerous outliers, where the much simpler $\sigma$-clipping approach can converge to the wrong solution.

\subsection{Results for the PL/PW relations}

%The PL relations were obtained in all the bands listed in the first column of Table~\ref{ass}. As for the Wesenheit magnitudes, we calculated the PW relations for the five quantities defined in Table~\ref{tab:coeffwes}, where the coefficient that multiplies the color term was obtained from the wavelength-dependent absorption values listed in the second column of Table~\ref{ass}. 
%As for the $Gaia$ bands, which have a wider bandwidth compared to Johnson bands, we used the empirically-based Wesenheit coefficient equal to 1.90 according to \citet{2019A&A...625A..14R}. 
%As for the $PLC$, we considered the same magnitude-colour combinations which are at the base of the Wesenheit magnitudes shown in Table~\ref{tab:coeffwes}, where, obviously, the colour coefficients are not fixed but free to vary.
Fig~\ref{pllmc} shows the observational data used for deriving the PL relations for the LMC in the NIR bands. Likewise, all the data available for computing the PL and the PW relations in the LMC and SMC are available in Fig.~\ref{plsmc} in the Appendix~\ref{figure}. Throughout these figures, the two types of pulsators are distinguished using different colours. A visual examination of the figures points out several distinct features that will be quantitatively validated later:

\begin{itemize}

\item For the LMC PL relations, the dispersion of the data decreases from the optical to NIR bands. As expected, this trend is not observed in the SMC PL relations because the dispersion is dominated by the large depth of the galaxy along the line of sight  \citep[e.g.][]{2015subramanian, Ripepi2017vmc}. 

\item The PW relations are in all cases tighter than the PLs' \citep[see e.g.][]{2022ApJS..262...25D}, except for the  PL relations in the $J$ and $K_s$ bands, which in the LMC have dispersion comparable to that of the PW relations.

\item Both in the LMC and in the SMC, the slopes followed by the F and the 1O pulsators appear the same, while the respective zero points differ, as expected.

%\item Despite the position of the SMC with respect to our line of sight, the apparent dispersion of the PL/PW is an intrinsic dispersion and not wavelength-dependent. 

\end{itemize}

\begin{figure*}
    \vbox{
    \hbox{
    \includegraphics[width=0.5\textwidth]{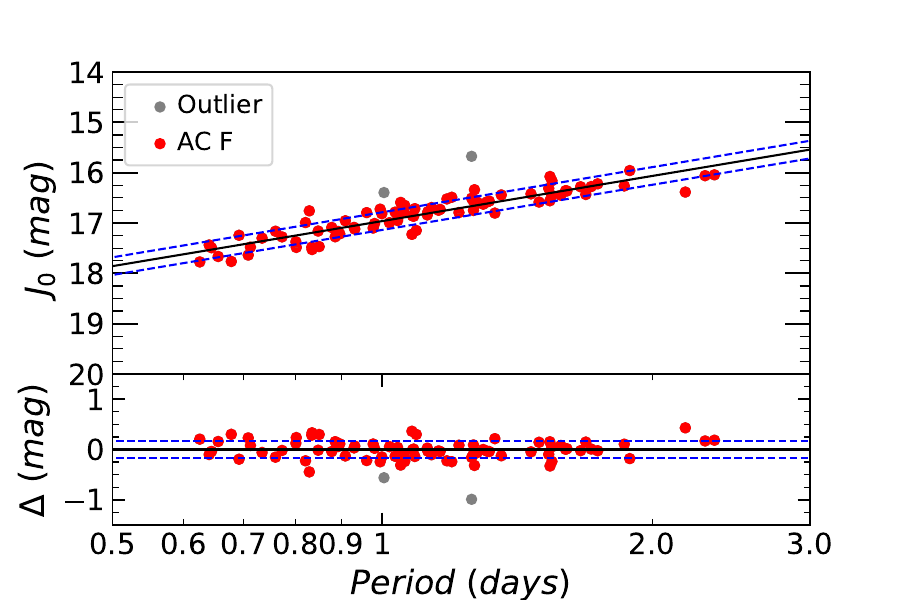}
    \includegraphics[width=0.5\textwidth]{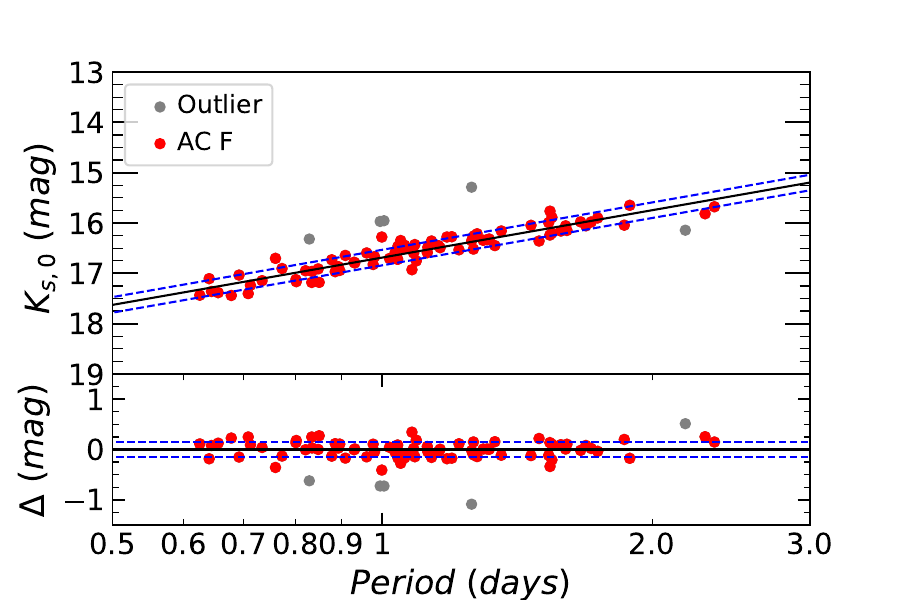}
    }
    \hbox{
    \includegraphics[width=0.5\textwidth]{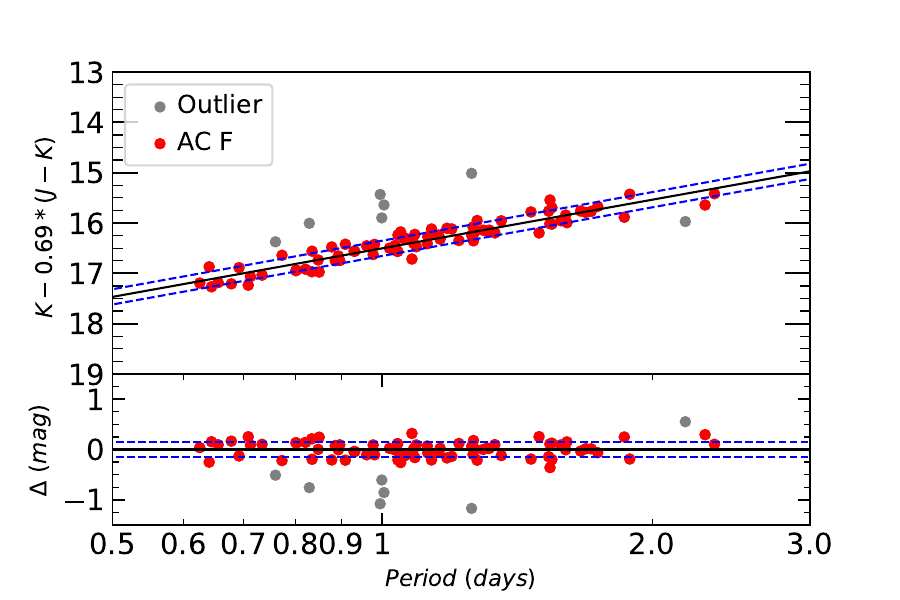}
    \includegraphics[width=0.5\textwidth]{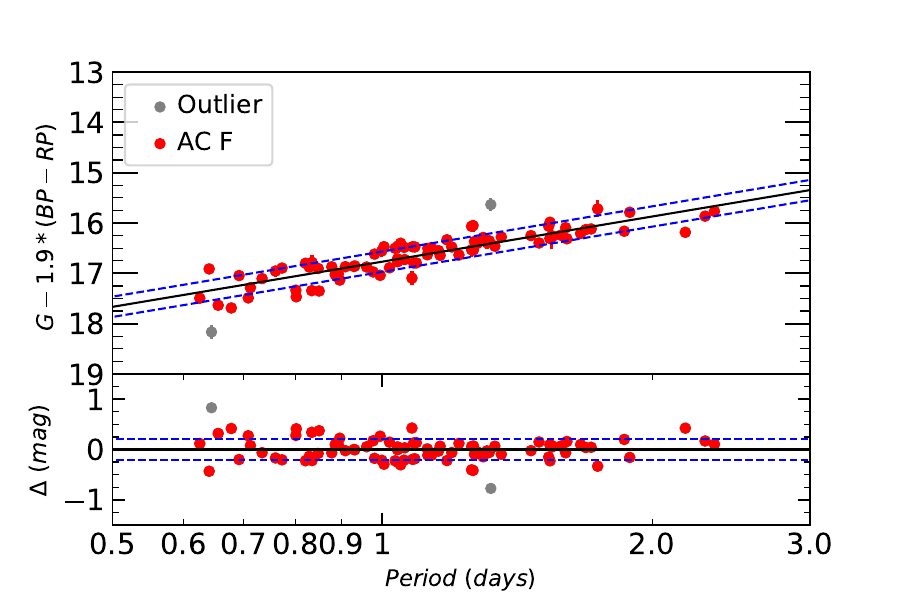}
    }
    }
    \caption{ Example best fits for the F ACs in the LMC. Left to right we show in the top panels the $PLJ$ and the $PLK_s$ relations, and I the bottom panels the $PWJK_s$ and the $PWG$ relations, respectively. The red and grey filled circles are the data used for the fit and the outliers. The solid black line is the best fit to the data, while the dashed blue lines show the $\pm 1 \sigma$ levels. Upper panles and lower panels in each figure contain the fits and the residuals of the fit, respectively. The fits for the other bands are in the Appendix~\ref{appfigure}. }	
	\label{esempifit2}
    \end{figure*}

We have derived the coefficients of the PL and PW relations for F and 1O ACs, first separately, to quantify the differences in the slopes between the different modes of pulsation, and then together (see Sect.~\ref{fundsect}) to assess the possibility of combining the two sub-samples. Fig.~\ref{esempifit2} displays selected tight PL and PW relations for the F mode AC in the LMC. The fit results for all the considered PL and PW relations are reported in Table~\ref{tab:lmcsmcAC}, while the figures showing the fitted PL/PW relations are in Appendix~\ref{appfigure} (Fig.~\ref{fit1lmc} and Fig.~\ref{fit1smc}) for the LMC and the SMC, respectively.
In the subsequent analyses, we will use the relations that present the smallest dispersion: PL relations in the $J$ and $K_s$ bands and the PW relations in $K_s, J-K_s$ and $G, G_{BP}-G_{RP}$. These are highlighted in Table~\ref{tab:lmcsmcAC}.

\begin{table*}
%\begin{adjustwidth}{-2cm}{}
\footnotesize\setlength{\tabcolsep}{3pt} 
\caption{Coefficients of the PL and PW relations for ACs in the LMC and the SMC. }
  \label{tab:lmcsmcAC}
\begin{center}
\begin{tabular}{ll|llllll|llllll}
 \hline  
 \noalign{\smallskip} 
       \multicolumn{1}{c}{Rel.}   &\multicolumn{1}{c|}{Mode} &\multicolumn{1}{|c}{ $\alpha $} &  \multicolumn{1}{c}{$\sigma_{\alpha}$} & \multicolumn{1}{c}{$\beta$} & \multicolumn{1}{c}{$\sigma_{\beta}$} & \multicolumn{1}{c}{RMS} & \multicolumn{1}{c|}{s/o}& \multicolumn{1}{|c}{$\alpha$} &  \multicolumn{1}{c}{$\sigma_{\alpha}$} & \multicolumn{1}{c}{$\beta$} & \multicolumn{1}{c}{$\sigma_{\beta}$} & \multicolumn{1}{c}{RMS}& \multicolumn{1}{c}{s/o} \\
   &  & mag & mag &  mag/dex & mag/dex  &mag && mag & mag &  mag/dex   &  mag/dex &   mag & \\                     
(1)  & (2)  & (3) & (4) & (5) & (6) & (7) &(8) &(9)& (10)& (11)  & (12)  & (13) & (14) \\   
 \hline  
 & & 
\multicolumn{6}{|c|}{LMC} & 
\multicolumn{6}{|c}{SMC} \\
\hline
PLI   & F & 17.330 & 0.028 & $-$3.01 & 0.15 & 0.171 & 79/5   & 17.729 & 0.024 & $-$2.70 & 0.16 & 0.128 & 45/3\\
  PLI   & 1O & 16.706 & 0.032 & $-$3.28 & 0.24 & 0.166 & 34/1   & 17.152 & 0.043 & $-$3.81 & 0.41 & 0.212 & 28/0\\
  PLI   & All & 17.287 & 0.026 & $-$2.95 & 0.13 & 0.179 & 113/6   & 17.737 & 0.031 & $-$2.97 & 0.17 & 0.181 & 75/1\\
 PLV   & F & 17.801 & 0.035 & $-$2.82 & 0.19 & 0.223 & 80/4   & 18.214 & 0.029 & $-$2.43 & 0.19 & 0.155 & 46/2\\
  PLV   & 1O & 17.096 & 0.041 & $-$3.31 & 0.32 & 0.221 & 35/0   & 17.568 & 0.054 & $-$3.78 & 0.52 & 0.266 & 28/0\\
  PLV   & All & 17.735 & 0.035 & $-$2.74 & 0.17 & 0.242 & 115/4   & 18.193 & 0.040 & $-$2.65 & 0.23 & 0.235 & 76/0\\
   PLG   & F & 17.784 & 0.037 & $-$2.99 & 0.20 & 0.238 & 83/1   & 18.148 & 0.030 & $-$2.60 & 0.20 & 0.158 & 46/2\\
  PLG   & 1O & 17.054 & 0.038 & $-$3.27 & 0.29 & 0.202 & 34/1   & 17.486 & 0.049 & $-$3.85 & 0.47 & 0.240 & 27/1\\
  PLG   & All & 17.708 & 0.036 & $-$2.85 & 0.18 & 0.253 & 117/2   & 18.125 & 0.038 & $-$2.80 & 0.21 & 0.219 & 75/1\\
  PLBP   & F & 17.871 & 0.041 & $-$2.75 & 0.22 & 0.266 & 82/2   & 18.288 & 0.034 & $-$2.37 & 0.23 & 0.179 & 47/1\\
  PLBP   & 1O & 17.201 & 0.048 & $-$3.05 & 0.38 & 0.258 & 35/0   & 17.652 & 0.047 & $-$3.28 & 0.46 & 0.224 & 25/1\\
  PLBP   & All & 17.812 & 0.041 & $-$2.62 & 0.20 & 0.288 & 118/1   & 18.239 & 0.041 & $-$2.46 & 0.23 & 0.237 & 74/0\\
  PLRP   & F & 17.356 & 0.034 & $-$2.78 & 0.18 & 0.214 & 81/3   & 17.763 & 0.030 & $-$2.66 & 0.20 & 0.158 & 47/1\\
  PLRP   & 1O & 16.758 & 0.036 & $-$3.13 & 0.29 & 0.195 & 35/0   & 17.196 & 0.043 & $-$3.55 & 0.44 & 0.208 & 26/0\\
  PLRP   & All & 17.313 & 0.030 & $-$2.76 & 0.15 & 0.208 & 115/4   & 17.742 & 0.034 & $-$2.76 & 0.19 & 0.194 & 74/0\\
  PLY   & F & 17.109 & 0.036 & $-$2.80 & 0.19 & 0.226 & 81/1   & 17.475 & 0.037 & $-$2.62 & 0.23 & 0.192 & 48/0\\
  PLY   & 1O & 16.415 & 0.026 & $-$3.61 & 0.20 & 0.133 & 32/3   & 16.962 & 0.043 & $-$3.55 & 0.44 & 0.210 & 26/0\\
  PLY   & All & 17.074 & 0.030 & $-$2.82 & 0.15 & 0.210 & 115/2   & 17.487 & 0.035 & $-$2.84 & 0.20 & 0.201 & 74/0\\
   \rowcolor{lightgray}
   PLJ   & F & 16.962 & 0.028 & $-$2.98 & 0.14 & 0.176 & 81/2   & 17.357 & 0.030 & $-$2.75 & 0.20 & 0.159 & 47/1\\
  PLJ   & 1O & 16.370 & 0.025 & $-$3.25 & 0.20 & 0.134 & 34/1   & 16.839 & 0.041 & $-$3.70 & 0.42 & 0.201 & 26/0\\
  \rowcolor{LightBlue}
  PLJ   & All & 16.924 & 0.025 & $-$2.95 & 0.12 & 0.174 & 116/2   & 17.380 & 0.031 & $-$3.01 & 0.18 & 0.178 & 73/1\\
  \rowcolor{lightgray}
  PLK   & F & 16.686 & 0.025 & $-$3.12 & 0.13 & 0.155 & 78/5   & 17.073 & 0.032 & $-$2.92 & 0.21 & 0.171 & 47/1\\
  PLK   & 1O & 16.126 & 0.024 & $-$3.41 & 0.18 & 0.127 & 33/2   & 16.590 & 0.046 & $-$3.75 & 0.44 & 0.223 & 27/0\\
  \rowcolor{LightBlue}
  PLK   & All & 16.667 & 0.022 & $-$3.14 & 0.11 & 0.152 & 112/6   & 17.09 & 0.03 & $-$3.08 & 0.17 & 0.174 & 72/3\\
  PWVI   & F & 16.601 & 0.024 & $-$3.14 & 0.13 & 0.148 & 80/4   & 16.942 & 0.026 & $-$3.04 & 0.17 & 0.121 & 44/4\\
  PWVI   & 1O & 16.240 & 0.015 & $-$2.61 & 0.11 & 0.041 & 19/16   & 16.379 & 0.023 & $-$4.29 & 0.23 & 0.079 & 22/6\\
  PWVI   & All & 16.589 & 0.021 & $-$3.07 & 0.11 & 0.117 & 112/7   & 17.02 & 0.032 & $-$3.41 & 0.18 & 0.164 & 75/1\\
   \rowcolor{lightgray}
   PWG   & F & 16.768 & 0.031 & $-$2.98 & 0.16 & 0.202 & 82/2   & 17.144 & 0.033 & $-$3.18 & 0.21 & 0.158 & 46/2\\
  PWG   & 1O & 16.238 & 0.027 & $-$3.23 & 0.22 & 0.143 & 33/2   & 16.664 & 0.045 & $-$3.84 & 0.45 & 0.213 & 26/0\\
  \rowcolor{LightBlue}
  PWG   & All & 16.759 & 0.027 & $-$3.05 & 0.13 & 0.175 & 116/3   & 17.161 & 0.030 & $-$3.28 & 0.18 & 0.161 & 69/5\\
  PWVK   & F & 16.541 & 0.022 & $-$3.17 & 0.12 & 0.136 & 75/8   & 16.950 & 0.024 & $-$3.14 & 0.16 & 0.126 & 42/6\\
  PWVK   & 1O & 15.991 & 0.026 & $-$3.42 & 0.20 & 0.136 & 33/2   & 16.452 & 0.050 & $-$3.77 & 0.48 & 0.243 & 27/0\\
  PWVK   & All & 16.537 & 0.021 & $-$3.21 & 0.10 & 0.136 & 109/9   & 16.962 & 0.037 & $-$3.25 & 0.21 & 0.216 & 75/0\\
  PWYK   & F & 16.484 & 0.023 & $-$3.30 & 0.12 & 0.141 & 76/6   & 16.888 & 0.026 & $-$3.14 & 0.17 & 0.134 & 43/5\\
  PWYK   & 1O & 15.944 & 0.027 & $-$3.41 & 0.21 & 0.140 & 33/2   & 16.307 & 0.023 & $-$3.78 & 0.26 & 0.093 & 19/7\\
  PWYK   & All & 16.470 & 0.021 & $-$3.29 & 0.10 & 0.139 & 109/8  & 16.915 & 0.038 & $-$3.40 & 0.22 & 0.222 & 74/0\\
 \rowcolor{lightgray}
 PWJK   & F & 16.504 & 0.025 & $-$3.21 & 0.13 & 0.152 & 76/7   & 16.874 & 0.029 & $-$3.04 & 0.19 & 0.148 & 44/4\\
  PWJK   & 1O & 15.951 & 0.028 & $-$3.50 & 0.21 & 0.146 & 33/2   & 16.386 & 0.048 & $-$4.12 & 0.49 & 0.235 & 26/0\\
  \rowcolor{LightBlue}
  PWJK   & All & 16.494 & 0.022 & $-$3.25 & 0.11 & 0.148 & 109/9   & 16.892 & 0.035 & $-$3.21 & 0.20 & 0.200 & 72/2\\
\noalign{\smallskip}
\hline  
\noalign{\smallskip}
\end{tabular}      
\end{center}
\tablefoot{The PL and the PW relations have the form $y=\alpha +\beta\cdot \log x$, where x, and y are the period and magnitude, respectively. The different columns report: $(1)$ the type of relationship and the band of interest; $(2)$ the pulsating mode; $(3)$--$(4)$ and $(9)$--$(10)$ the $\alpha$ coefficient (relative intercept) and relative uncertainty; $(5)$--$(6)$ and $(11)$--$(12)$  the $\beta$ coefficient (slope) and relative uncertainty; $(7)$ and $(13)$ the Root Mean Square (RMS) of the relation; $(8)$ and $(14)$ the number of stars used in the fit and the outliers excluded from the fit, respectively. Grey and cyan shaded rows are PL and PW relations with the smallest dispersion for F mode and F+1O mode ACs, respectively. }
\end{table*}

\subsection{Fundamentalization of first overtone ACs} 
\label{fundsect}
In the previous section, we calculated PL and PW relations for F and 1O ACs separately. However, it is sometimes convenient to determine such relations for both modes of pulsation together, especially when the number of objects is not large. The so-called "fundamentalization" of the 1O periods has been extensively carried out in the literature for CCs \citep[e.g.][]{Feast1997} and RRL \citep[e.g.][]{2015marconi} variables, but not for ACs thus far. 

To fundamentalize the 1O mode AC pulsators in the LMC, considering that they cover a similar period to the RRL pulsators and belong to the same central helium burning evolutionary phase, having ignited Helium in degenerate conditions, we first tried to use the standard value from the literature for 1O RRL stars (RRc type), namely $\log(1/R)=0.127$ \citep[see e.g.][and references therein]{2015marconi}. However, the resulting distribution in the PL and PW planes seems to suggest that a higher correction should be applied to 1O mode ACs with respect to the RRc pulsators.
This difference should be ascribed to the different envelope structures of ACs, which are relatively more massive and expanded than RR Lyrae stars, thus modifying the relation between the mean density and the periodicities of the first two radial modes, as well as their difference.
Indeed the value of $R$ that minimizes the dispersion of the PL and PW relations for the combined F+1O mode AC sample in the LMC is $\log(1/R)=0.145$. Since the fundamentalization of the 1O mode pulsators in the PL and PW relations is independent of temperatures and luminosities \citep[see e.g.][]{pilecki2024fundamentalization}, the $R$ value is first obtained using the PL relations in the NIR bands, which are the tightest ones, and then verified in all other PL and PW relations. In the following analysis, we use this revised value of  $\log~(1/R)$ to generate a collective sample of F+10 mode AC pulsators.

%We used the LMC ACs data to calculate the period ratio $R=P_{1O}/P_{F}$ for the fundamentalization of 1O mode periods. Since ACs span a small range of periods similar to RRL variables, we first tried to use the standard value from the literature for 1O RR Lyrae stars (RRc type), namely $log(1/R)=0.127$ \citep[see e.g.][and references therein]{2015marconi}. We found that the adoption of this value leads to discontinuities between F and 1O in the PL/PW planes. Therefore, we searched for a new value of $R$ that minimizes the dispersion of the PW relations for the combined F+1O AC sample in the LMC. The procedure started using the PL relations in the NIR bands, which are the tightest ones. The value of $R$ obtained in this way was then verified on all other PL and PW relations. At the end of this procedure, we found that the optimal value to fundamentalize 10 AC periods is $R=0.716$ ($\log(1/R)=0.145$). In the following analysis, we use this value to generate a collective sample of F+10 AC pulsators. 

\begin{figure*}
    \vbox{
%    \hbox{
%    \includegraphics[width=0.45\textwidth]{PLI_lmc.pdf}
 %   \includegraphics[width=0.45\textwidth]{PLG_lmc.pdf}
%    }    
    \hbox{
    \includegraphics[width=0.5\textwidth]{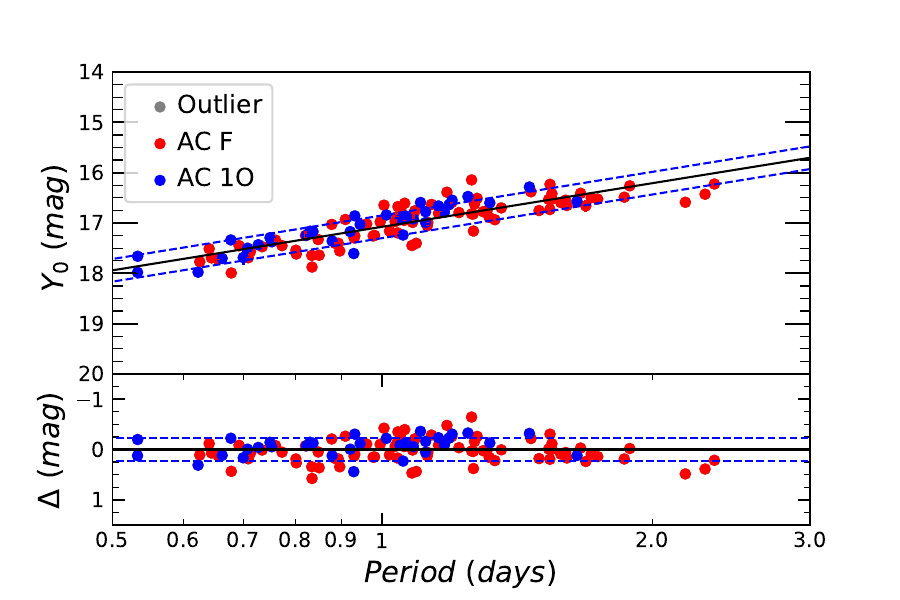}
    \includegraphics[width=0.5\textwidth]{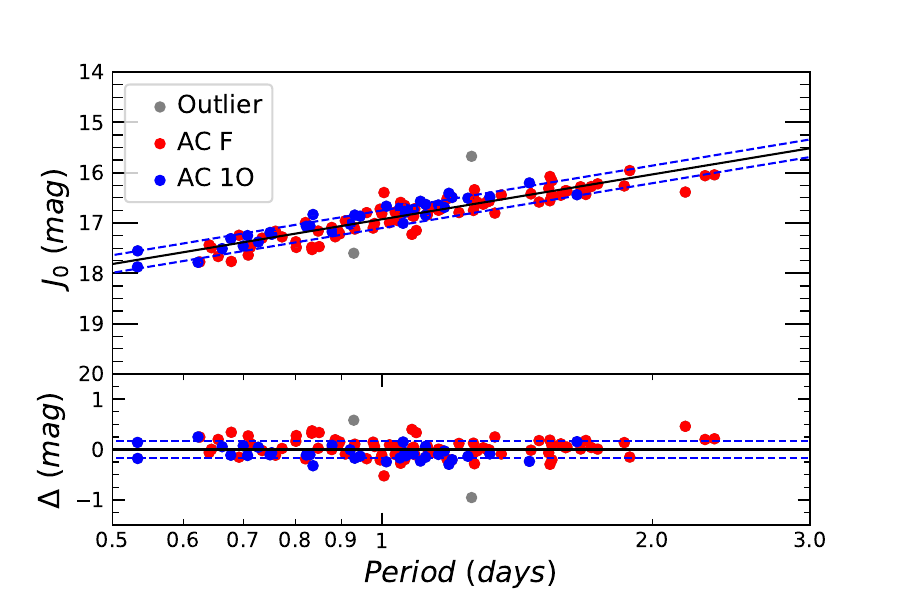}
    }
    \hbox{
    \includegraphics[width=0.5\textwidth]{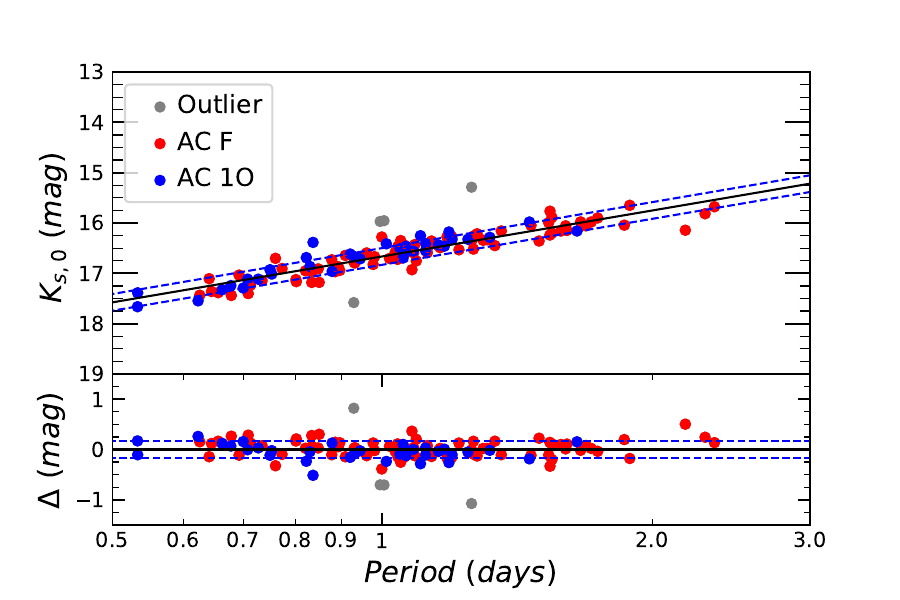}
    \includegraphics[width=0.5\textwidth]{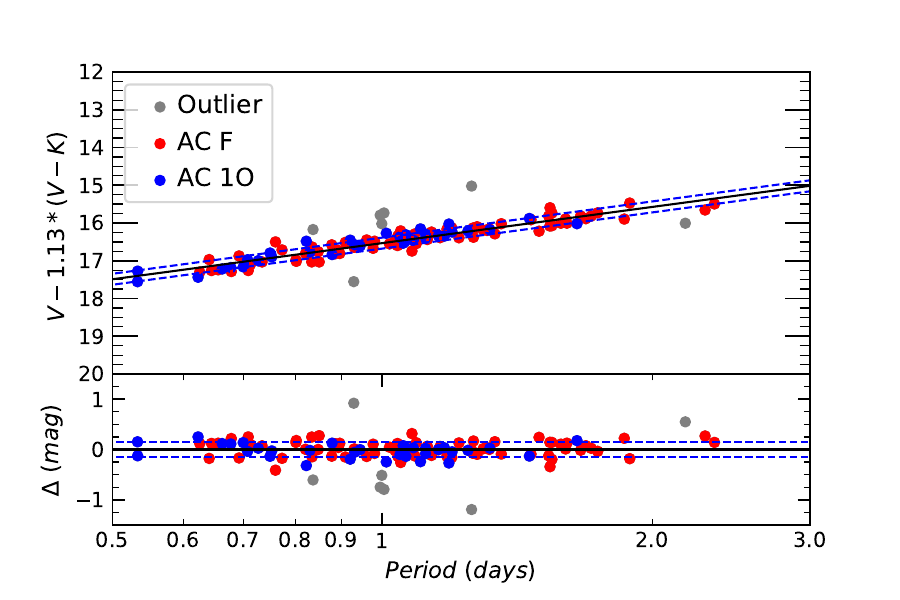}
    
    }
    \hbox{
    \includegraphics[width=0.5\textwidth]{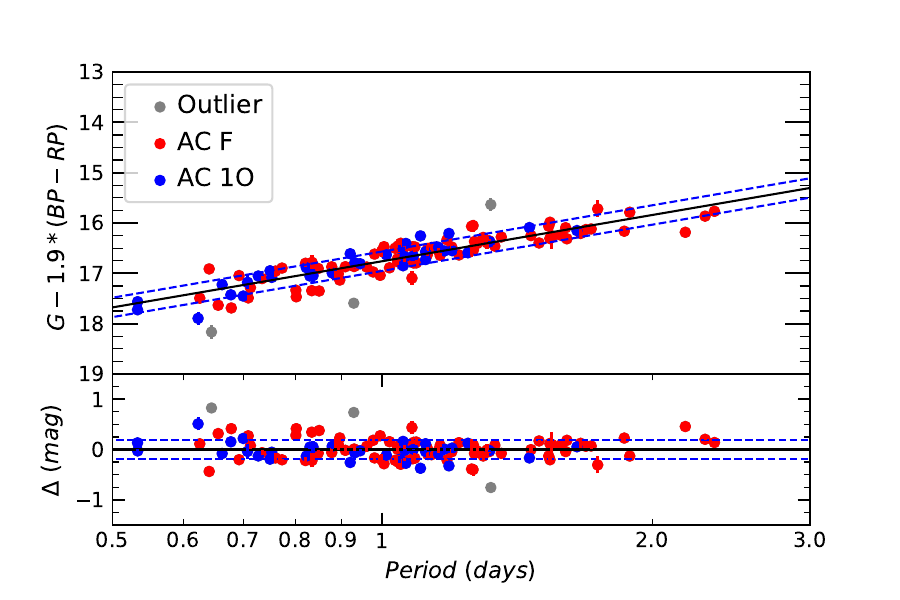}
    \includegraphics[width=0.5\textwidth]{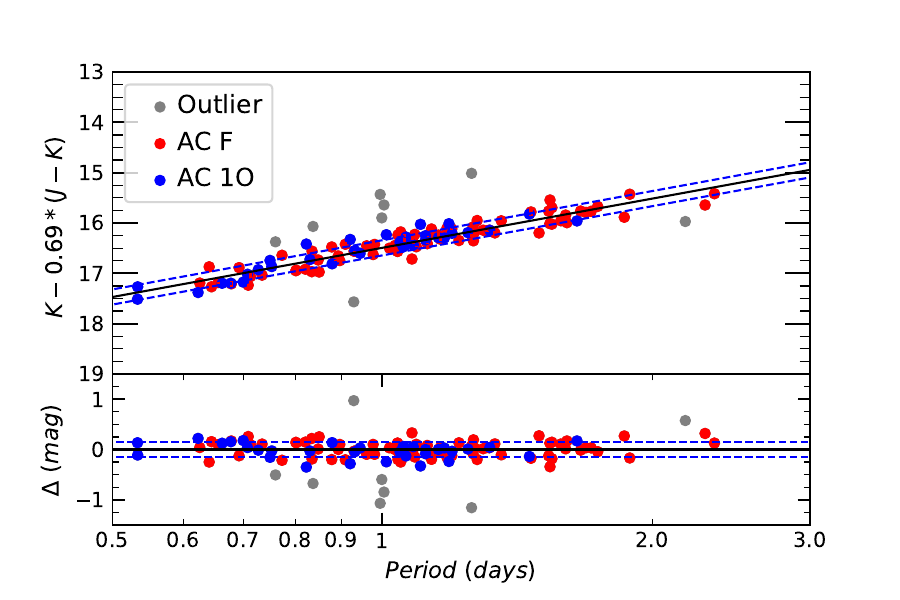}
    }
    }
    \caption{Example of fits for the F and 1O ACs in the LMC, after 1O ACs were fundametalized by adding 0.145 to the logarithm of their periods. Each row contains two figures, displayed in the left and right panels, respectively, in the following order: $PLY$, $PLJ$, $PLK_s$, $PWVI$, $PWVK_s$, $PWG$, $PWYK$, and $PWJK_s$. For each figure, the upper sub-panel shows the best-fitting line, while the lower sub-panel reports the fit residuals. Red and blue-filled circles represent F and 1O mode ACs respectively. The grey-filled circles are outliers objects not included in the fit. The solid black line is the best fit to the data, while the dashed blue lines show the $\pm 1 \sigma$ levels. The best fits in additional bands can be found in Appendix~\ref{appfigure}}.	
    \label{fund}
\end{figure*}
Fig.~\ref{fund} displays examples of the selected PL/PW relations fits for the combined F+1O sample after fundamentalization of the 1O mode periods, while the results of the best fitting procedure are listed in Table~\ref{tab:lmcsmcAC}. The figure shows that also for the combined sample, the tightest relations are the PL relations in the NIR and PW relations in all the bands. 
%The R factor does not taking into account the different extents of the IS for the 2 different pulsators.   

\section{Results}
%{\bf vedi comment sulle acep di omega cen appendix a: https://www.aanda.org/articles/aa/olm/2015/05/aa24838-14/aa24838-14.html}

In this section, we analyze the PL/PW relationships calculated in this work.  Initially, we explore how the PL/PW coefficients vary with wavelength. Subsequently, we compare our findings with the predictions of pulsation theory and empirical literature studies. Finally, after calibrating our most accurate relationships using an assumed LMC distance, we calculate the distances to specific MW globular clusters and nearby galaxies containing AC variables. An additional calibration of the ACs' PW relations, based on selected Galactic field AC stars having reliable multi-band photometric data and $Gaia$ parallaxes, is discussed at the end of this section.

\subsection{Wavelength dependence of the PL coefficients}

The extensive set of PL and PW relations shown in Table~\ref{tab:lmcsmcAC} allows us to investigate how the slopes and dispersion of these relations vary wavelength and to choose the best relationships to use in further analysis. 
The slopes and dispersion of the PL relations increase (in absolute value) and decrease, respectively, as the wavelength increases. While this trend is already known in the literature for the CCs \citep{1991PASP..103..933M} and T2Cs (S24), this work provides its first clear confirmation for the ACs. The physics behind this anti-correlation was noted for the CCs by \citet{1991PASP..103..933M} and was explained physically by \citet{madore2012}. This feature is due to the different surface brightness dependence on the effective temperature at different wavelengths. 
%Notably, in S24 we checked that these trends can be found also for T2Cs. 

\begin{figure}
\sidecaption
\vbox{
    \includegraphics[width=\hsize]{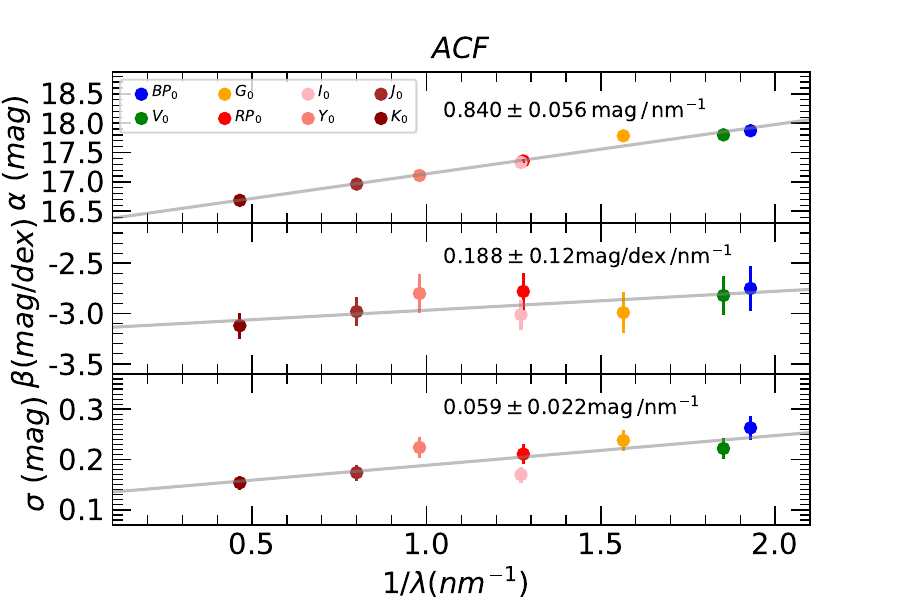}
    \includegraphics[width=\hsize]{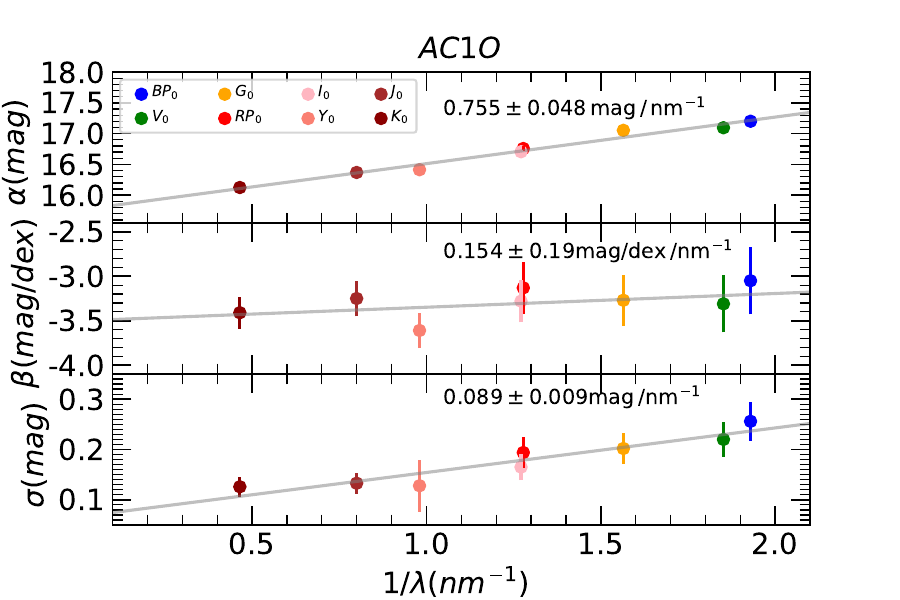}
    \includegraphics[width=\hsize]{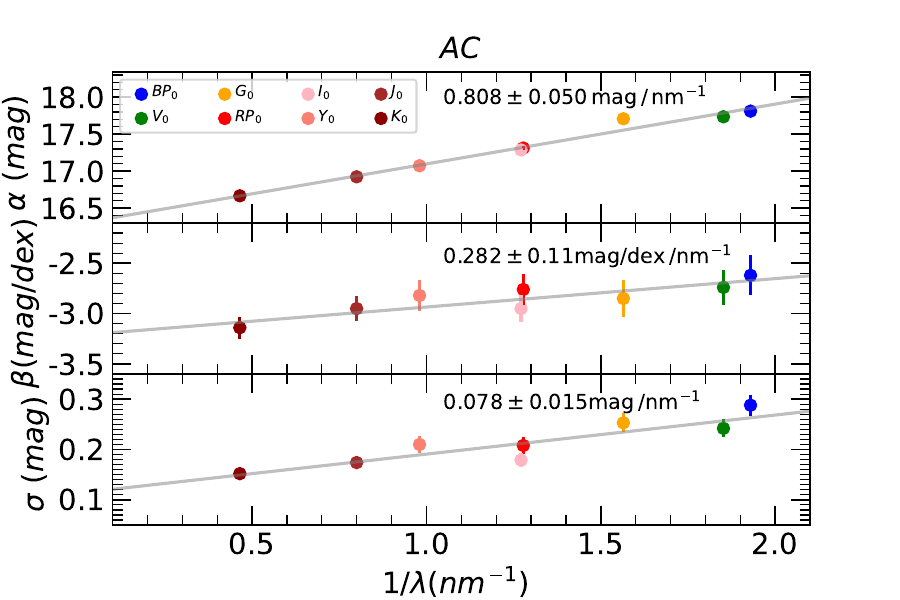}
    }
    \caption{From top to bottom: dependency of the LMC PL relation zero point, slope, and dispersion, respectively, as a function of 1/wavelength. }
    \label{coeffpllmc}
\end{figure}

Fig.~\ref{coeffpllmc} shows the result of this investigation for the PL relations for the LMC F and 1O mode ACs separately and F+1O ACs (using fundamentalized periods for the 1O).  Looking at the panels displaying the $\alpha$ (relative zero point), $\beta$ (slope), and $\sigma$ (dispersion) coefficients in each figure, the above-mentioned trends stand out clearly, especially for the F+1O case.

\begin{figure}
\sidecaption
\vbox{
    \includegraphics[width=\hsize]{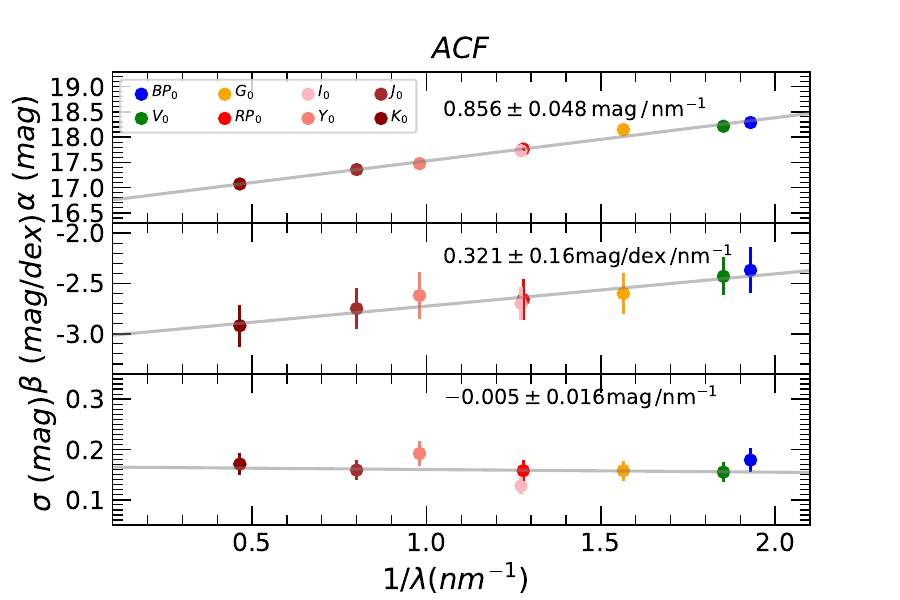}
    \includegraphics[width=\hsize]{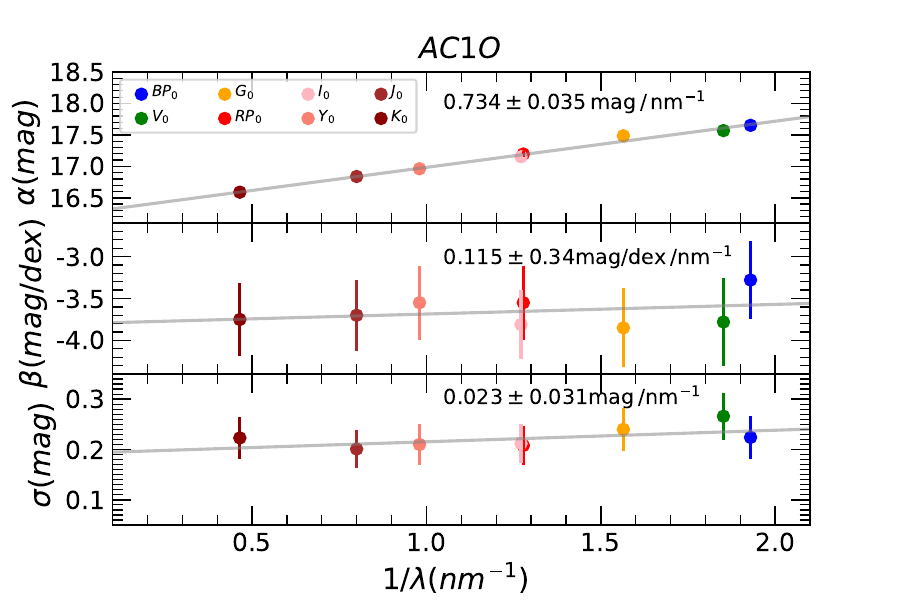}
    \includegraphics[width=\hsize]{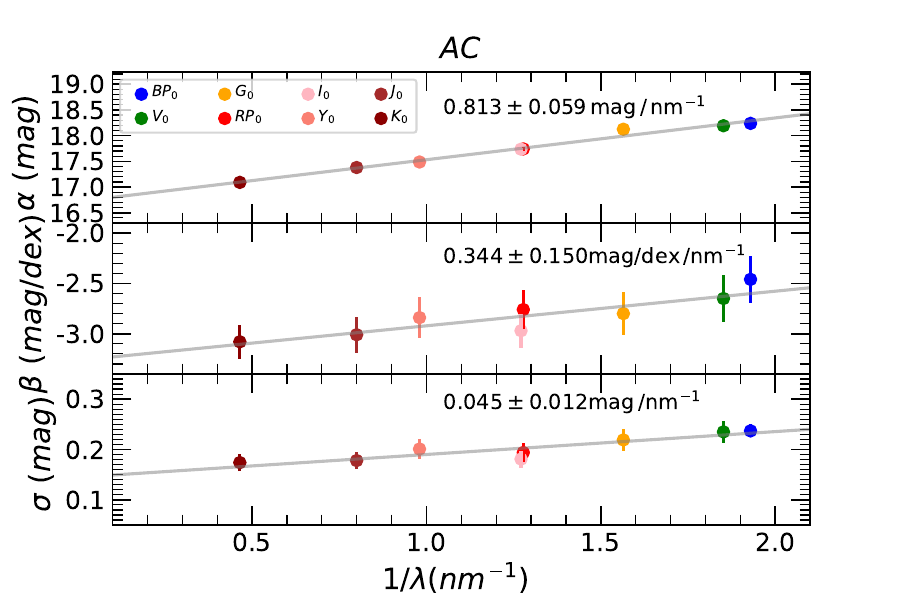}
    }
    \caption{As in Fig.~\ref{coeffpllmc}, but for the SMC.}
    \label{coeffplsmc}
\end{figure}

The dependences of the slope and dispersion on the wavelength are slightly less evident in Fig.~\ref{coeffplsmc} for the SMC. Indeed, the significant depth along the line of sight \citep[$\sim 0.17$ mag, e.g.][]{Ripepi2017vmc} produces an additional dispersion due to the geometry of the system that can be larger than the intrinsic width of the IS. The influence of the SMC depth is more evident in the case of the F-mode ACs (Fig.\ref{coeffplsmc}). 
%In any case, it is important to remark that the resulting correlations, as reported in Figs.~\ref{coeffpllmc} and \ref{coeffplsmc}, are directly linked to the $\sigma$ clipping performed during the fitting procedure (see Sect.\ref{ltsfit}).

%For the PW relations (see Figs. ~\ref{coeffpllmc2} and ~\ref{coeffplsmc2} ), the general trend is the same as for the PLs but with a less clear behavior. This is because the inclusion of a colour term in the Wesenheit magnitudes tends to mitigate the effects on the PL of the different IS shapes and widths in the optical and NIR bands. 
 
\subsection{Absolute calibration of the PL/PW relations using the geometric distance to LMC}

Calibrating the PL and PW zero points (intercepts) in absolute terms is mandatory to compare them with literature values as well as to calculate distances. To carry out this crucial procedure, we decided first to use the geometric distance modulus of the LMC ($\mu_{LMC}=18.477\pm0.026$ mag) as measured by P19 based on a sample of eclipsing binaries.

The calibrated PL and PW relations are listed in Table~\ref{tab:confrontolmct2}, while Table~\ref{diffzero} in App.~\ref{diffext} shows the coefficients of the PWJK in the LMC starting from different extinction laws. Whereas the relative zero points based on three different extinction laws \citep[][]{Wang_2023,1989ApJ...345..245C,fitzpatrick1999} agree within 2 $\sigma$, the impact on the absolute zero points, where also the uncertainty of the geometric distance modulus of the LMC is taken into account, is less than 1 $\sigma$.

\subsection{Comparison with the literature}

The relations with absolute intercepts as presented in the previous section and the ones listed in Table~\ref{tab:lmcsmcAC} (with relative intercepts) can now be compared to those available in the literature for the LMC and the theory.

Table~\ref{tab:confrontolmct2} shows the comparison of the values derived in this paper and a variety of PL and PW relations values from the literature \citep[R14,][]{marconi2004, Groenewegen2017,iwanek2018}. \citet{marconi2004} presented theoretical PL relations for ACs in optical bands based on nonlinear convective pulsation models. R14 exploited a subset of VMC data in 11 tiles in the LMC and derived PL and PW relations in the V, I and $K_s$ bands. \citet{Groenewegen2017}, using the data collected in OGLE III survey, published PL and PW relations for the ACs in the LMC, while \citet{iwanek2018} used the complete OGLE ACs sample in the LMC and the SMC.

\begin{table*}[h]
  \footnotesize\setlength{\tabcolsep}{4pt} 
  \caption{Comparison between present results for ACs in the LMC, the SMC, and the literature. } 
  \label{tab:confrontolmct2}
 \begin{center}
  \begin{tabular}{llllllllllll} 
  \hline  
  \noalign{\smallskip}   
        Relation & Galaxy & Mode & $\alpha $ &  $\sigma_{\alpha}$& $\alpha_0 $ &  $\sigma_{\alpha_0}$ & $\beta$ & $\sigma_{\beta} $ & RMS & n.stars & Author \\
    & &  & mag & mag & mag & mag & mag/dex & mag/dex &mag & &   \\                                        
  \noalign{\smallskip}
  \hline  
  \noalign{\smallskip}  
  %PLI & LMC+GC & F & - & - & $-$0.72 & 0.05 & $-$2.03 & 0.59 & 0.18 & 5 & N22\\
  \rowcolor{lightgray}
  %PLI & LMC & F & 17.33 & 0.03 & $-$1.15 & 0.04 & $-$3.01 & 0.15 & 0.17 & 79 & TW\\
  %PLG & LMC+GC & F & - & - & $-$0.39 & 0.04 & $-$2.36 & 0.99 & 0.36 & 5 & N22\\
  \rowcolor{lightgray}
  PLG & LMC & F & 17.78 & 0.04 & $-$0.69 & 0.05 & $-$2.99 & 0.20 & 0.24 & 83 & TW\\
  %PLV & theory & F & - & - & $-$0.71 & 0.03 & $-$2.64 & 0.17 & - & - & P02\\
  PLV & theory & F & - & - & $-$0.69 & 0.14 & $-$2.73 & - & 0.14 & - & \citet{marconi2004}\\
  PLV & LMC & F & - & - & $-$0.56 & 0.04 & $-$3.21 & 0.21 & 0.20 & 32 & \citet{2014ripepianomalous}\\
  \rowcolor{lightgray}
  PLV & LMC & F & 17.80 & 0.04 & $-$0.68 & 0.05 & $-$2.82 & 0.19 & 0.22 & 80 & TW\\
  %PLV & theory & 1O & - & - & $-$1.61 & 0.07 & $-$3.74 & 0.20 & 0.25 & - & P02\\
  PLV & theory & 1O & - & - & $-$1.56 & 0.25 & $-$2.95 & - & 0.25 & - & \citet{marconi2004}\\
  PLV & LMC & 1O & - & - & $-$1.26 & 0.09 & $-$3.14 & 0.37 & 0.23 & 10 & \citet{2014ripepianomalous}\\
  \rowcolor{lightgray}
  PLV & LMC & 1O & 17.10 & 0.04 & $-$1.38 & 0.05 & $-$3.31 & 0.32 & 0.22 & 35 & TW\\
  %PLJ & Draco & F & 17.94 & 0.03 & - & - & $-$4.03 & 0.15 & 0.11 & 4 & B24\\
  %\rowcolor{lightgray}
  %PLJ & LMC & F & 16.96 & 0.03 & $-$1.52 & 0.04 & $-$2.98 & 0.11 & 0.17 & 81 & TW\\
  %PLJ & LMC & 1O & 16.37 & 0.03 & -2.11 & 0.04 & -3.25 & 0.20 & 0.13 & 34/1 & TW\\
  PLK & LMC & F & 16.74 & 0.02 & - & - & $-$3.54 & 0.15 & 0.10 & 32 & \citet{2014ripepianomalous}\\
  %PLK & Draco & F & 17.64 & 0.03 & $-$ & $-$ & $-$3.83 & 0.21 & 0.09 & 4 & B24\\
  \rowcolor{lightgray}
  PLK & LMC & F & 16.69 & 0.03 & $-$1.79 & 0.04 & $-$3.12 & 0.13 & 0.16 & 78 & TW\\
  PLK & LMC & 1O & 16.06 & 0.07 & - & - & $-$4.18 & 0.33 & 0.10 & 10 & \citet{2014ripepianomalous}\\
  \rowcolor{lightgray}
  PLK & LMC & 1O & 16.13 & 0.02 & $-$2.35 & 0.04 & $-$3.41 & 0.18 & 0.13 & 33 & TW\\
  PWVI & theory & F & - & - & $-$1.93 & - & $-$3.34 & - & 0.20 & - & \citet{marconi2004}\\
  PWVI & LMC & F & 16.59 & 0.02 & - & - & $-$3.41 & 0.16 & 0.15 & 32 & \citet{2014ripepianomalous}\\
  PWVI & LMC & F & 16.61 & 0.02 & - & - & $-$3.16 & 0.14 & 0.15 & 62 & \citet{Groenewegen2017}\\
  PWVI & LMC & F & 16.59 & 0.02 & - & - & $-$2.96 & 0.12 & - & 94 & \citet{iwanek2018}\\
  \rowcolor{lightgray}
  PWVI & LMC & F & 16.60 & 0.02 & $-$1.88 & 0.04 & $-$3.12 & 0.12 & 0.14 & 79& TW\\
  PWVI & SMC & F & 16.95 & 0.02 & - & - & $-$2.89 & 0.14 & - & 72 & \citet{iwanek2018}\\
  \rowcolor{lightgray}
  PWVI & SMC & F & 16.95 & 0.03 & - & - & $-$3.03 & 0.17 & 0.14 & 44 & TW\\
  PWVI & LMC & 1O & 16.03 & 0.06 & - & - & $-$3.37 & 0.25 & 0.14 & 19 & \citet{Groenewegen2017}\\
  PWVI & LMC & 1O & 16.05 & 0.05 & - & - & $-$3.44 & 0.22 & 0.13 & 10 & \citet{2014ripepianomalous}\\
  PWVI & LMC & 1O & 16.04 & 0.04 & - & - & $-$3.30 & 0.20 & - & 39& \citet{iwanek2018}\\
  \rowcolor{lightgray}
  PWVI & LMC & 1O & 16.10 & 0.03 & $-$2.38 & 0.03 & $-$3.10 & 0.21 & 0.15 & 32 & TW\\
  PWVI & SMC & 1O & 16.55 & 0.05 & - & - & $-$3.69 & 0.28 & - & 39 & \citet{iwanek2018}\\
  \rowcolor{lightgray}
  PWVI & SMC & 1O & 16.38 & 0.02 & - & - & $-$4.30 & 0.23 & 0.10 & 22 & TW\\
  PWVK & theory & F & - & - & $-$1.71$-$1.83logM & - & $-$2.93 & - & 0.04 & - & \citet{marconi2004}\\
  PWVK & LMC & F & 16.58 & 0.02 & - & - & $-$3.58 & 0.15 & 0.10 & 32 & \citet{2014ripepianomalous}\\
  \rowcolor{lightgray}
  PWVK & LMC & F & 16.54 & 0.02 & $-$1.94 & 0.04 & $-$3.17 & 0.12 & 0.14 & 75 & TW\\
  PWVK & LMC & 1O & 15.93 & 0.07 & - & - & $-$4.14 & 0.33 & 0.10 & 10 & \citet{2014ripepianomalous}\\
  \rowcolor{lightgray}
  PWVK & LMC & 1O & 15.99 & 0.03 & $-$2.49 & 0.04 & $-$3.42 & 0.20 & 0.14 & 33 & TW\\
  %PWJK & Draco & F & 17.42 & 0.04 & - & - & $-$3.69 & 0.22 & 0.10 & 4& B24\\
  %\rowcolor{lightgray}
  %PWJK & LMC & F & 16.50 & 0.03 & $-$1.97 & 0.04 & $-$3.21 & 0.13 & 0.15 & 76 & TW\\
  \noalign{\smallskip}
  \hline  
  \noalign{\smallskip}
  \end{tabular}
  \tablefoot{ The different columns report: $(1)$ the type of relationship and the band of interest; $(2)$ the galaxy or the theory where the relation was tested; $(3)$ the pulsating mode; $(4)$--$(5)$ the $\alpha$ coefficient (relative intercept) and relative uncertainty; $(6)$--$(7)$ the $\alpha_0$ coefficient (absolute intercept) and relative uncertainty; $(8)$--$(9)$ the $\beta$ coefficient (slope) and relative uncertainty; $(10)$ the Root Mean Square (RMS) of the relation; $(11)$ the number of stars used in the fit; $(12)$ the autors. "TW" (\textit{gray shading}) refers to this work .}
  \end{center}
\end{table*}

An inspection of the table allows us to affirm that the slopes and the intercepts derived in this work for the PL and PW relations are in agreement with the literature studies within 1-2 $\sigma$ of the reported errors. The most difference is observed for the PL coefficients in the $V$ band and for the slopes of the PL in the $K_s$ band and the PW in the $VK_s$ combination with R14. Our PL/PW relations have been derived from the largest sample of stars for each pulsation mode and filter combination. In particular, our ACs sample is more than doubled with respect to the R14 sample.

%The intercept of F mode ACs for the PL relations in the $I$ band is not in agreement with \citet{ngeow2022} while the slope is in agreement within 2 $\sigma$. The same behavior is repeated for the $G$ band. In the $V$ band, for the F mode ACs, our coefficients are in agreement within 1 $\sigma$ with the theory and within 2 $\sigma$ with \citet{2014ripepianomalous}; whereas the 1O mode coefficients are in agreement within 1 $\sigma$ with both \citet{marconi2004} and \citet{2014ripepianomalous}. In the $J$, $K$, and PW in the $J$ and $K_s$ bands, the comparison with \citet{bhardwaj2024draco} shows a significant discrepancy. This may be explained by the fact that the Draco sample is composed of only four F and one 1O AC, respectively. Moreover, the AC formation channel in this galaxy and the LMC could be different, leading to different PL and PW relations. An additional possibility is the different metallicity between the two galaxies since our relations do not take into account this parameter. Note that our LMC sample is very different from the Draco sample.
The comparison with the literature is encouraging: the agreement between different studies based both on theory and observations demonstrates that the ACs are reliable and precise distance indicators when comparing them within the same stellar systems. 
In Section~\ref{application} we apply these to the MW GCs and LG galaxies, where additional factors such as metallicity may come into play.

\subsection{The LMC distance modulus through absolute calibration of the PW relations using $Gaia$ parallaxes}

An independent calibration of selected PW relations for ACs can be derived using $Gaia $ parallaxes for a sample of Galactic ACs. In this way, we can use the ACs to obtain the distance modulus to the LMC instead of assuming it. To carry out this exercise, we decided to use two of the best PW relations, namely those based on the $Gaia$ and the $J,\, K_s$ bands. The procedure is composed of several steps. The first step was to determine average magnitudes in $J,\,Ks$ bands for Galactic ACs (for the detailed method see Appendix~\ref{averagemag}). The second step is the calibration of the PW relations based on the average magnitudes.

To calibrate the Galactic AC PW relations, we cannot just invert the $Gaia$ parallaxes \citep[e.g.][]{Luri2018}. Instead, as in S24, we adopted the photometric parallax \citep[][]{Feast1997} to carry out the calculations. The photometric parallax (in mas) is defined as:

\begin{equation}
    \varpi_{phot}= 10^{-2(m-M-10)} ,
\end{equation}

\noindent
where $m$ is the apparent magnitude (the apparent Wesenheit magnitude in our case); $M$ is the absolute magnitude (or absolute Wesenheit magnitude), defined as:

\begin{equation}\label{mabseq}
    M = \alpha + \beta \times \log_{10}P ,
\end{equation}

\noindent where $\beta$ is the slope that we obtained in two ways.

In the first case, we fixed $\beta$ to the values obtained for the LMC (listed in Table~\ref{tab:lmcsmcAC}\footnote{The relations used refer to the LMC ACs with the fundamentalised 1O mode periods, i.e. those with "Mode = All".}). 
The value of $\alpha$ is obtained by minimizing the following $\chi^2$ expression:
\begin{equation} \label{chiquadro}
    \chi^2 = \sum \frac{(\varpi_{EDR3}-\varpi_{phot})^2}{\sigma^2},
\end{equation}
\noindent
where $\varpi_{EDR3}$ are the parallaxes from $Gaia$ EDR3 corrected individually with the recipe provided by \citet{2021A&A...649A...4L}. A comprehensive description of the fitting process can be found in \citet{Ripei2022_PW_Gaia}, and will not be reiterated here. 

We calculated the PW relations adopting three values of the counter zero-point-offset to be applied to the $Gaia$ parallaxes: $\varpi_{ZP} =$ 0 mas, $-$0.014 mas and $-$0.022 mas, i.e. no counter offset and progressively larger values, according to \citet{2021ApJ...908L...6R} and \citet{molinaro2023}, respectively. 
The resulting values for the intercept $\alpha$ are listed in Table~\ref{dislmc}. 

\begin{table}[h]
%\begin{adjustwidth}{-2cm}{}
  \tiny\setlength{\tabcolsep}{4pt} 
  \caption{ Distances to the LMC for the fixed slope of the PW relations for MW ACs. }
  \label{dislmc}
 \begin{center}
  \begin{tabular}{llllllll}
  \hline  
  \noalign{\smallskip}   
         $\alpha_{MW} $ &  +$\sigma_{\alpha_{MW}}$ &$-\sigma_{\alpha_{MW}}$ & $\varpi_{ZP}$ & $\mu_{LMC} $ & +$\sigma_{\mu_{LMC}}$ &  $-\sigma_{\mu_{LMC}}$ & MW AC \\
          mag & mag &mag & mas & mag & mag & mag& \\                                        
  \noalign{\smallskip}
  \hline  
  \noalign{\smallskip} 
\multicolumn{8}{c}{PWG  $\beta_{LMC}=-3.05\pm0.13$} \\
  \hline  
  \noalign{\smallskip} 
$-$1.96  & +0.08 & $-$0.08& 0& 18.17 &  +0.09&  $-$0.09 & 215\\
$-$2.15 & +0.08 & $-$0.09 & $-$0.014 & 18.36 & +0.09 & $-$0.09 & 215\\
$-$2.27 & +0.09 & $-$0.10 & $-$0.022 & 18.48 &  +0.09&$-$0.11 & 215\\
  \noalign{\smallskip}
  \hline  
  \noalign{\smallskip}
\multicolumn{8}{c}{PWJK $\beta_{LMC}=-3.25\pm0.11$} \\
  \hline  
  \noalign{\smallskip} 
$-$2.15 & +0.09 &$-$0.11 &  0 & 18.06   & +0.01 & $-$0.11 & 215\\
$-$2.40 & + 0.11  & $-$0.11 & $-$0.014 & 18.31  & +0.11 & $-$0.11 & 215\\
$-$2.55 & + 0.11 & $-$0.11  & $-$0.022   & 18.46 &  + 0.12 & $-$0.11  &  215\\
  \noalign{\smallskip}
  \hline  
  \noalign{\smallskip}
  \end{tabular}
  %\tablefoot{}
\end{center}
\end{table}

In the second case, we left both $\alpha$ and $\beta$ free to vary while minimizing Eq.~\ref{chiquadro}. Again, we repeated the calculation for the three values of the parallax counter-correction discussed above. The results for $\alpha$ and $\beta$ and their errors are shown in Table~\ref{dislmc1}. The $\sigma$ term at the denominator of Eq.~\ref{chiquadro}
is the sum in quadrature of the uncertainties on both $\varpi_{EDR3}$ and $\varpi_{phot}$.

%Before going on with the discussion on results, it is remarkable to note that the total error present in Equation~\ref{chiquadro} depends on the uncertainties related to $\varpi_{EDR3}$ and $\varpi_{phot}$. In the case of the ACs, the two main contributors to the total error are the parallaxes errors and the photometric ones.  

A comparison between Tables~\ref{dislmc} and ~\ref{dislmc1} reveals that not fixing the slope of the PW relations leads to large errors on the $\beta$ values.  The large uncertainties do not allow us to firm any conclusion from this occurrence. The differences in the values of $\alpha$ are less important and they agree within 1 $\sigma$.

\begin{figure}
\sidecaption
    \includegraphics[width=\hsize]{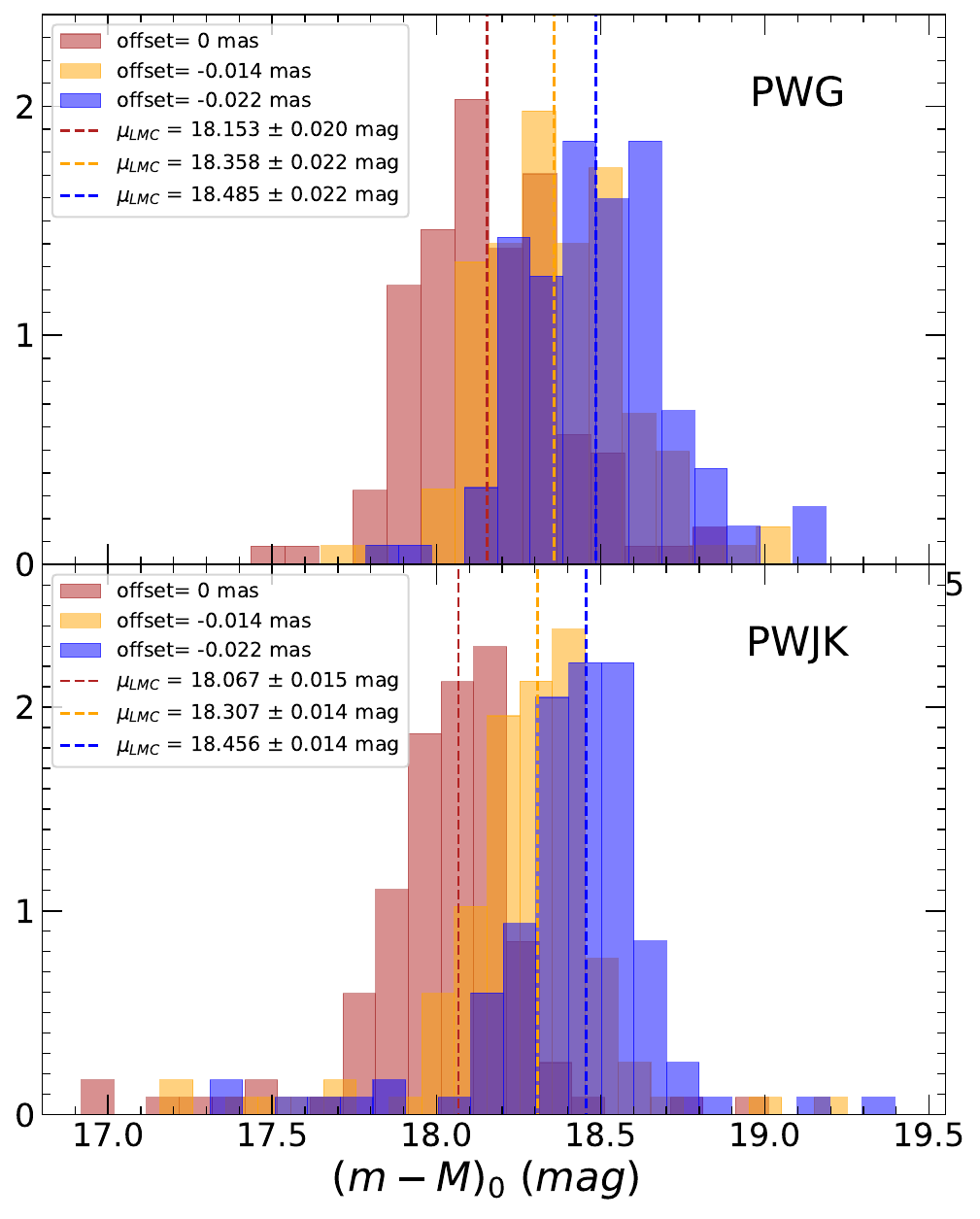}
    \caption{Distribution of the inferred LMC absolute distance moduli calibrated through $Gaia$ DR3 MW ACs parallaxes. The red histogram corresponds to the values obtained using  MW ACs parallaxes with no counter zero-point offset, while the red vertical line is the median of the distribution. The orange and blue histograms and vertical lines are the same as the red ones but derived using, respectively, the counter zero-point-offsets $\varpi_{ZP}=-0.014~mas$ as suggested by \citet{2021ApJ...908L...6R} and $\varpi_{ZP}=-0.022~mas$ as suggested by \citet{molinaro2023}.}
    \label{lmc_mw}
\end{figure}

In Tables~\ref{dislmc} and ~\ref{dislmc1} we have several PW relations. To verify which is the most accurate one, we can again use the geometric distance modulus of the LMC by P19 as a reference. From the periods of the LMC ACs in our sample, we can calculate the absolute Wesenheit magnitudes for each star. Applying in the formula $\mu=m-M$ these absolute Wesenheit values and the corresponding apparent magnitudes calculated in this paper, we calculated the individual distance moduli of the ACs in the LMC. 
Figure~\ref{lmc_mw} shows the distributions of the inferred distance moduli for the LMC ACs. The different colours refer to the different parallax counter zero-point offsets applied to $Gaia$ DR3 parallaxes before minimizing Eq.~\ref{chiquadro}.  

The average LMC distance moduli ($\mu_{LMC}$) and the relative errors (rms of the mean) are listed both in Table~\ref{dislmc} and Table~\ref{dislmc1},respectively, for the three parallax counter-offset values. The MW ACs parallaxes seem to need a counter zero-point offset as large as $\varpi_{ZP}=-0.022$~mas to obtain an average LMC distance modulus in agreement within 1 $\sigma$ with the LMC geometric value of 18.477 $\pm$ 0.026 mag (P19) and the canonical distance modulus of 18.49 $\pm$ 0.09 mag as suggested by \citet{2014degrijs}. However, the PW relations at the basis of our LMC distance modulus do not take into consideration possible metallicity dependence. This additional term may play a role in the absolute calibration of the PW relation through the MW ACs $Gaia$ parallaxes.

\begin{table*}[h]
%\begin{adjustwidth}{-2cm}{}
  \footnotesize
  \setlength{\tabcolsep}{4pt} 
  \caption{ Distances to the LMC, without fixing the slope of the PW relations for the MW ACs. }
  \label{dislmc1}
 \begin{center}
  \begin{tabular}{llllllllll}
  \hline  
  \noalign{\smallskip}   
       $\alpha_{MW} $ &  $+\sigma_{\alpha_{MW}}$ &  $-\sigma_{\alpha_{MW}}$ &  $\beta_{MW} $ &  $+\sigma_{\beta_{MW}}$ &  $-\sigma_{\beta_{MW}}$ & $\varpi_{ZP}$ &$\mu_{LMC} $ &  $\sigma_{\mu_{LMC}}$  & MW ACs \\
      mag & mag& mag&  mag/dex & mag/dex & mag/dex & mas & mag & mag &\\                                        
  \noalign{\smallskip}
  \hline  
  \noalign{\smallskip} 
\multicolumn{10}{c}{PWG} \\
  \hline  
  \noalign{\smallskip} 
$-$2.01  & +0.09 &  $-$0.11 & $-$3.47 & +0.49 & $-$0.47  & 0&  18.153&  0.020& 215\\
 $-$2.22  & +0.12 &  $-$0.14 &$-$3.54 &+0.59 & $-$0.64 &   $-$0.014 & 18.358 & 0.022 &215\\
$-$2.37  & +0.14 & $-$0.16  &  $-$3.65 & +0.66 &  $-$0.72 & $-$0.022& 18.485 & 0.022 &215\\ 
  \noalign{\smallskip}
  \hline  
  \noalign{\smallskip}
\multicolumn{10}{c}{PWJK} \\
  \hline  
  \noalign{\smallskip} 
$-$2.19  & +0.12 &  $-$0.13 & $-$3.51 & +0.63 & $-$0.60  & 0&  18.067 &  0.015& 215\\
 $-$2.43  & +0.14 &  $-$0.15 &$-$3.52 &+0.68 & $-$0.72 &   $-$0.014 & 18.307 & 0.014 &215\\
$-$2.58  & +0.15 & $-$0.16  &  $-$3.56 & +0.70 &  $-$0.76 & $-$0.022& 18.456 & 0.014 &215\\ 
  \noalign{\smallskip}
  \hline  
  \noalign{\smallskip}
  \end{tabular}
  %\tablefoot{}
\end{center}
\end{table*}

\subsection{The LMC--SMC relative distance modulus}

Given the agreement among almost all the slopes for the LMC and the SMC (see Table \ref{tab:lmcsmcAC}), it is possible to calculate the relative distance of the two Magellanic Clouds.
\begin{figure}
\sidecaption
     \includegraphics[trim=0cm 0cm 0.3cm 0cm, clip,width=0.9\hsize]{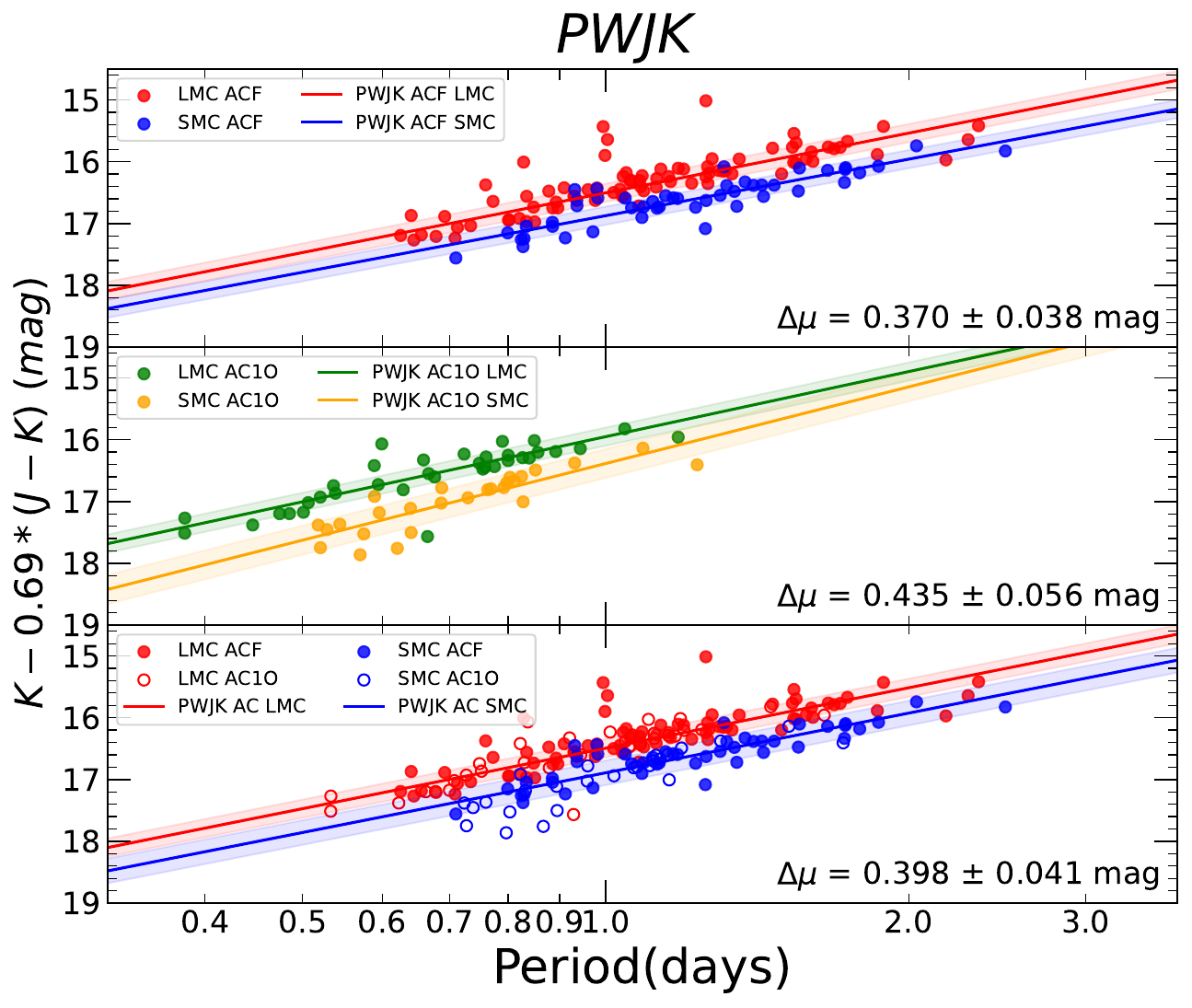}
    \caption{Comparison between the LMC and SMC PWJK relation to determine the relative distances (see labels). Top panel: red and blue filled circles are, respectively, the LMC and SMC F mode ACs, while the red and the blue solid lines are the fitted PL relations in the corresponding Cloud. Middle panel: green and yellow filled circles are, respectively, the LMC and SMC F mode ACs, while the green and the yellow solid lines are the fitted PL relations in the corresponding Cloud. Bottom panel: red filled and empty circles are, respectively, the LMC F mode and fundamentalized 1O mode ACs; blue filled and empty circles are, respectively, the SMC F mode and fundamentalized 1O mode ACs, while the red and the blue solid lines are the fitted PWJK relations in the corresponding Cloud.}
    \label{lmc_smc}
\end{figure}

Figure~\ref{lmc_smc} illustrates the adopted technique for the case of the PL in the $K_s$ band. In practice, we compare the PL relations for the LMC and the SMC, calculating the $\Delta \mu $ as the difference of the zero points, for F mode (top panel), 1O mode (middle panel) and F mode plus fundamentalized 1O mode (bottom panel) ACs.
%of the selected relations, namely the PL relation in the K band and the PW relations in $J\,K_s$ and in the $aia$ bands, in the LMC and the SMC. 
%After several tries, we concluded that, differently from the CCs \citep[][]{ripepi2016vmc}, the slopes do not vary at different periods.
Similar figures for the comparison through PWJKs and PWG relations can be found in the Appendix~\ref{appfigure}.

\begin{table}[h]
%\begin{adjustwidth}{-2cm}{}
  \footnotesize\setlength{\tabcolsep}{4pt} 
  \caption{ Relative distances between the LMC and SMC, and the absolute SMC distance modulus calibrated through the LMC geometric distance modulus by P19. }
  \label{dissmc}
 \begin{center}
  \begin{tabular}{lllll}
  \hline  
  \noalign{\smallskip}   
       Mode & $\Delta \mu $ &  $+\sigma_{\Delta \mu}$ & $\mu_{SMC} $ &  $\sigma_{\mu_{SMC}}$   \\
      & mag& mag& mag & mag \\                                        
  \noalign{\smallskip}
  \hline  
  \noalign{\smallskip} 
\multicolumn{5}{c}{PLK} \\
  \hline  
  \noalign{\smallskip} 
F  & 0.387 &  0.041 &18.864 & 0.049\\
 1O  & 0.464 &  0.052 & 18.941&0.058\\
All  & 0.423 & 0.037  & 18.900  &0.045\\ 
\noalign{\smallskip}
  \hline  
  \noalign{\smallskip}
\multicolumn{5}{c}{PWJK} \\
  \hline  
  \noalign{\smallskip} 
F  & 0.370 &  0.038 & 18.847 & 0.046\\
 1O  & 0.435 &  0.056 & 18.912&0.062\\
All  & 0.398 & 0.041  & 18.875 &0.049\\ 
  \noalign{\smallskip}
  \hline  
  \noalign{\smallskip}
  \multicolumn{5}{c}{PWG} \\
  \hline  
  \noalign{\smallskip} 
F  & 0.376 &  0.045 & 18.853 & 0.052\\
 1O  & 0.426 &  0.052 &18.903 &0.058\\
All  & 0.402 & 0.040  & 18.879 &0.048\\ 
  \noalign{\smallskip}
  \hline  
  \noalign{\smallskip}
  \end{tabular}
  %\tablefoot{}
\end{center}
\end{table}
Table~\ref{dissmc} reports the relative distance between the LMC and the SMC, and the SMC absolute distance obtained by adding to the $\Delta\mu$ the geometric LMC distance \citep[][]{2019Natur.567..200P}.
Our absolute SMC distances are in agreement within 2 $\sigma$  with the SMC geometric distance (18.977 $\pm$ 0.032 mag) as provided by \citet{Graczyk_2020} and the mean value of 18.96 $\pm$ 0.02 mag according to \citet{2015AJdegrijs}.
The difference between our SMC distance modulus and the literature might be attributed to a metallicity dependence, which is not included in our study due to a lack of spectroscopic data for ACs. Although we know that in the Milky Way ACs are metal-poor \citep{2024ripepi}, in the LMC and SMC their metallicity is not known. These galaxies present significant metallicity spread \citep[][]{2023povicklmc, 2023apovicksmc} and therefore spectroscopy is needed to establish appropriate [Fe/H] values for their ACs. 
%Given the metallicity span both in the LMC \citep[][]{2023povicklmc} and the SMC \citep[][]{2023apovicksmc}, ACs can show different metallicities in the two galaxies, through they are know to be metal-poor \citep[][]{2024ripepi}.  
Furthermore, as suggested by \citet{Breuval2024}, a metallicity term needs to be added to the CCs PW relations to retrieve the SMC distance modulus in agreement with the geometric distance modulus \citep[][]{Graczyk_2020}.

\section{Application of PL/PW in the MW and the Local Group}\label{application}

To test the PW relations derived in this work, we applied them to GGCs and dSph galaxies hosting ACs. 
Accurate GGC distances can be derived from a variety of methods \citep[see e.g.][and references therein]{2021MNRAS.505.5957B}. On the other hand, although the dSphs in the LG are rich in ACs, NIR photometry is available only for the Draco dSph (see below) which limits the possibility of testing the PW relations. 

 \subsection{ACs in GGCs}

Up to now, only two ACs are confirmed members in GGCs, namely V7 in M92 and V19 in NGC\,5466 \citep[see e.g.][]{ngeow2022}. For M92 V7 we adopted the accurate NIR photometry by \citet{delprincipe2005}, while for NGC\,5466 V19, we only have 2MASS single-epoch photometry. 

Two additional variable stars, V68 and V84, in omega Centauri ($\omega$Cen) have been indicated as candidate ACs in the literature \citep[see e.g.][and reference therein]{navarrete2015}. To verify whether these two stars could be ACs belonging to $\omega$ Cen, we first checked if their proper motions (PMs) from $Gaia$ are consistent with the PMs of the bulk of $\omega$ Cen members. This allowed us to immediately reject V84 as a field star, while we kept V68 for further analysis. To this aim we adopted the accurate NIR photometry by \citet{braga2016}.
%not only for this star but also for the RRLs and T2Cs belonging to $\omega$ Cen.  
\citet{}
Recently \citet{reyes2025variable} using $Gaia$ DR3 data have identified seven new candidate ACs in GGCs (not including the two stars in $\omega$ Cen discussed above). These cluster members, previously known as T2Cs, were reclassified as ACs based on their positions close to or on the PW relation in the $Gaia$ bands. We searched the literature for NIR photometry for these objects. We were not able to find it for NGC\,2419 V19 or NGC\,6388 V29. The 2MASS data are not useful, because the former cluster is very distant, with respect to other GGCs distances, whereas the latter has extremely crowded central regions. For the remaining five stars the source of the NIR photometry is reported in Table~\ref{check_acep} along with data for the three ACs in M92, NGC\,5466 and $\omega$ Cen. 

For M92 V7, Terzan\,1 V4 and $\omega$ Cen V68 we have found in the literature average NIR magnitudes from multi-epoch data, as specified in Table~\ref{check_acep}. For the remaining five stars we have found only single-epoch photometry either from the 2MASS or VHS DR5 \citep[VISTA Hemisphere Survey Data Release 5][]{VHS2013} surveys. For these objects we computed the intensity-averaged magnitudes using the template technique as described in Appendix.~\ref{averagemag}. 
However, period and epoch of maximum light are available only for NGC\,6752 V1 based on ASAS-SN data \citep[All-Sky Automated Survey for Supernovae][]{Shappee2014,Christy2023} and for M22 V11 in the $Gaia$ catalog \citep[see][]{Ripepi2023}. The $J$ and $K_s$ magnitudes of these two objects in Table~\ref{check_acep} are intensity-averaged magnitudes obtained from the template fitting technique. 
For NGC\,6388 V18 and NGC\,2808 V29 we adopted the single-epoch $J$ and $K_s$ magnitudes from VHS DR5 and included a conservative error of 0.15 mag to account for the random phase. 

\begin{figure}
\sidecaption
    \includegraphics[trim=0.8cm 2cm 2cm 0.8cm, clip,width=0.9\hsize]{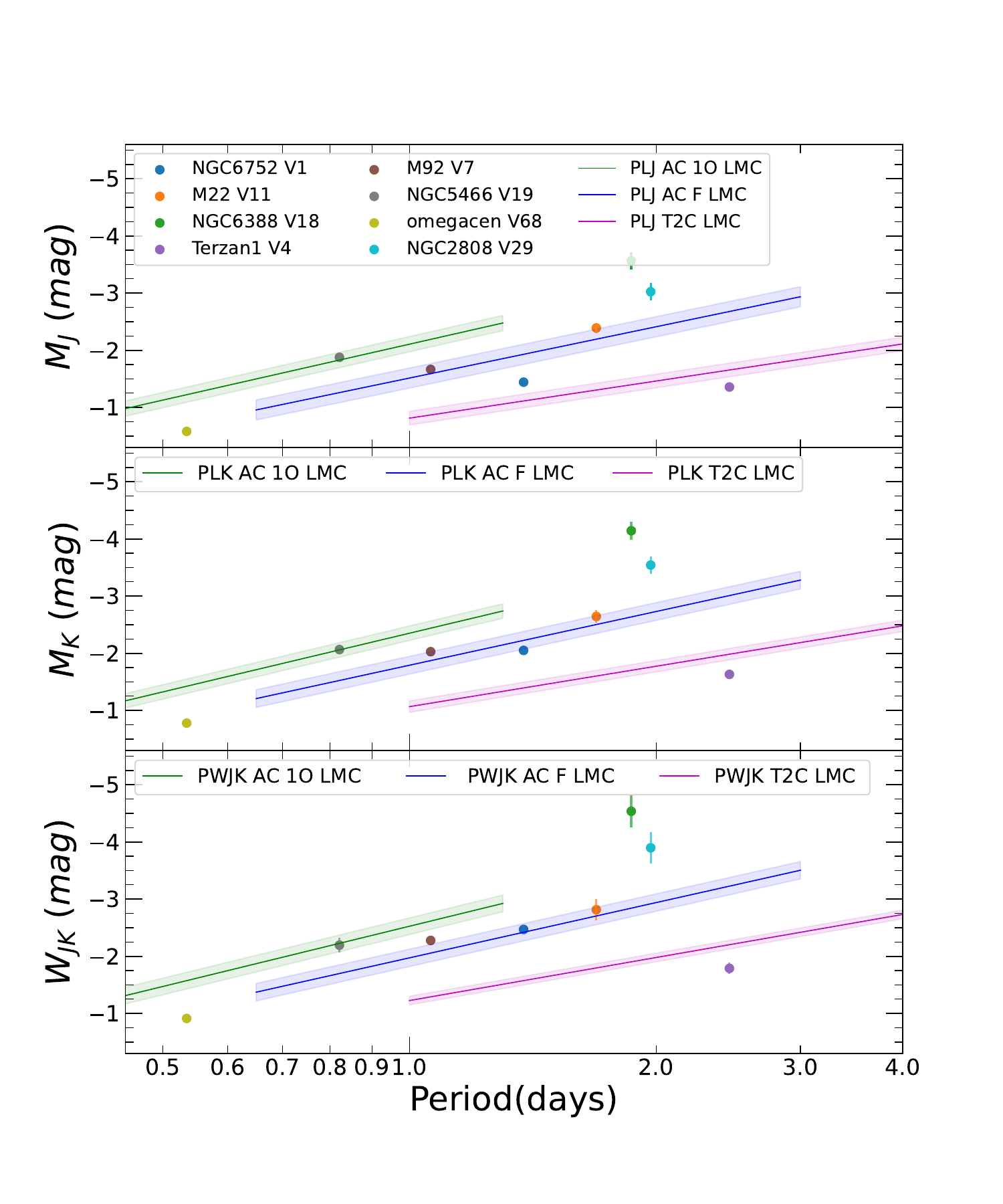}
    \caption{Position of the known and candidate ACs in GGCs from \citet{reyes2025variable} in the PL and PW relations for ACs and T2Cs in the LMC. Top panel: Green, blue, and magenta lines are the 1O mode, F mode ACs, and T2Cs PLJ relations in the LMC, calibrated with the LMC distance modulus by \citet{2019Natur.567..200P}, respectively. Blue, orange, green, violet, brown, grey, light green and cyan-filled circles are the known and candidate ACs in $\omega$cen, Terzan1, NGC6752, NGC6388, NGC5466, NGC2808, M92, and M22. Middle and bottom panels: as for the top panel but for the PL$K_s$ and PW$JK_s$ relations, respectively. NIR photometry for M92 is from \citet{delprincipe2005}.}
    \label{check}
\end{figure}

This NIR photometry allows us to use our best PL/PW relations (namely the PW$JK_s$, $PLJ$, and $PLK_S$), calibrated with the geometric distance of the LMC, to verify the AC nature of these GGC stars.
Figure~\ref{check} shows the results. The confirmed and candidate ACs in GGCs are plotted with different colours and compared with the PL/PW relations for F and 1O-mode ACs derived in this work and for T2Cs as published by S24. In all the cases these relations have the zero point anchored to the geometric distance modulus of the LMC by P19. To calculate the absolute magnitudes for the GGC ACs, we adopted the distances by \citet{2021MNRAS.505.5957B} and the reddening values by \citet{harris2010}, as reported in Table~\ref{check_acep}.  

Inspection of this figure leads to the following observations and conclusions:

\begin{itemize}
    \item 
    The known ACs in GGCs, namely M92 V7 and NGC\,5466 V19 lie on top of the LMC PL/PW relations for F and 1O-mode ACs, respectively. This means that their AC nature is confirmed and also that the distances of the two clusters agree within 1 $\sigma$ with the distance scale of ACs devised in this work (in the figure the shaded regions report the 1 $\sigma$ uncertainty of the PL/PW relations). 
    \item 
    The candidate AC M22 V11 lies on all the F-mode AC PL/PW relations of Fig.~\ref{check} and therefore can be considered the third confirmed AC in a GGC. Also, the distance of M22 appears in agreement with the LMC distance scale. 
    \item 
    Our data provide non-conclusive results for NGC\,6752 V1. In the PL$J$ it is in agreement with the T2C relationship, in the PL$K_s$ it is placed in between the T2C and AC F-mode relationships, while in the PW$J\,K_s$ it lies on the AC F-mode locus. In conclusion, we cannot prove or disprove the AC nature of this star.  
    \item 
    The star Terzan\,1 V4  is not an AC but a T2C as previously reported in the literature. On the other hand, the extremely high reddening in the direction of this cluster makes it difficult to estimate the dereddened magnitude in the optical ($Gaia$) bands. Indeed, in this regime, even the Wesenheit magnitude could include significant uncertainty, due to the possible non-standard value of the total to selected extinction ratio $R_V$.    
    \item 
    NGC\,6388 V18 and NGC\,2808 V29 are located at a significantly higher luminosity compared to all AC PL / PW relations of Fig.~\ref{check}. 
    %This is not the case in the optical $Gaia$ bands (Cruz Reyes et al. private communication). 
    This could be explained by each of the two variable stars being significantly blended with a red source. Such blending would not be surprising, because the two clusters are severely crowded in their central regions and the main contaminants are red giant stars. 
    In any case, this prevents us from drawing a firm conclusion regarding the nature of these stars.
    \item 
    $\omega$ Cen V68 has a period which makes it unlikely to be a F-mode AC but is too faint to be a 1O-mode. It is therefore likely that this star is an evolved RRL. 
\end{itemize}

\begin{table*}[h]
%\begin{adjustwidth}{-2cm}{}
  \footnotesize\setlength{\tabcolsep}{4pt} 
  \caption{ Candidate ACs in GGCs. "Flag NIR" indicates the source of the NIR photometry, "ID" lists the $Gaia$ DR3 identification according to \citet{gaiadr32023}.}
  \label{check_acep}
 \begin{center}
  \begin{tabular}{lllllllllllll}
  \hline  
  \noalign{\smallskip}   
       Name & ID Gaia DR3& Ra & Dec & P & $\langle J \rangle$ & $\sigma_{\langle J \rangle}$& $\langle K_\mathrm{s} \rangle$ & $\sigma_{\langle K_\mathrm{s} \rangle}$& $E(B-V)$ & $\mu$ & $\sigma_{\mu}$ & Flag NIR \\
        &  & deg &  deg  &  days& mag& mag &mag &mag & mag & mag &mag&  \\                                        
  \noalign{\smallskip}
  \hline  
  \noalign{\smallskip} 

 $\omega$ Cen V68 & 6083718920208854528 & 201.55333 & $-$47.32658 & 0.54 & 13.20 & 0.01 & 12.94 & 0.01 & 0.12 & 13.67 & 0.02 & BR16\\
  Terzan1 V4 & 4055452878472660480 & 263.98826 & $-$30.49802  & 2.46 & 14.27 & 0.01 & 12.91  & 0.02 & 1.99 & 13.77 & 0.07 & VVVDR2\\
  NGC6752 V1 & 6638377937476917376 & 287.85367 & $-$59.94009 & 1.38 & 11.67 & 0.05 & 11.04 & 0.03 & 0.04 & 13.08 & 0.01 & 2MASS\\
  NGC6388 V18 & 5955269337621694592 & 264.06388 & $-$44.73400 & 1.87 & 12.02 & 0.15 & 11.24& 0.15 & 0.37 & 15.24 & 0.03 & VSH\\
  NGC5466 V19 & 1452625254531322752 & 211.41857 & +28.48674  & 0.82  & 14.16 & 0.08 & 13.97  & 0.07 & 0.00 & 16.04 & 0.02  & 2MASS\\
  NGC2808 V29 & 5248759277665209472 & 138.00335 & $-$64.86095 & 1.97 & 12.19 & 0.15 & 11.56 & 0.15 & 0.22 & 15.01 & 0.02  & VHS\\
  M92 V7 &1360405396085147008  &259.28171  & +43.12267 & 1.06 & 13.00 & 0.03 & 12.63 & 0.03 & 0.02 & 14.65 & 0.02 & DP05\\
  M22 V11 & 4077588452718318080 & 279.09468  & $-$23.90256  & 1.69  & 10.52  & 0.09 & 10.08  & 0.10 & 0.34 & 12.59  & 0.02  & 2MASS\\
  \noalign{\smallskip}
  \hline  
  \noalign{\smallskip}
  \end{tabular}
   \tablefoot{Ra and Dec refer at Epoch=J2016.0. "VVVDR2" refers to \citet{vvvdr2}. "VSH\_DR5" refers to \citet{vshdr5}. "BR16" refers to \citet{braga2016}. "2MASS" refers to \citet{2MASS}."DP05" refers to \citet{delprincipe2005}. $E(B-V)$ values are from \citet{harris2010}, while $\mu$ values are from \citet{2021MNRAS.505.5957B}}
\end{center}
\end{table*}
%, used with the GGC distances from \citet{2021MNRAS.505.5957B}. 

\subsection{The distance to Draco}

The Draco dSph hosts 9 ACs as suggested by \citet[][]{muraveva2020MNRAS.499.4040M}. However, only five have been observed in the NIR. \citet[][hereinafter B24]{bhardwaj2024draco} published accurate NIR photometry for 4 F and 1 1O-mode ACs, as well as for 336 RRLs in Draco. This allows us to make an independent comparison of the RRLs and AC distance scales.   

\begin{figure}
\sidecaption
    \includegraphics[width=0.48\textwidth]{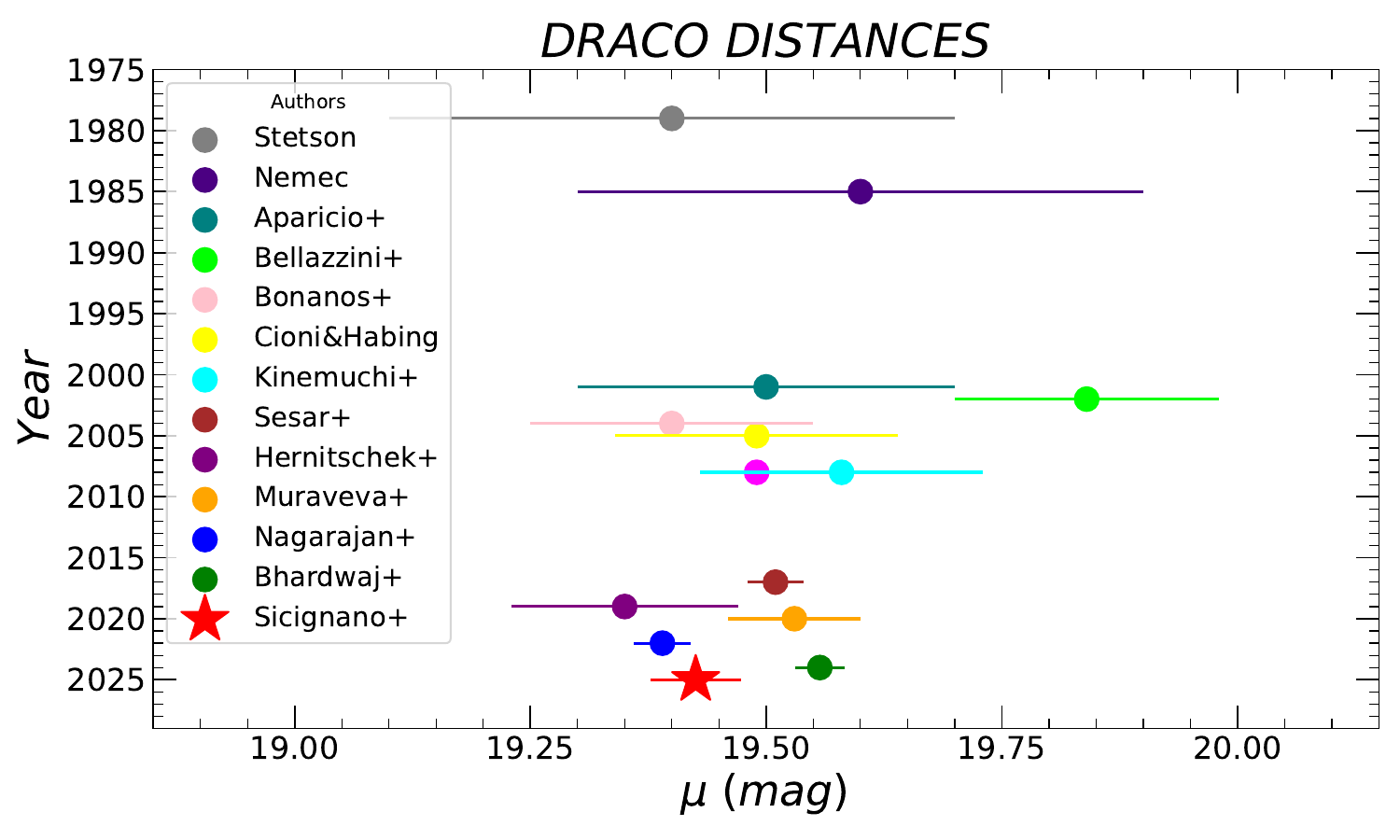}
    \caption{Comparison of different Draco distance moduli. The reference values are taken from \citet{stetson1979,nemec1985,aparicio2001AJ....122.2524A,bellazini2002,bonanos2004,cioni2005,kinemuchi2008,sesar2017AJ....153..204S,Hernitschek2019ApJ...871...49H,muraveva2020MNRAS.499.4040M,nagarajan2022ApJ...932...19N,bhardwaj2024draco}. A red star shows the distance modulus to Draco obtained in the present work. }
    \label{dracocomp}
\end{figure}

We first fundamentalised the unique AC 1O-mode pulsator to obtain a total sample of five pulsators, which we then use to derive the distance modulus of Draco. Table~\ref{draco} lists our final estimates of Draco distance modulus, while Figure~\ref{dracocomp} compares our best value, based on the PWJK relation, with literature ones.
Our estimate for the Draco distance modulus is in agreement within 2 $\sigma$ with all the recent and past distance estimates \footnote{An extensive explanation about all the methods can be found in \citet{bhardwaj2024draco}} and within 1 $\sigma$ with \citet{nagarajan2022ApJ...932...19N,muraveva2020MNRAS.499.4040M,Hernitschek2019ApJ...871...49H} and B24.
%Our Draco distance is in perfect agreement with the AC-based distance estimate by \citet{bhardwaj2024draco}, which \marina{was} derived using \marina{an older} PL relation \marina{from} \citet[][]{2014ripepianomalous}. 
More in detail, assuming our new AC PL relations instead of R14 ones, the difference between RRLs and AC distance moduli of Draco reduces from 0.17 mag, according to B24, to 0.13 mag.
%we find a difference of 0.13 mag between the RRLs distance modulus suggested by B24 and our ACs distance moduli, whereas B24 found a difference of 0.17 mag  between RRL and ACs (based on PL relations by R14). 
B24 suggested that assuming a metallicity difference of $\Delta$[Fe/H]=0.5 dex between LMC and Draco AC abundances, the entire distance discrepancy could be explained with a metallicity dependence of the PL and PW relations of $\sim - $0.34 mag/dex, a value typical for CCs \citep[see e.g.][]{2024Riess,2024trentin}. This value would slightly decrease to $\sim - $0.26 mag/dex in our case. However, current pulsation models \citep[e.g.][]{marconi2004} do not predict a significant metallicity dependence of the AC PL relations and the AC PL and PW relations presented in this work and those by R14, as already discussed above, do not include this parameter.

The hypothesis of a significantly different average metallicity between Draco and the LMC is however plausible, considering that different metallicities can be linked to very different star formation histories of these two galaxies \citep[e.g.][for Draco and LMC, respectively]{aparicio2001AJ....122.2524A, Mazzi2021}. Furthermore, we showed that LMC and SMC distance moduli based on AC PW relations need a posteriori correction to align them with the corresponding geometric distance moduli. 

Also, according to the bottom panel of Figure~\ref{lmc_smc}, the LMC-- SMC relative distance modulus needs an increase of 0.10 $\pm$ 0.04 mag to recover the LMC--SMC relative distance modulus as derived by using eclipsing binaries. The metallicity difference is larger between the LMC \citep[][]{2021choudhurylmc} and Draco \citep[][]{2020choudhurysmc} than the one between the LMC and the SMC \citep[][]{2020choudhurysmc} and supports the evidence of a larger scatter in the comparison of the Draco distances. 
%of the metallicities Draco \citep[-1.7 dex, according to][]{bellazini2002} is, in average, more metal poor than the LMC \citet{2021choudhurylmc} and the SMC \citep[][]{2020choudhurysmc}.
%The previous statement and the difference between the Draco distance modulus based on AC and RRLs, 0.13 $\pm$ 0.06 mag, support the possibility to correct for a metallicity term since the metallicity scatter is more consistent
This calls for further empirical tests based on spectroscopic data to verify the dependence of the AC PL and PW relations on metallicity.

\begin{table}[h]
%\begin{adjustwidth}{-2cm}{}
  \footnotesize\setlength{\tabcolsep}{4pt} 
  \caption{ Draco distance moduli based on our PL and PW relations and \citet{bhardwaj2024draco} NIR photometry. }
  \label{draco}
 \begin{center}
  \begin{tabular}{lll}
  \hline  
  \noalign{\smallskip}   
       Relation & $\mu $ &  $\sigma_{\mu}$ \\
     & mag & mag  \\  
  \hline  
  \noalign{\smallskip} 
   PLK & 19.458 & 0.057 \\
   PLJ & 19.532 & 0.089  \\
   PWJK & 19.425 & 0.048 \\
  \noalign{\smallskip}
  \hline  
  \noalign{\smallskip}
  \end{tabular}
\end{center}
\end{table}

\section{Summary and conclusions}

In this study, we present and exploit the $Y$, $J$, and $K_s$ time-series photometry from the ESO public survey, the "VISTA survey of the Magellanic Clouds system" for a sample of approximately 200 ACs located in the LMC and SMC. The VMC data were complemented with optical data from the OGLE IV survey and the $Gaia$ mission. From these surveys, we obtained the identifications and positions of the pulsators, their periods, their modes, and their photometry (and amplitudes where possible) in the $V$, $I$, $G$, $G_{BP}$, and $G_{RP}$ bands.

After selecting the stars with the best-quality VMC time series, we constructed a set of light-curve templates for F-mode and 1O-mode ACs. These templates were employed to derive accurate intensity-averaged magnitudes (and amplitudes, when possible) for all pulsators at our disposal through a rigorous pipeline previously developed in the context of VMC studies of CCs and T2Cs.

The final $Y$, $J$, and $K_s$ photometry, along with the aforementioned optical data, was used to create a large set of PL and PW relations for both F and 1O-mode ACs. 
The same relations were also calculated for the global F+1O-mode sample. This was achieved by fundamentalising the 1O-mode period with a value of $P_{1O}/P_F$=0.716. This ratio has been found here for the first time for ACs. 
The possibility of using the total sample of ACs increases the statistics of the PL and PW relations in all stellar systems hosting both F and 1O-mode ACs.

The dependence of the PL relations — namely the slopes and the dispersions — on the adopted wavelength was investigated for the first time. Similarly to the case for CCs and T2Cs, our findings demonstrate that also for the ACs, the slope increases and the dispersion decreases as we move towards longer wavelengths.

In general, our PL and PW relations are in agreement with the predictions of the pulsation theory, while the coefficients of the PL and PW relations are consistent within better than 2 $\sigma$ for both the LMC and the SMC compared to the literature. These evidences suggest that ACs are precise and reliable distance indicators.

We calibrated the PL and PW relations using a sample of Galactic field ACs for which 2MASS single epoch observations were available. Thanks to our set of light-curve templates, the single-epoch magnitudes were transformed into mean magnitudes. To calculate the absolute distance to the LMC, we followed two procedures: fixing the slopes to those of the LMC and not fixing the slopes. In both cases, we calculated the zero points of the PL and PW relations using the $Gaia$ EDR3 parallaxes. As a result, we obtained a $\mu_{LMC}$ that is within 1 $\sigma$ of the geometric estimate provided by P19, particularly when we adopt a counter zero-point offset for the $Gaia$ parallaxes of about $-$22 $\mu$as. Alternatively, an additional metallicity term in the PL and PW relations might compensate for such a counter zero-point offset.

The tightest relations derived in this work, namely the PL in the $K_s$ band and the PW relations in the $J$, $K_s$, and $Gaia$ bands in the LMC and the SMC, served to calculate the relative distance between the Clouds, which we find to be within 2$\sigma$ of the literature values. 
%Such distance although is in agreement with estimates based on different methods suggests the possibility to take into account the metallicity dependence. 
%In particular, it aligns with values derived from different age pulsating stars, demonstrating the reliability of ACs as distance indicators within the LG.

Our relative tightest relations for the LMC ACs were calibrated using the LMC geometric distance modulus. These absolute relations were used to study the ACs in GGCs and in Draco dSph. Firstly, we examined new candidate ACs in GGCs, confirming that M22 V11 is an F-mode AC. No firm conclusion can be reached for NGC\,6752 V1, NGC\,6388 V29 and  NGC\,2808 V29. Likely the photometry of the latter two objects is highly contaminated by close luminous red sources.
Terzan\,1 V4 and $\omega$ Cen are a T2C and an RRL, respectively. 
Concerning the known ACs in GGCs, M92 V7 and NGC\,5466 V19 we confirm their AC F and 1O-mode nature. Their location on the PL/PW relations calibrated on the LMC allows us to verify that the distance scale of GGCs by \citet{2021MNRAS.505.5957B} agrees within 1 $\sigma$ with our ACs based distances.

The application of our NIR PL/PW relations to the five ACs in the Draco dSph (the only dSph for which NIR photometry is currently available) allowed us to determine its distance modulus which is $\mu_{Draco}=19.425\pm0.048$ mag (from PW$_{J\,K_s}$), in agreement with B24. We confirm that the distance modulus to Draco calculated from RRL is about 0.1 mag larger than that obtained from ACs. In agreement with B24 we suggest that this may be due to a metallicity difference between Draco and LMC ACs.

Proprietary UVES (\href{https://www.eso.org/sci/facilities/paranal/instruments/uves.html}{Ultraviolet and Visual Echelle Spectrograph}) data (P.I. V. Ripepi) for a few tens Galactic ACEPs and the advent of large spectroscopic surveys, such as those anticipated with 4MOST (\href{https://www.eso.org/sci/facilities/develop/instruments/4MOST.html}{4-metre Multi-Object Spectroscopic Telescope}) and WST (\href{https://www.wstelescope.com/}{Wide-field Spectroscopic Telescope}), along with the availability of more accurate parallaxes due to the forthcoming $Gaia$ DR4, will undoubtedly provide us with the information necessary to draw firmer conclusions regarding the distance scale of ACs and in particular to verify if metallicity affects their PL/PW relations.

%In conclusion, ACs have allowed us to calculate the accurate and precise distances among several objects in the Local Group. As intermediate age pulsating stars, they serve as reliable distance indicators, comparable to the well-known RRL and CCs.
%It is not easy to interpret these conflicting results.
%One possibility is to invoke a significant effect of the metallicity ---which we neglected in this paper-- based on the theoretical and empirical results of \citet{ripepi2015vmc} and \citet[and references therein]{ngeow2022}, who do not report a significant metallicity dependence of T2C PL and PW relations. However, even a significant metallicity effect seems insufficient to simultaneously reconcile the distances of the LMC (and SMC) and the GGCs, given the large range of abundances spanned by the latter.   
%A more likely possible explanation is that the spatial distributions of the late-type ECBs adopted by \citet{pietrzynski2019distance} and \citet{graczyk2020} to derive the geometric distances of the LMC and the SMC differ significantly from those spanned by the T2Cs in our samples. This occurrence could account for at least part of the observed discrepancy.   

\begin{acknowledgements}
The authors thank the anonymous referee for their comments, which helped improve the quality of the manuscript. We thank Mauricio Cruz Reyes for kindly sharing the results of his paper in submission. We acknowledge useful comments and discussions with Matteo Monelli, Giuliana Fiorentino, and Santi Cassisi.
Based on data products created from observations collected at the European Organisation for Astronomical Research in the Southern Hemisphere under ESO program 179.B-2003. This research has used the SIMBAD database operated at CDS, Strasbourg, France ad data from the European Space Agency (ESA) mission Gaia, processed by the Gaia Data Processing and Analysis Consortium. Funding for the DPAC has been provided by national institutions, in particular, the institutions participating in the Gaia Multilateral Agreement.  We acknowledge funding from: INAF GO-GTO grant 2023 “C-MetaLL - Cepheid metallicity in the Leavitt law” (P.I. V. Ripepi); PRIN MUR 2022 project (code 2022ARWP9C) 'Early Formation and Evolution of Bulge and Halo (EFEBHO),' PI: Marconi, M., funded by the European Union – Next Generation EU; Large Grant INAF 2023 MOVIE (P.I. M. Marconi).  This research was supported by the Munich Institute for Astro-Particle and BioPhysics (MIAPbP), funded by the Deutsche Forschungsgemeinschaft under Germany´s Excellence Strategy – EXC-2094 – 390783311, by the International Space Science Institute (ISSI) in Bern/Beijing through ISSI/ISSI-BJ International Team project ID \#24-603 - “EXPANDING Universe” (EXploiting Precision AstroNomical Distance INdicators in the Gaia Universe), and by INAF-ASTROFIT fellowship. T.S. and G.D.S. thank the Istituto Nazionale di Fisica Nucleare (INFN), Naples section, for support through specific initiatives QGSKY. Finally, this paper is based on work supported by COST Action CA21136, "Addressing Observational Tensions in Cosmology with Systematics and Fundamental Physics (CosmoVerse)," funded by COST (European Cooperation in Science and Technology). 
\end{acknowledgements}


   \bibliographystyle{aa} % style aa.bst
   \bibliography{myBib} % your references Yourfile.bib

\begin{appendix}
\section{Notes about individual stars}\label{samplean}
OGLE-LMC-ACEP-112 is classified as 1O mode AC according to OGLE IV. Once we found out this star was an outlier in the NIR PL relations, we double-checked its parameters in VSA. The distance between OGLE IV and VMC coordinates was found to be about 0\farcs7. In VSA the wrong light curve is associated to OGLE-LMC-ACEP-112: indeed the light curve is not likely a pulsating star light curve. Unfortunately, no close variable star is available in the VSA.

We carefully examined all the stars with larger distance between OGLE and VMC  coordinates and found also an issue with OGLE-LMC-ACEP-039(sourceID=558386833752). This star is also an outlier in the PL relation in the $K_s$ for the F mode ACs. Its amplitude in the K band is too small compared to the I band one. It is probably blended, and was thus removed from the sample. 

OGLE-LMC-ACEP-100 stands out in the PL relation in the $K_s$ for F mode ACs, although its light curve is typical of a variable star.

DR35278444889019947136 was among the outliers in the PLK relation. We checked its properties in the Gaia database and decided to change its pulsation mode to 1O.

DR34655256851096632448 was removed from the sample due to blending. 

OGLE-SMC-ACEP-072 showed a non-variable light curve.

\section{Templates fitting} \label{apptemplates}
The tables in this Appendix list the coefficients of the templates adopted to fit the observed light curves in the $Y,\,J,$ and $\,K_s$ bands.
\begin{table}
%\begin{adjustwidth}{-2cm}{}
\tiny\setlength{\tabcolsep}{3pt} 
\caption{Fourier parameters of the light curves templates in the $Y$ band. The entire table is available electronically.}
  \label{templatesy}
  \begin{tabular}{rlrrrrrrrrr}
 \hline  
 \noalign{\smallskip} 
         Type & Period & $A_1$ &  $\Phi_1$ & $A_2$ & $\Phi_2$ & ... & $A_9$ & $\Phi_9$ & $A_{10}$ & $\Phi_{10}$  \\
  & days  & mag  & rad &mag  & rad  & & mag & rad & mag  & rad    \\    
 \noalign{\smallskip}
 \hline  
 \noalign{\smallskip}
1O & 0.570 & 0.467 & 3.279 & 0.130 & 2.907 & ...& 0.001 & 2.047 & 0.001 & 5.634\\
  1O & 0.575 & 0.460 & 2.533 & 0.124 & 3.768 & ...& 0.000 & 1.123 & 0.000 & 4.663\\
  1O & 0.857 & 0.416 & 2.829 & 0.118 & 3.512 & ...& 0.001 & 3.961 & 0.00 & 3.692\\
  F & 0.980 & 0.410 & 2.355 & 0.163 & 2.978 & ...& 0.002 & 2.462 & 0.002 & 3.143\\
  F & 1.343 & 0.410 & 2.508 & 0.1467 & 2.985 & ... & 0.001 & 2.720 & 0.000 & 4.676\\
  F & 1.710 & 0.361 & 2.621 & 0.196 & 3.508 & ...& 0.000 & 5.473 & 0.000 & 3.003\\
\hline  
\noalign{\smallskip}
\end{tabular}
%\end{adjustwidth}
\end{table}

\begin{table}
%\begin{adjustwidth}{-2cm}{}
\tiny\setlength{\tabcolsep}{3pt} 
\caption{Same as Table \ref{templatesy} but in the $J$ band.}
  \label{templatesj}
  \begin{tabular}{rlrrrrrrrrr}
 \hline  
 \noalign{\smallskip} 
      Type & Period & $A_1$ &  $\Phi_1$ & $A_2$ & $\Phi_2$ & ... & $A_9$ & $\Phi_9$ & $A_{10}$ & $\Phi_{10}$  \\
    & days & mag  & rad &mag  & rad  & & mag & rad & mag  & rad    \\    
 \noalign{\smallskip}
 \hline  
 \noalign{\smallskip}
  1O & 0.529 & 0.453 & 2.297 & 0.142 & 3.033 & ...& 0.002 & 1.810 & 0.000 & 3.092\\
  1O & 0.828 & 0.483 & 2.922 & 0.045 & 4.794 & ...& 0.000 & 1.677 & 0.000 & 5.003\\
  F & 0.830 & 0.372 & 2.293 & 0.160 & 3.112 & ...& 0.004 & 4.019 & 0.003 & 5.459\\
  F & 0.886 & 0.464 & 2.299 & 0.148& 3.310 & ...& 0.003 & 1.071 & 0.002 & 1.300\\
  F & 1.888 & 0.381 & 2.125 & 0.187 & 3.206 & ...& 0.002 & 0.785 & 0.001 & 0.109\\
\noalign{\smallskip}
\hline  
\noalign{\smallskip}
\end{tabular}
%\end{adjustwidth}
\end{table}

\begin{table}
%\begin{adjustwidth}{-2cm}{}
\tiny\setlength{\tabcolsep}{3pt} 
\caption{Same as Table \ref{templatesy}but in the $K_s$ band.}
  \label{templatesk}
  \begin{tabular}{rlrrrrrrrrr}
 \hline  
 \noalign{\smallskip} 
        Type & Period & $A_1$ &  $\Phi_1$ & $A_2$ & $\Phi_2$ & ... & $A_9$ & $\Phi_9$ & $A_{10}$ & $\Phi_{10}$  \\
    & days & mag & rad &mag  & rad  & & mag & rad & mag  & rad    \\    
 \noalign{\smallskip}
 \hline  
 \noalign{\smallskip}
  F & 0.799 & 0.450 & 1.902 & 0.154 & 3.301 & ...& 0.001 & 4.084 & 0.001 & 4.463\\
  1O & 0.840 & 0.509 & 3.069 & 0.047 & 4.992& ... & 0.000 & 6.183 & 0.000 & 3.351\\
  F & 1.033 & 0.448 & 2.848 & 0.106 & 4.350 & ...& 0.002 & 5.974 & 0.001 & 0.998\\
  F & 1.041 & 0.331 & 1.861 & 0.159 & 2.499 & ...& 0.001 & 6.137 & 0.001 & 3.560\\
  1O & 1.045 & 0.491 & 3.111 & 0.021 & 3.832& ... & 0.000 & 6.169 & 0.000 & 3.626\\
  F & 1.048 & 0.462 & 2.099 & 0.131 & 3.398 & ...& 0.002 & 4.789 & 0.001 & 5.422\\
  F & 1.538 & 0.446 & 2.795 & 0.129 & 4.764 & ...& 0.003 & 4.940 & 0.000 & 0.794\\
  F & 1.710 & 0.420 & 2.303 & 0.164 & 4.025 & ...& 0.001 & 0.172 & 0.001 & 5.359\\
  F & 1.888 & 0.391 & 1.471 & 0.185 & 2.643 & ...& 0.007 & 0.229 & 0.004 & 0.155\\
  F & 2.036 & 0.368 & 2.441 & 0.162 & 4.325 & ...& 0.001 & 4.699 & 0.002 & 2.335\\
  F & 2.347 & 0.485 & 3.251 & 0.092 & 0.369 & ...& 0.000 & 1.268 & 0.000 & 0.282\\
\noalign{\smallskip}
\hline  
\noalign{\smallskip}
\end{tabular}
%\end{adjustwidth}
\end{table}

%\FloatBarrier

The figures in this Appendix show examples of light curves with the best-fit template overplotted in red. Black dots are the outliers excluded from the fit.
\begin{figure*}
\begin{adjustwidth}{-0.75 cm}{}  
    \includegraphics[width=1.1\textwidth]{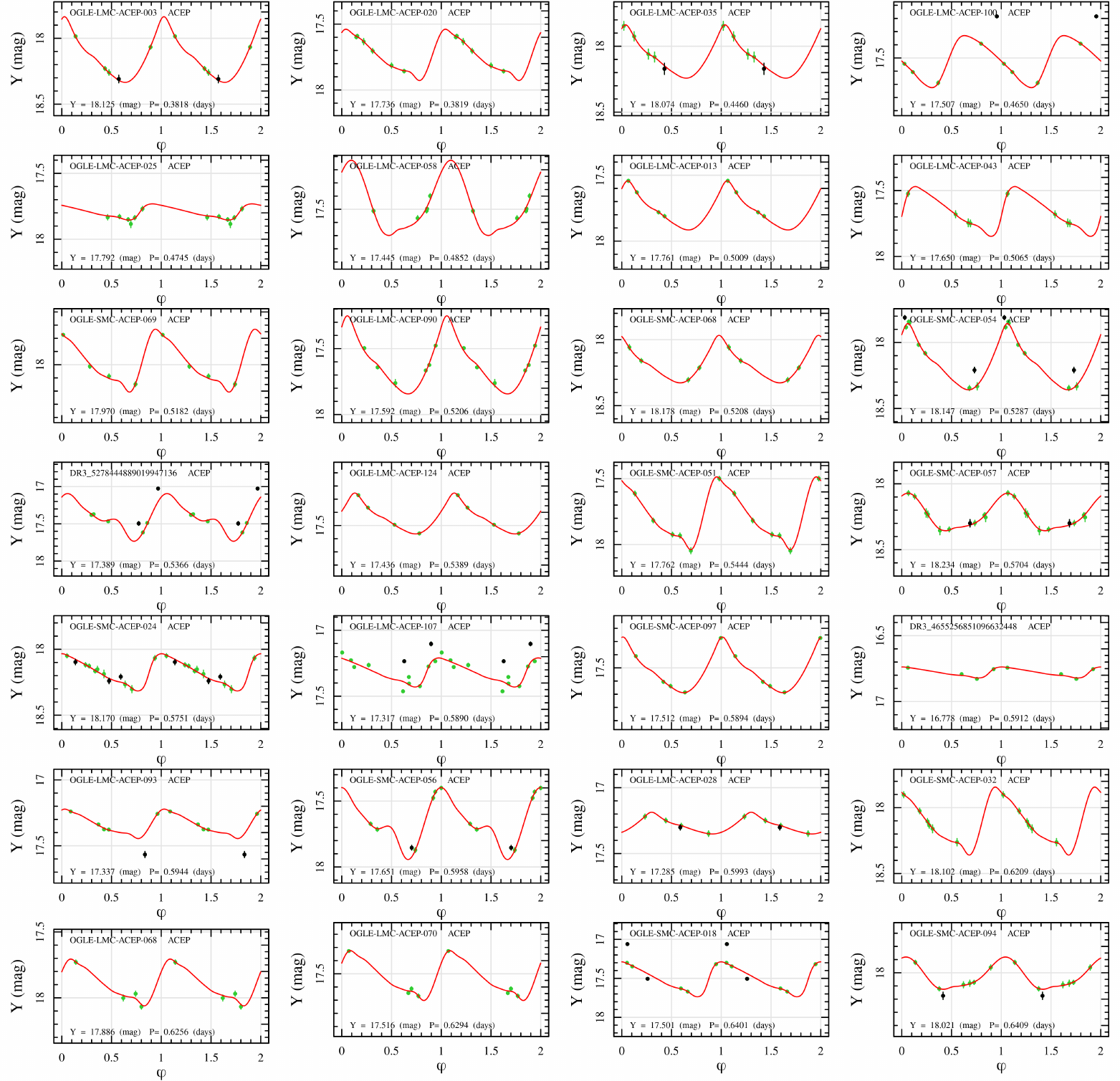}
    \caption{Examples of fitted templates for ACs light curves in the $Y$ band.}
    \label{fig:fittemplt2y}
    \end{adjustwidth}
\end{figure*}
\begin{figure*}
\begin{adjustwidth}{-0.75 cm}{}
    \includegraphics[width=1.1\textwidth]{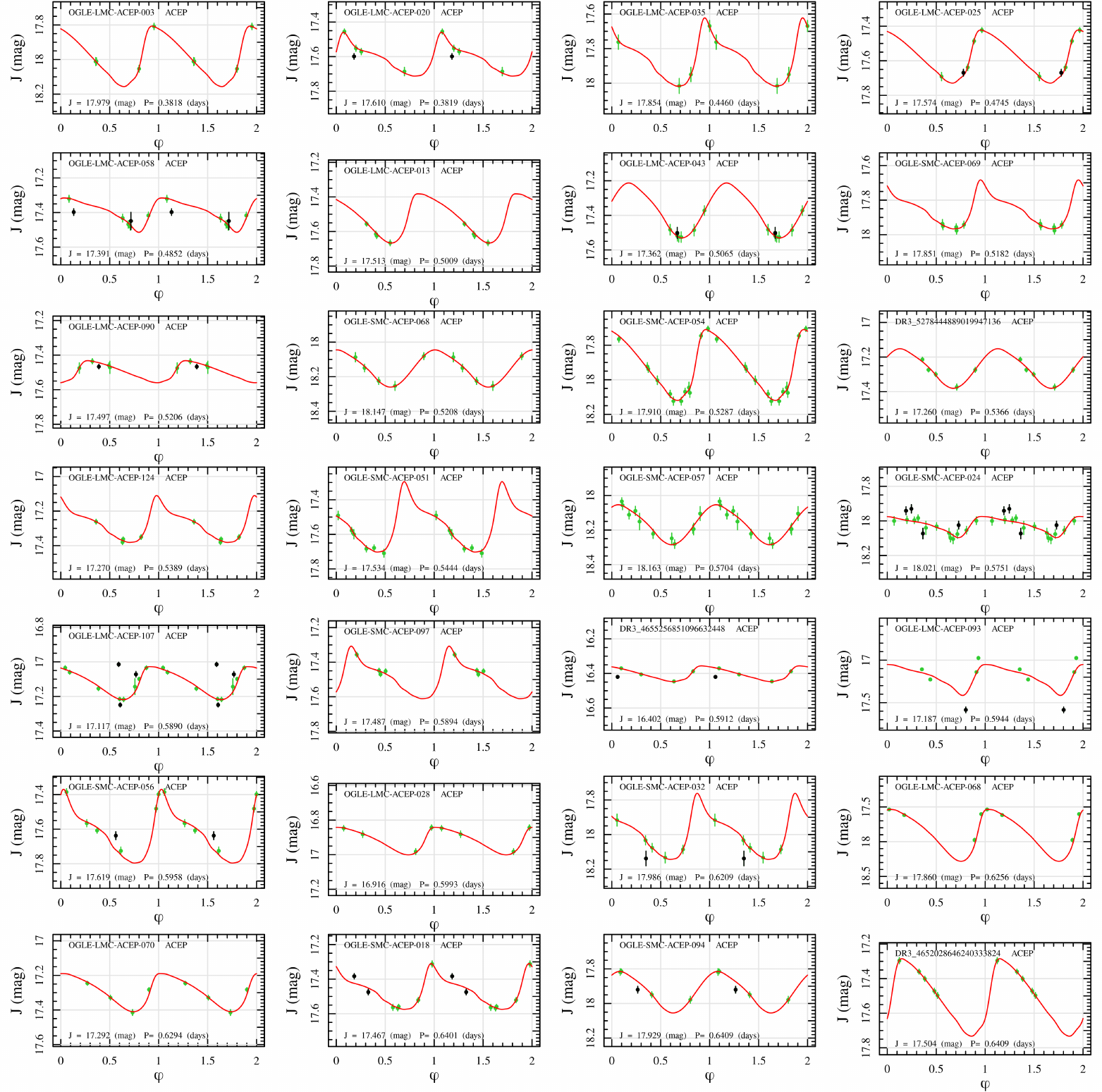}
    \caption{As for Figure~\ref{fig:fittemplt2y} but in the $J$ band.}
    \end{adjustwidth}
\end{figure*}
\begin{figure}
    \begin{adjustwidth}{-0.75 cm}{}
    \includegraphics[width=1.1\textwidth]{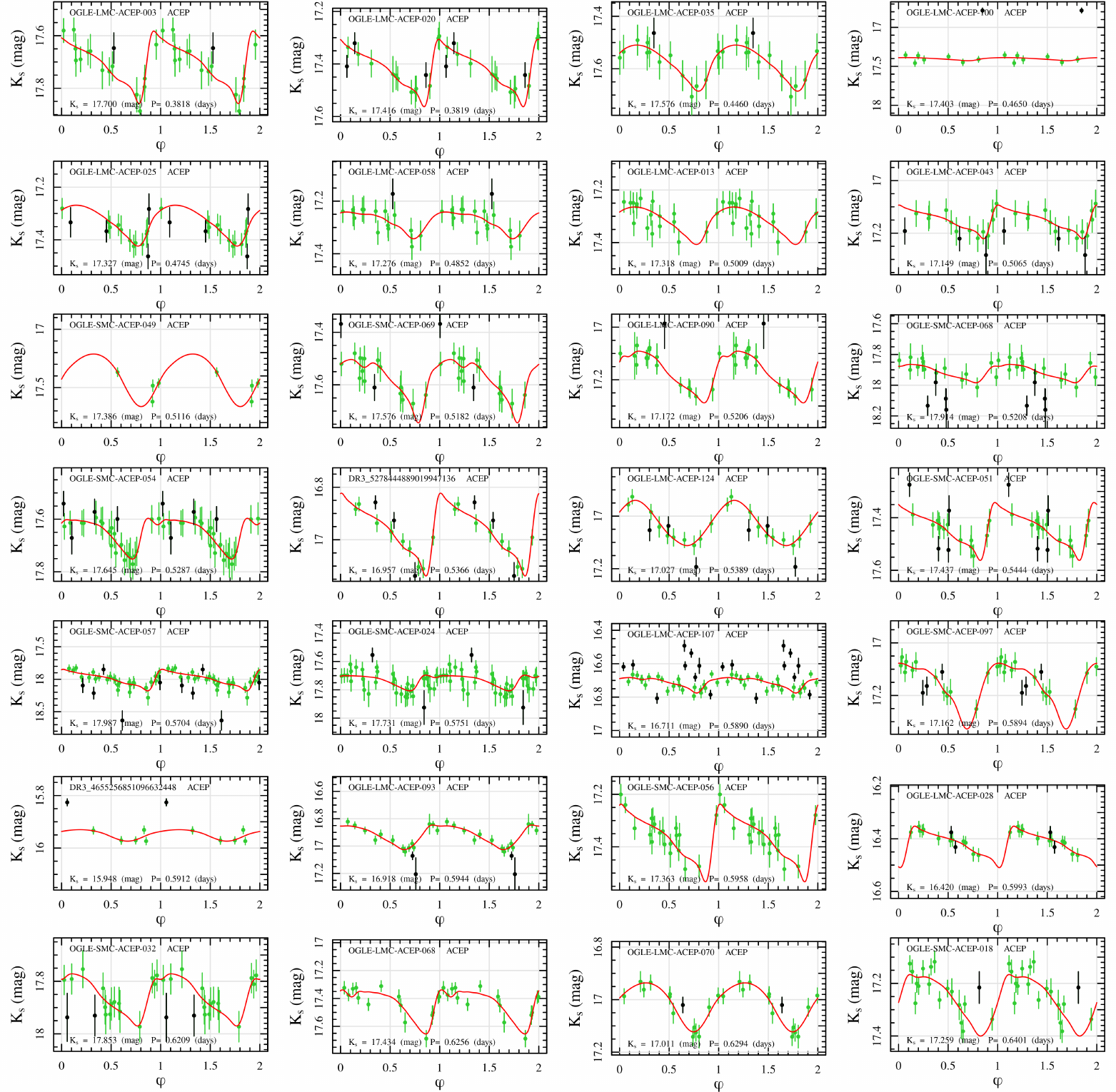}
    \caption{As for Figure~\ref{fig:fittemplt2y} but in the $K_S$ band.}
    \end{adjustwidth}
\end{figure}

\section{Comparison of relative zero points based on different extinction laws}\label{diffext}
Table \ref{diffzero} lists the coefficients of the PWJK for the LMC F mode and F+1O mode ACs obtained using different extinction laws \citep[][]{Wang_2023,1989ApJ...345..245C,fitzpatrick1999} and calibrated by adopting the geometric LMC distance as furnished by P19.

\begin{table*}[h]
  \footnotesize\setlength{\tabcolsep}{4pt} 
  \caption{Comparison between PWJK relations in the LMC based on different extinction laws. } 
  \label{diffzero}
 \begin{center}
  \begin{tabular}{lllllllll} 
  \hline  
  \noalign{\smallskip}   
         Mode & $\frac{A_{K_s}}{A_J-A_{K_s}}$ &$\alpha $ &  $\sigma_{\alpha}$& $\beta$ & $\sigma_{\beta} $ & RMS & n.stars & Ext. law \\
    &  & mag & mag & mag/dex & mag/dex  &mag & &   \\                                        
  \noalign{\smallskip}
  \hline  
  \noalign{\smallskip}  
  F& 0.52 &  16.564 & 0.024& $-$3.20 &0.13 &0.146 & 76& \citet{Wang_2023} \\
  F &0.69 & 16.504 & 0.025 & $-$3.21 & 0.13 & 0.152 & 76 & \citet{1989ApJ...345..245C} \\
  F & 0.74 & 16.492 & 0.025 & $-$3.21 & 0.13 & 0.152 & 76 & \citet{fitzpatrick1999} \\
 F+1O & 0.52 &  16.542 & 0.015 & $-$3.19 & 0.11  & 0.156 & 111& \citet{Wang_2023} \\
 F+1O & 0.69 &16.494 & 0.022 & $-$3.25 & 0.11 & 0.148 & 109& \citet{1989ApJ...345..245C} \\
 F+1O &0.74 &16.482 & 0.015 & $-$3.25 & 0.11 &0.148 & 109& \citet{fitzpatrick1999} \\

  \noalign{\smallskip}
  \hline  
  \noalign{\smallskip}
  \end{tabular}
  \tablefoot{The different columns report: $(1)$ the pulsating mode; $(2)$ the coefficient of the color term in the PWJK; $(3)$--$(4)$ the $\alpha$ coefficient (relative intercept) and relative uncertainty; $(5)$--$(6)$ the $\beta$ coefficient (slope) and relative uncertainty; $(7)$ the Root Mean Square (RMS) of the relation; $(8)$ the number of stars used in the fit; $(9)$ the corresponding extinction law.}
  \end{center}
\end{table*}

\section{Determination of the average magnitude in $J,\,Ks$ bands for Galactic ACs} \label{averagemag}
The identification of the Galactic AC sample was possible thanks to the OGLE IV and the Gaia surveys, resulting in 174 and 276 objects, respectively. From these surveys, we obtained periods and epochs of maximum and optical photometry in the $Gaia$ bands. For the objects for which $Gaia$ did not provide intensity-averaged magnitudes \citep[see][]{Ripepi2023}, i.e. those only present in the OGLE IV catalog, we adopted the arithmetic mean published in the main $Gaia$ source. Indeed, it has been shown that the Wesenheit magnitude in the $Gaia$ bands calculated from the arithmetic mean does not differ significantly from that obtained from intensity-averaged magnitudes \citep[e.g.][]{Ripei2022_PW_Gaia}. 

Concerning the NIR photometry, we only have a single-epoch photometry from the 2MASS  \citep[Two Micron All Sky Survey,][]{2MASS} survey. Since this could lead to errors as large as 0.2 mag (depending on the amplitude of the AC light curves), we decided to use a standard template-fitting technique to obtain mean magnitudes in the $J$ and $K_s$ bands. 
In principle, the adopted technique is similar to that used in Section 3.2. However, in this case, we only have one epoch of observation from 2MASS. Therefore we cannot rescale the amplitude of the star using the observations themselves but instead, we have to assume fixed amplitude ratios in the $J, K_s$ bands compared to amplitudes in the $I$ or $G$ bands, depending on the source of the data, i.e. OGLE IV or Gaia, respectively. 
%The first step is to put on the right phase the epoch of observation. Imagine a xy plane, where on the x-axis there are the phases and on the y-axis the magnitudes, the quantities needed are: the period of the observed star, epoch of observations, template (with amplitude=1 and mean magnitude=1 by definition), and amplitude.

We determine the average amplitude ratios using the ACs in the LMC and SMC from our sample for which we have well-determined amplitudes in the $J$ and $K_s$ bands and optical amplitudes from OGLE IV and Gaia. We verified that the amplitude ratios remained constant for both F and 1O pulsators with no dependence on the period up to P$\sim$1.8 days. The following equations list the different amplitude ratios in the various bands we adopted:
\begin{eqnarray} \label{eq:amplitudeScaling}
   \frac{amp~K_s}{amp~G}= 0.30 \pm 0.08~, \\
   \frac{amp~J}{amp~G}= 0.45 \pm 0.11~,\\
   \frac{amp~K_s}{amp~I}= 0.45 \pm 0.11~,\\
   \frac{amp~J}{amp~I}= 0.73 \pm 0.25~. \label{eq:amplfin}
\end{eqnarray}

In this way, from an amplitude in the optical band, we derived the amplitudes in the NIR bands. For each mode of pulsation, we used the template with a period closest to that of the analyzed star. 
Therefore for a given AC, the procedure foresees the following steps: i) use the period and epoch of maximum from OGLE IV or Gaia to calculate the phase of the single epoch 2MASS observation (only data with flag "AAA" were taken\footnote{AAA is the 2MASS quality flag indicating good photometry in the HJK bands.}); ii) rescale the amplitude of the assigned template using the appropriate quantity from Eqs.~\ref{eq:amplitudeScaling} to~\ref{eq:amplfin}; iii) add a $\Delta$ Mag to the rescaled template until it overlaps with the 2MASS phased observation; iv) calculate the intensity-averaged magnitude of the template modified as described in the previous two points.

%to use is selected through the period and of the pulsation mode and adapted to the selected star thanks to its amplitude and mean magnitude.  On the xy plane, where is drawn the template, the phase of the single epoch observation will indicate us the magnitude offset. This offset needs to be summed to the former magnitude to calculate the mean magnitude. 
The error in the mean magnitude was calculated using a bootstrap technique, taking into account the uncertainty in the amplitude and the 2MASS magnitude.
The mean magnitudes were then converted from the 2MASS photometric system to the VISTA one applying the transformations provided by \citet{2018MNRAS.474.5459G}. The final data sample is composed of 226 Galactic ACs (125 F and 101 1O mode, respectively). Through the R factor for fundamentalization calculated in Section~\ref{fundsect}, the MW ACs were studied as a unique sample.

\section{Figures} \label{appfigure}
The figures in this Appendix display the observed  and fitted PL and PW relations for all other filters and filter combinations, and the comparison between the LMC and SMC PWJK and PWG relations.
\begin{figure*}
    \vbox{
    \hbox{
    \includegraphics[width=0.34\textwidth]{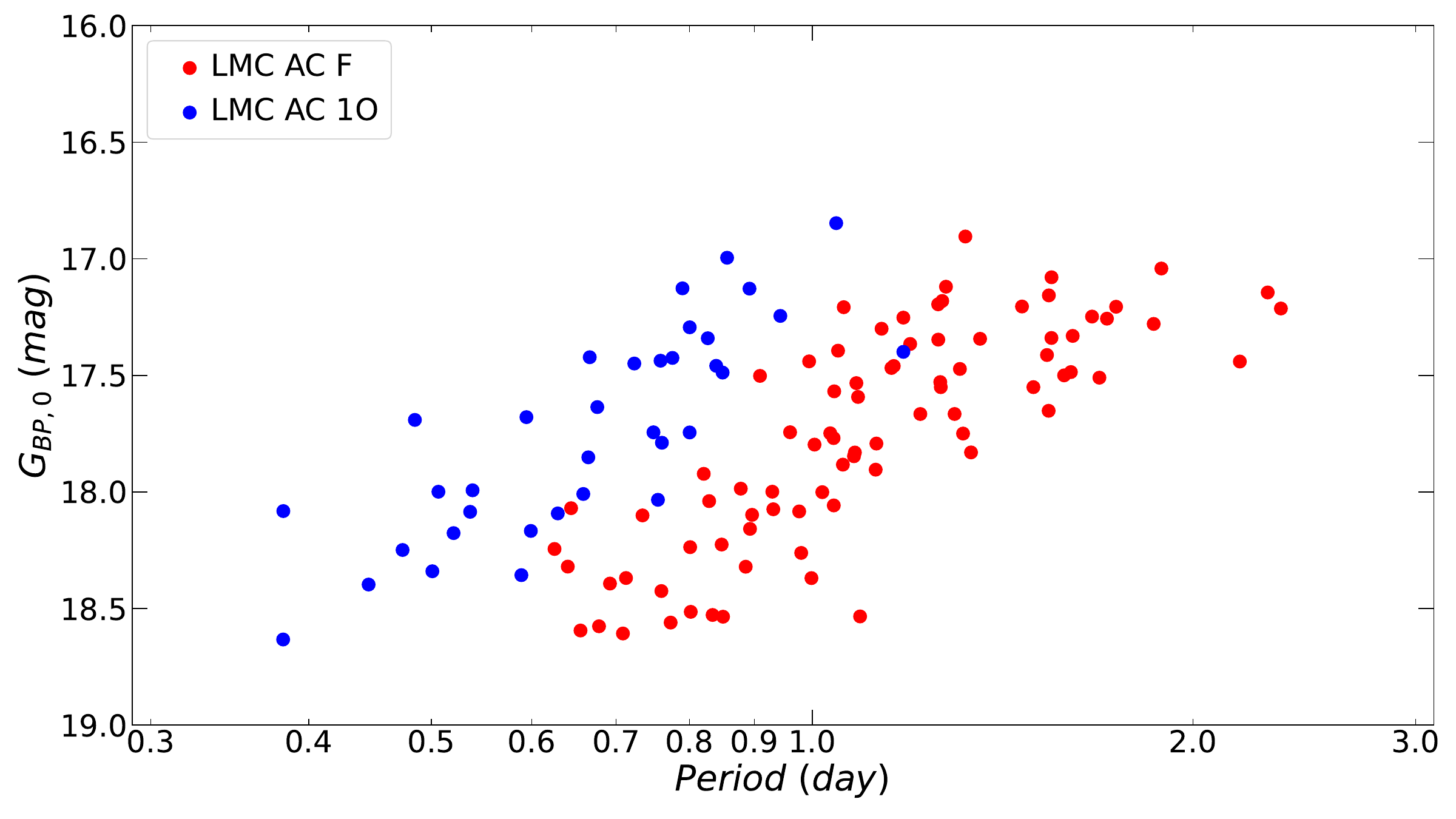}
    \includegraphics[width=0.34\textwidth]{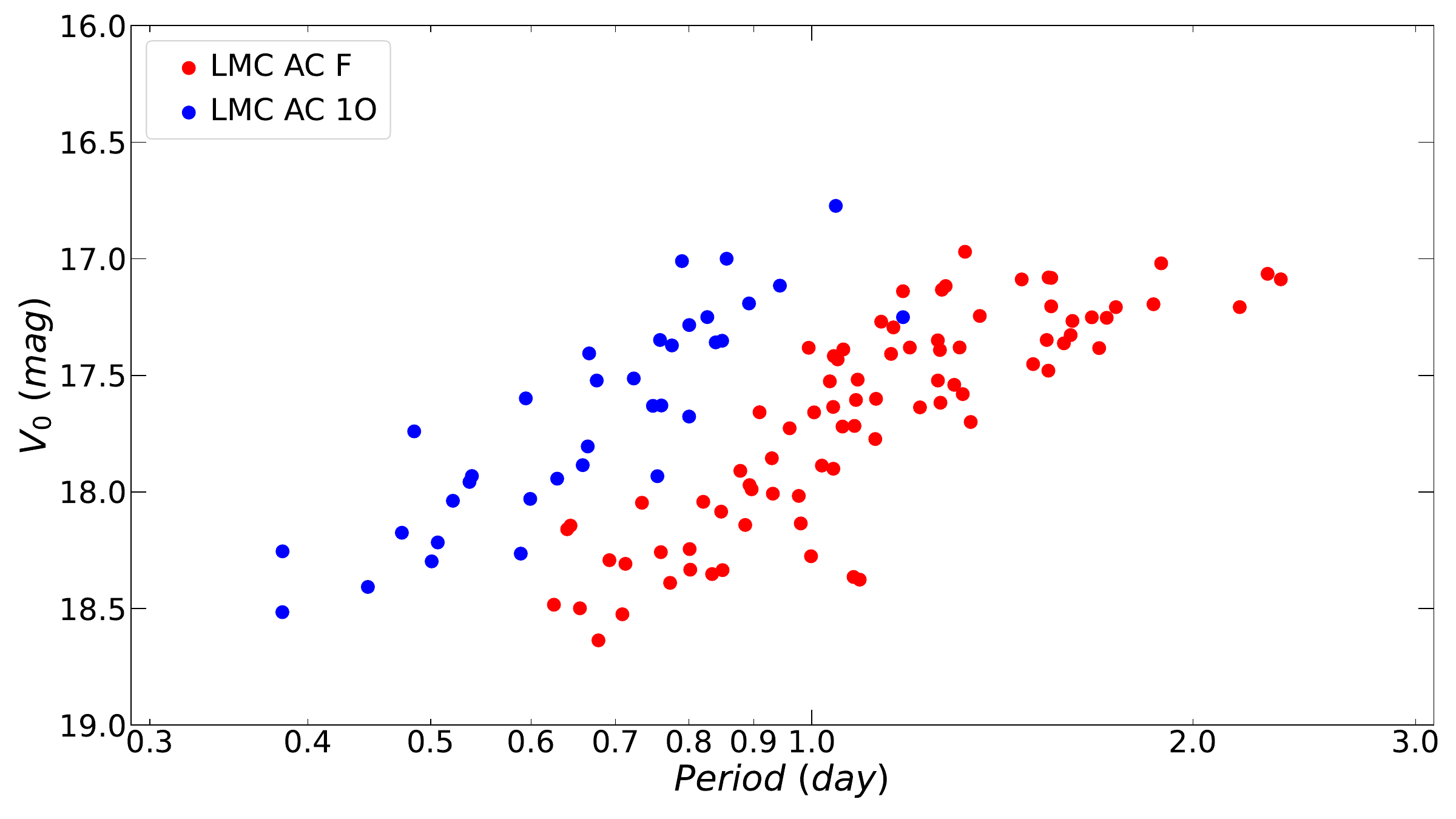}
    \includegraphics[width=0.34\textwidth]{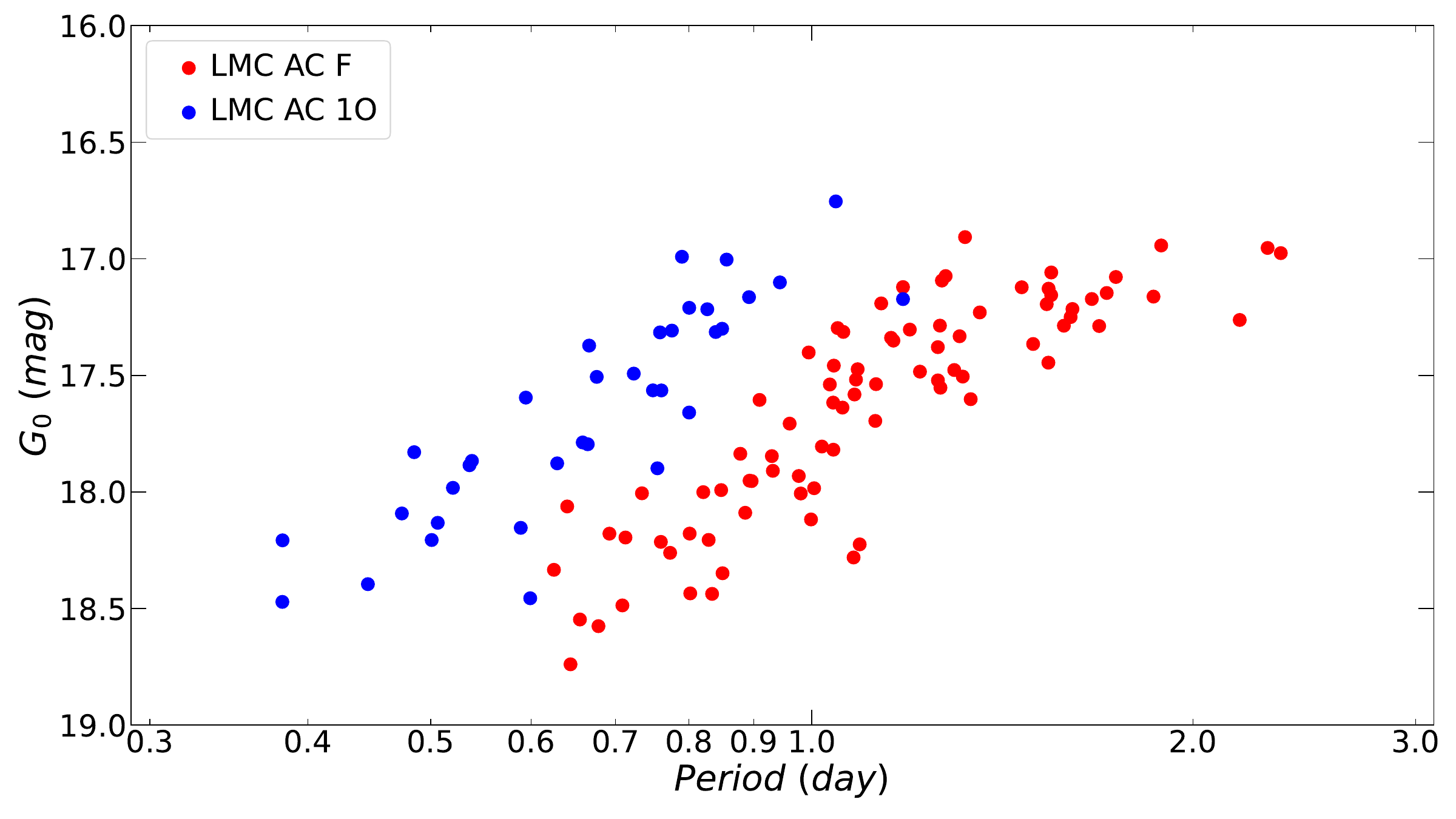}}
    \hbox{
    \includegraphics[width=0.34\textwidth]{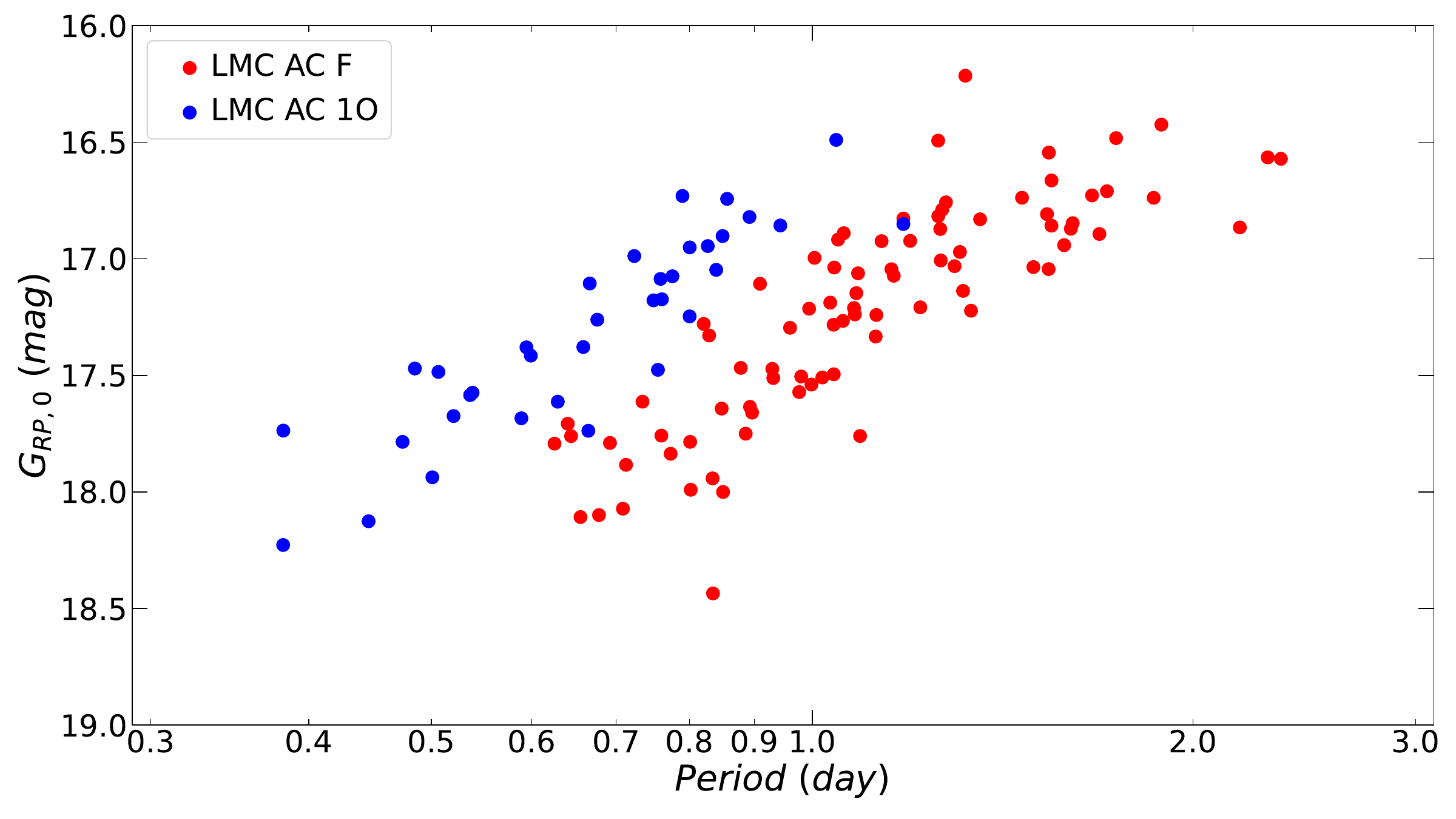}
    \includegraphics[width=0.34\textwidth]{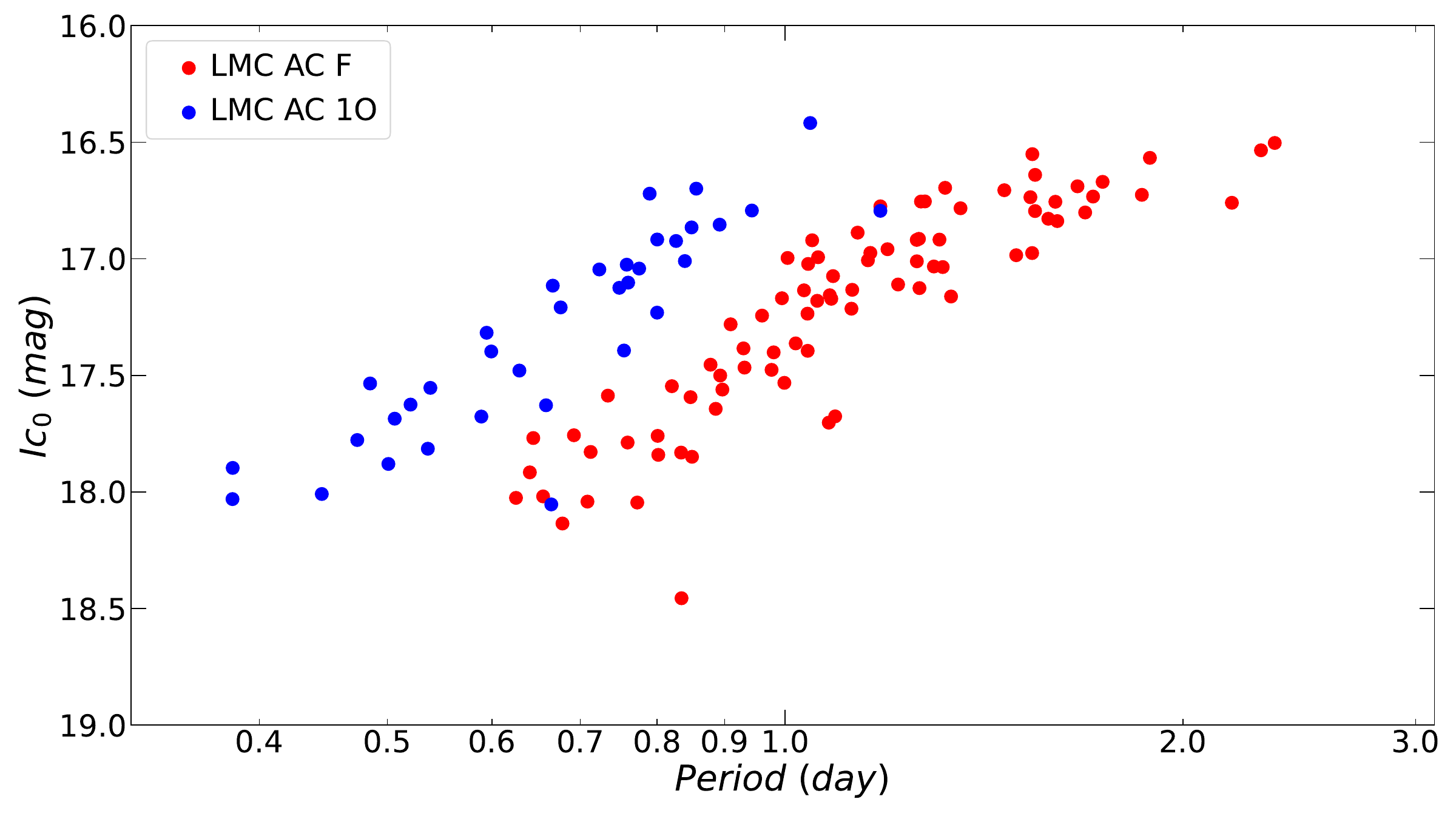}
    \includegraphics[width=0.34\textwidth]{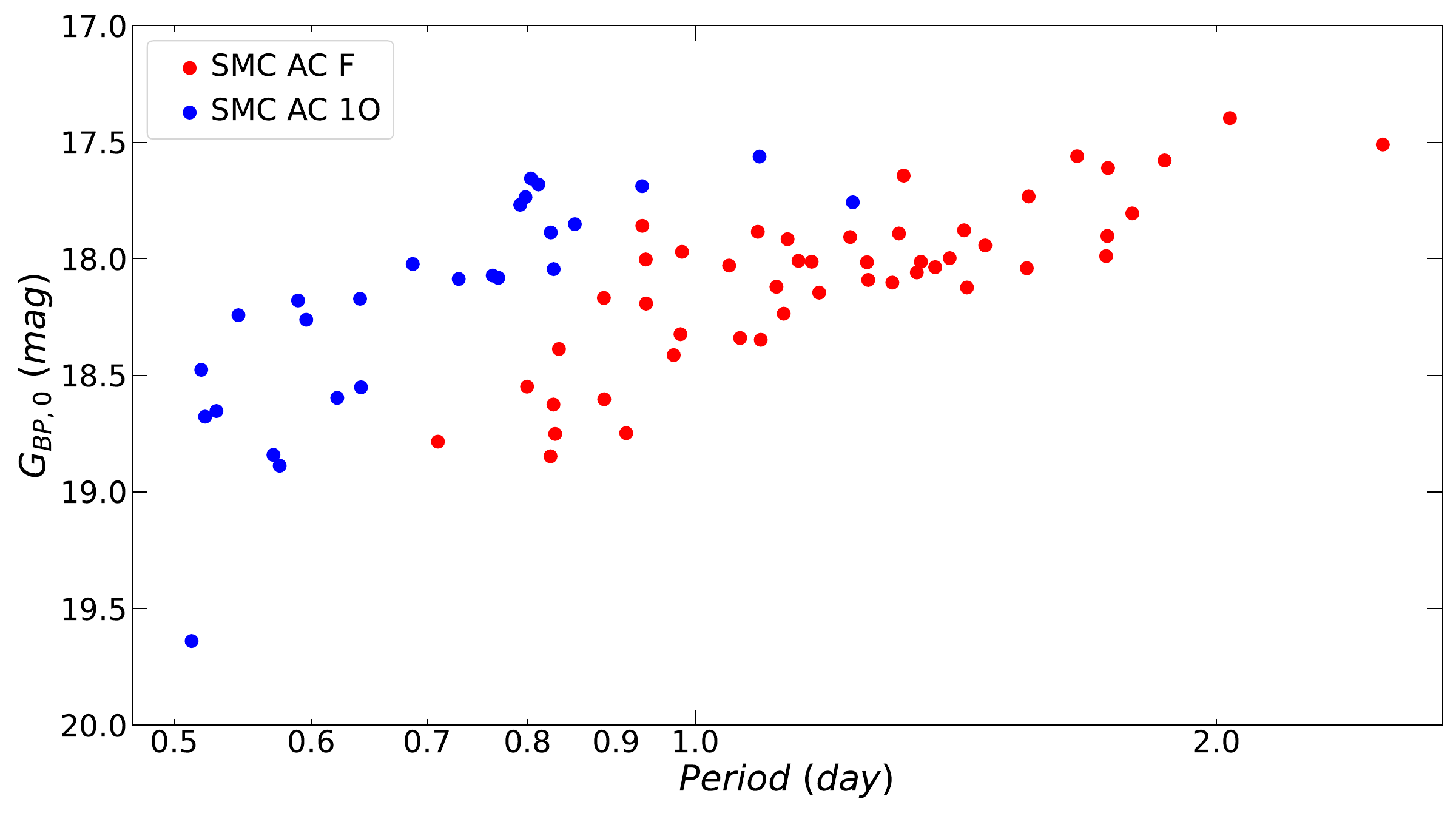}
    }
    \hbox{
    \includegraphics[width=0.34\textwidth]{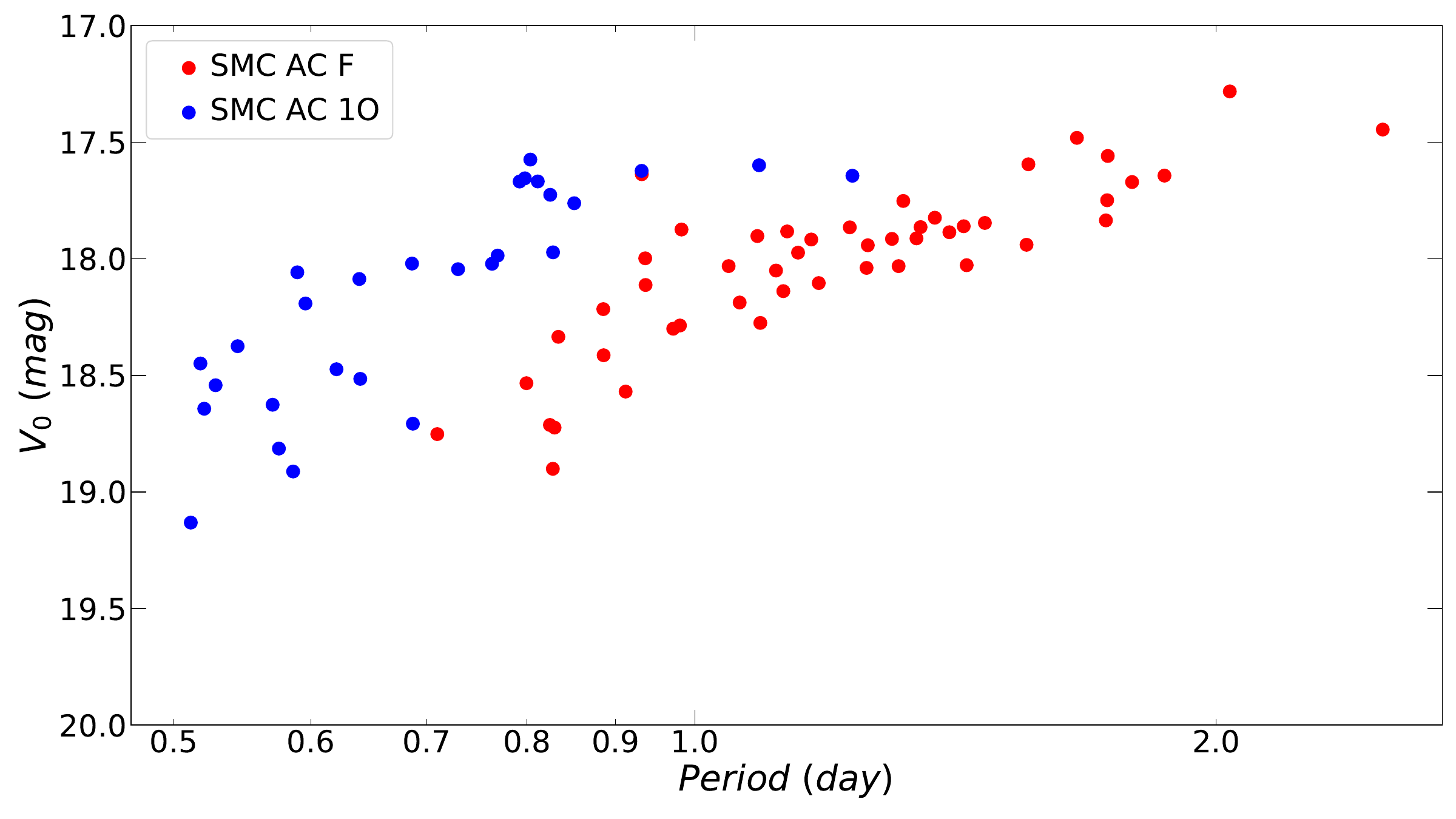}
    \includegraphics[width=0.34\textwidth]{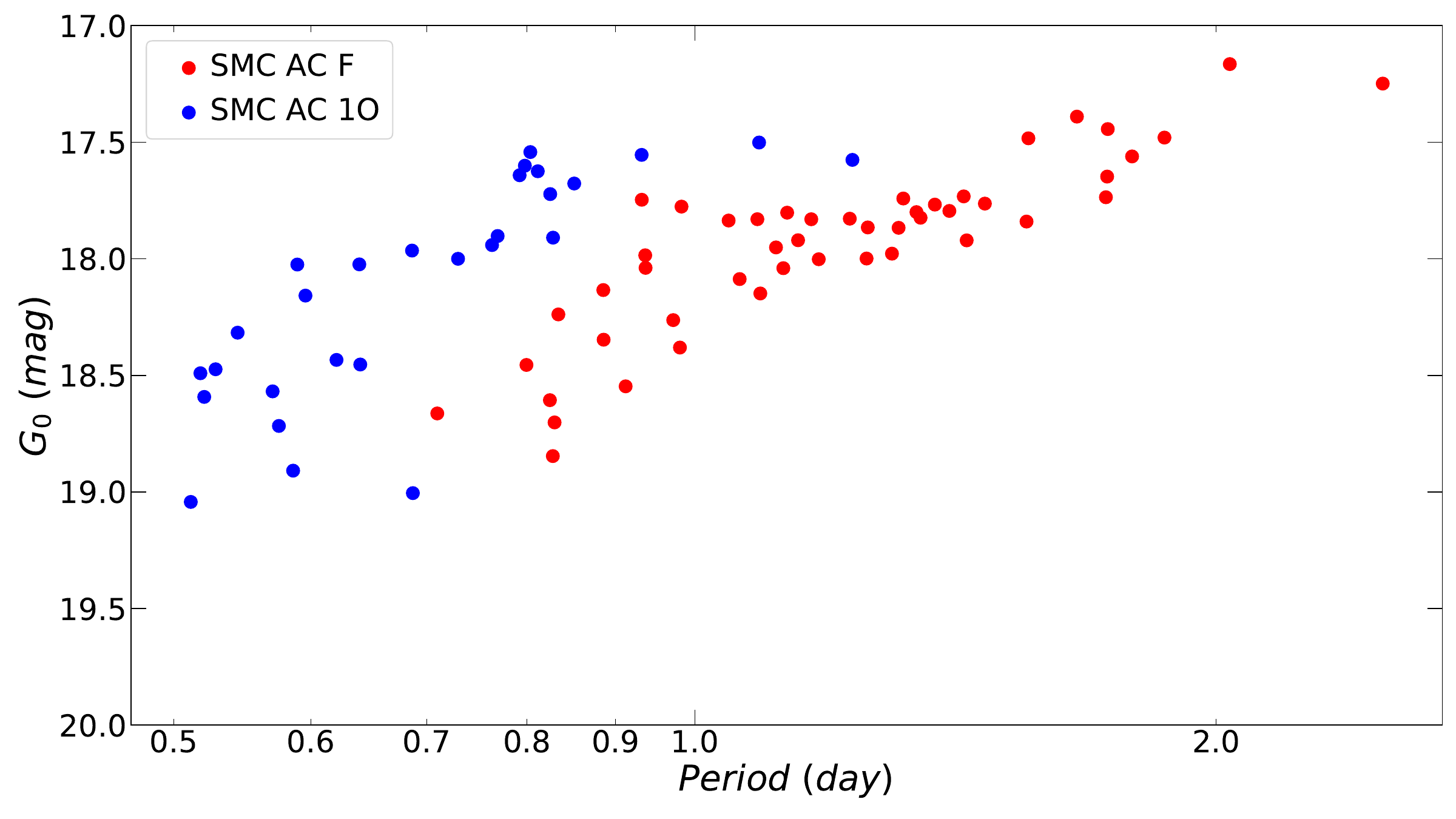}
    \includegraphics[width=0.34\textwidth]{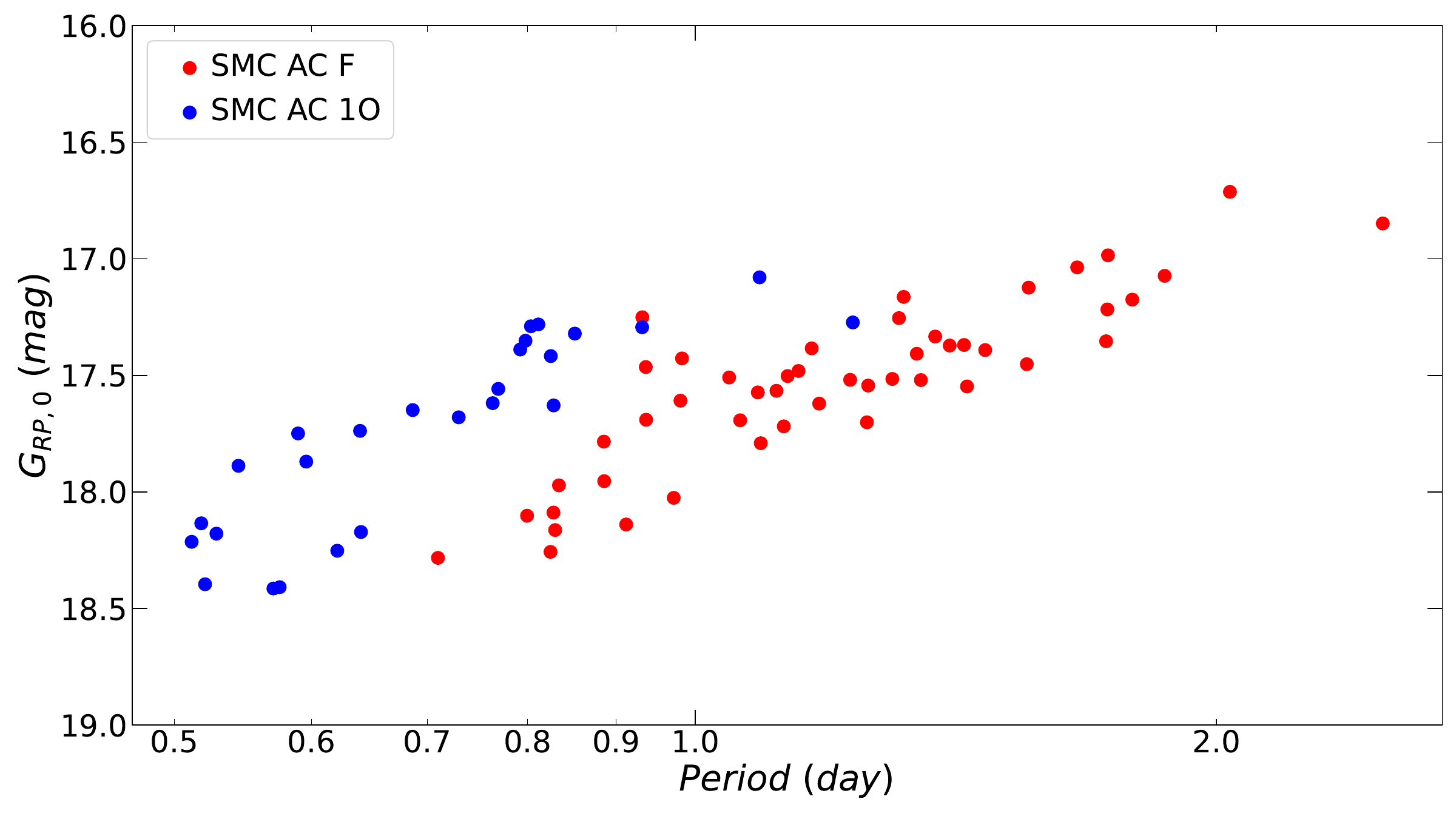}
    }
    \hbox{
    \includegraphics[width=0.34\textwidth]{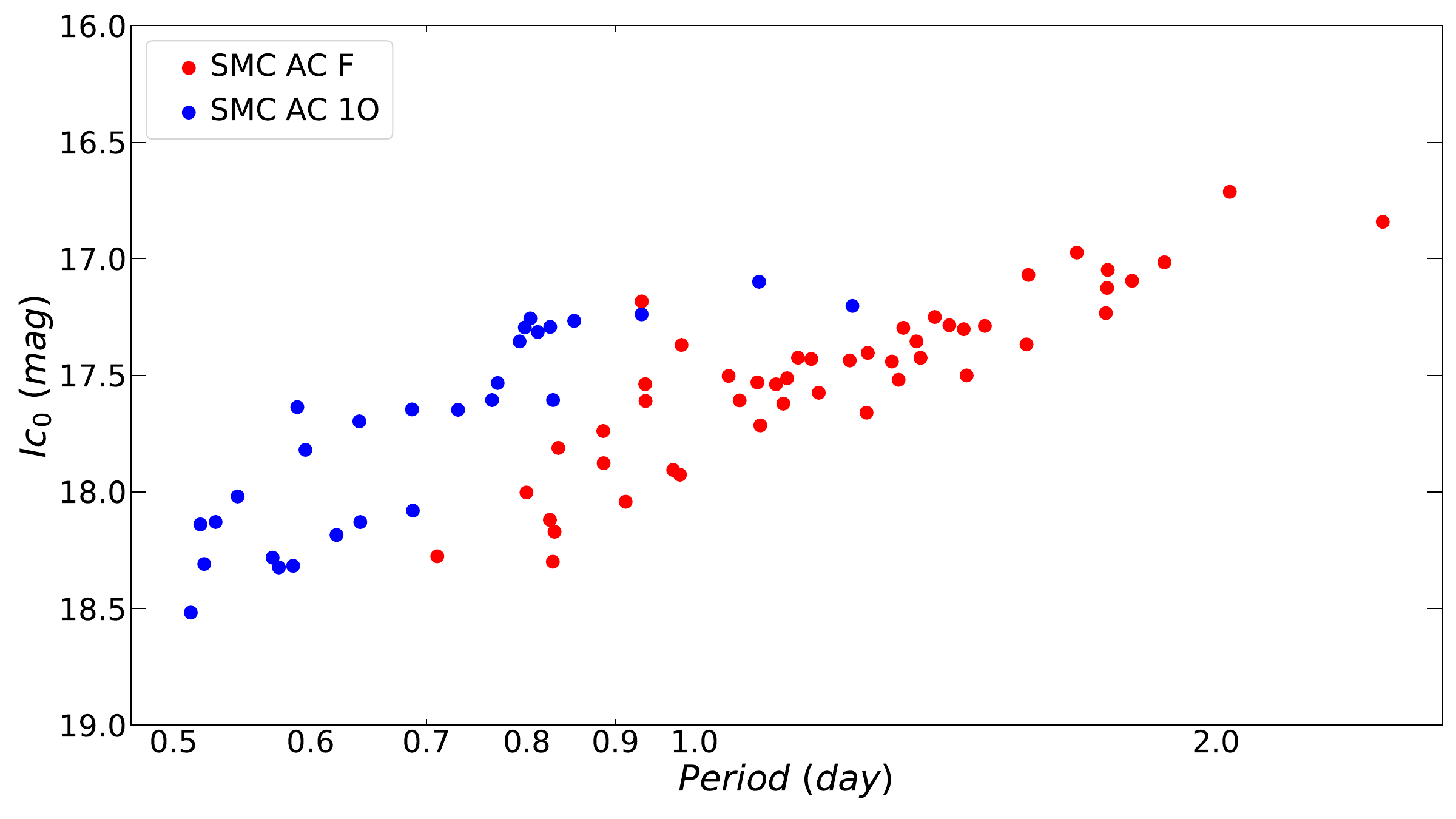}
    \includegraphics[width=0.34\textwidth]{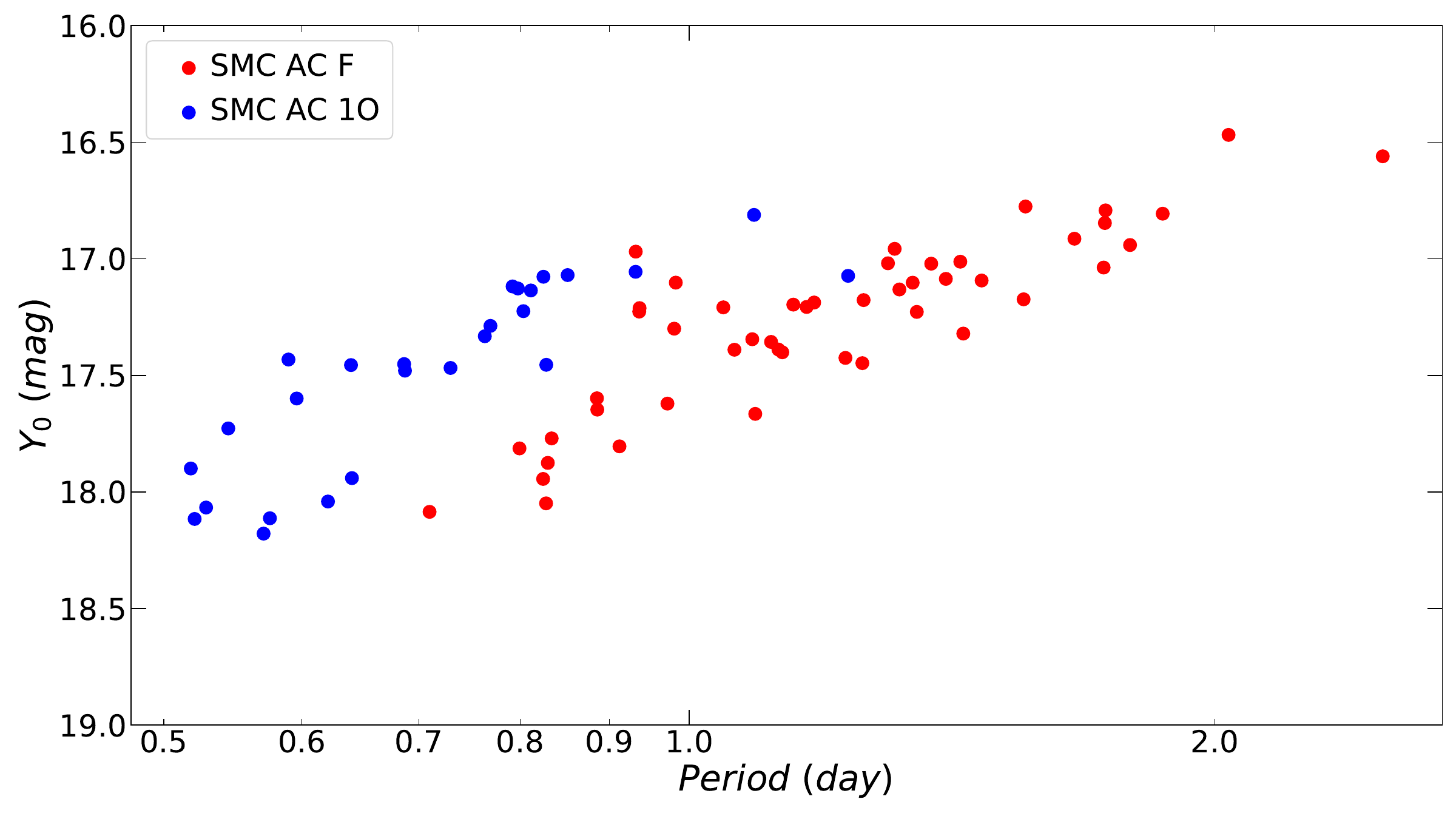}
    \includegraphics[width=0.34\textwidth]{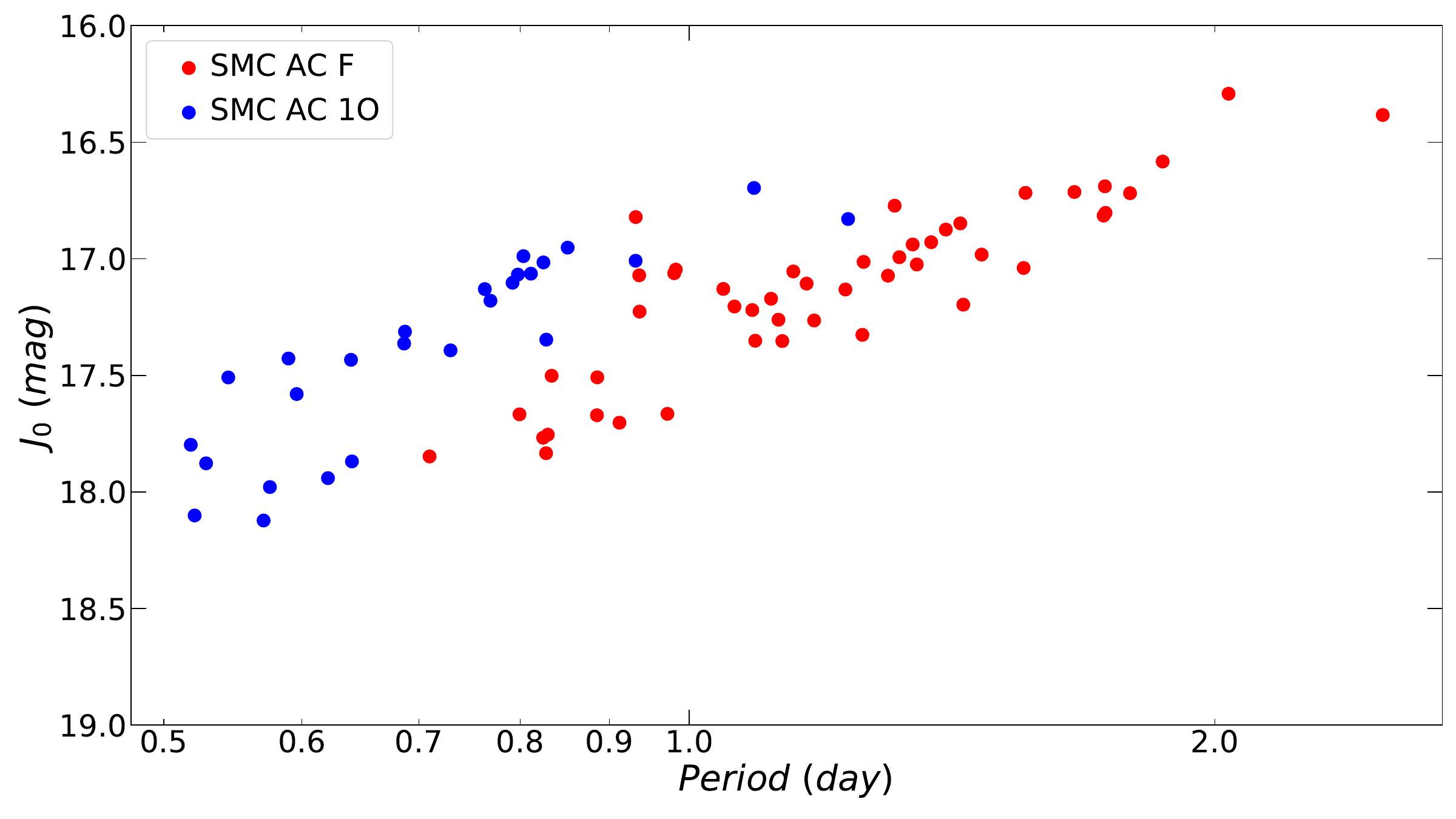}
    }
    \hbox{
    \includegraphics[width=0.34\textwidth]{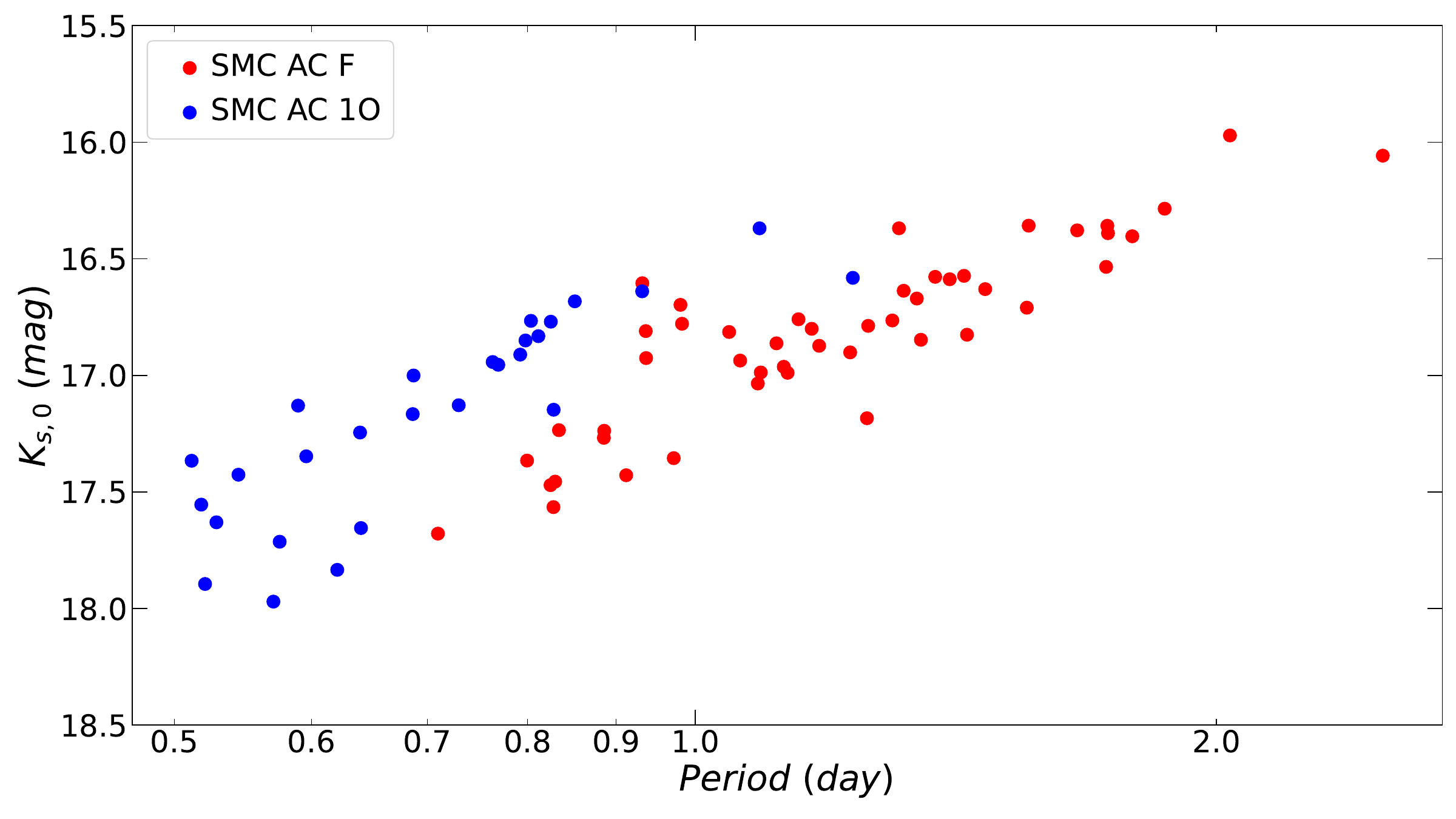}
    \includegraphics[width=0.34\textwidth]{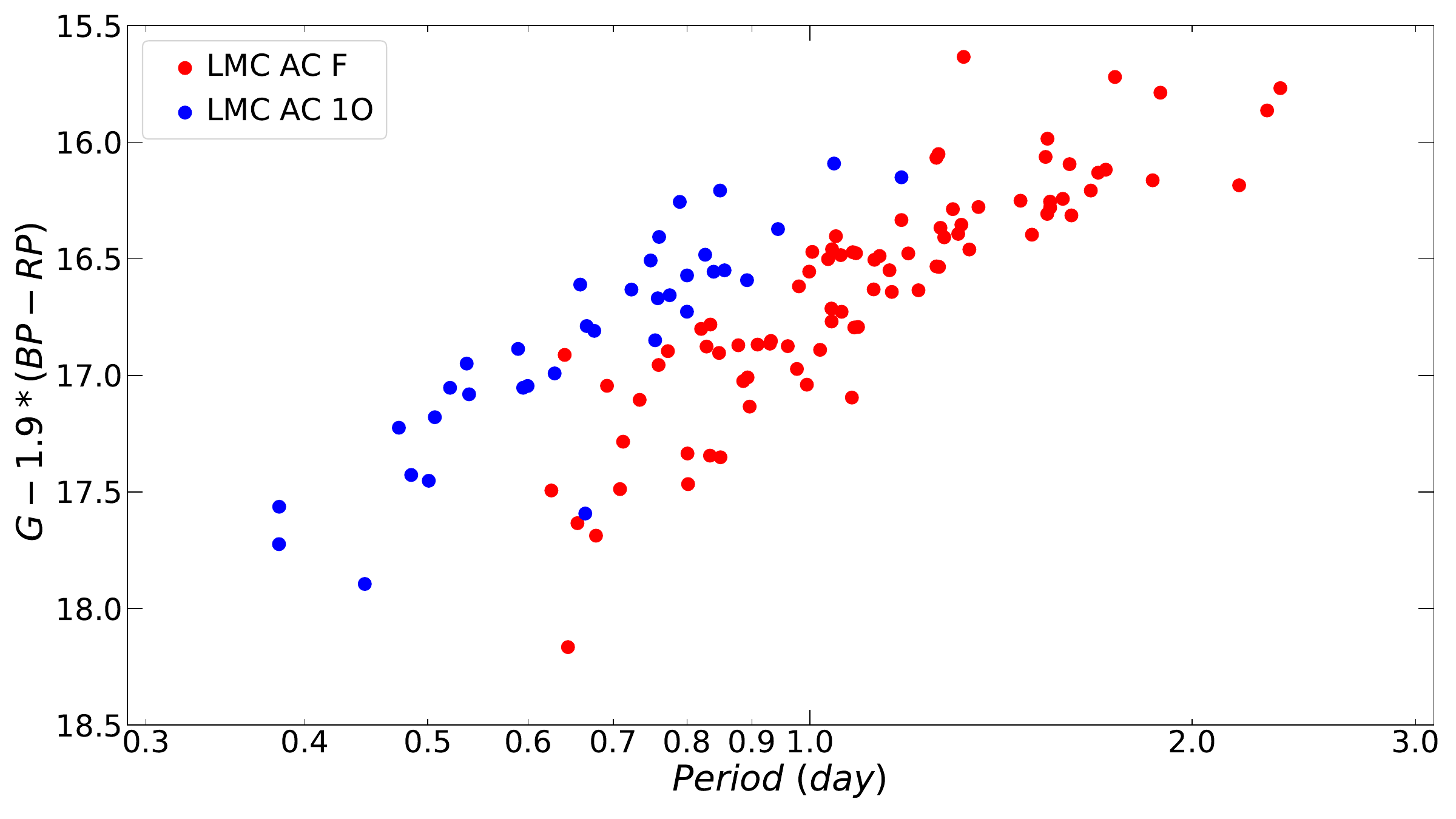}
    \includegraphics[width=0.34\textwidth]{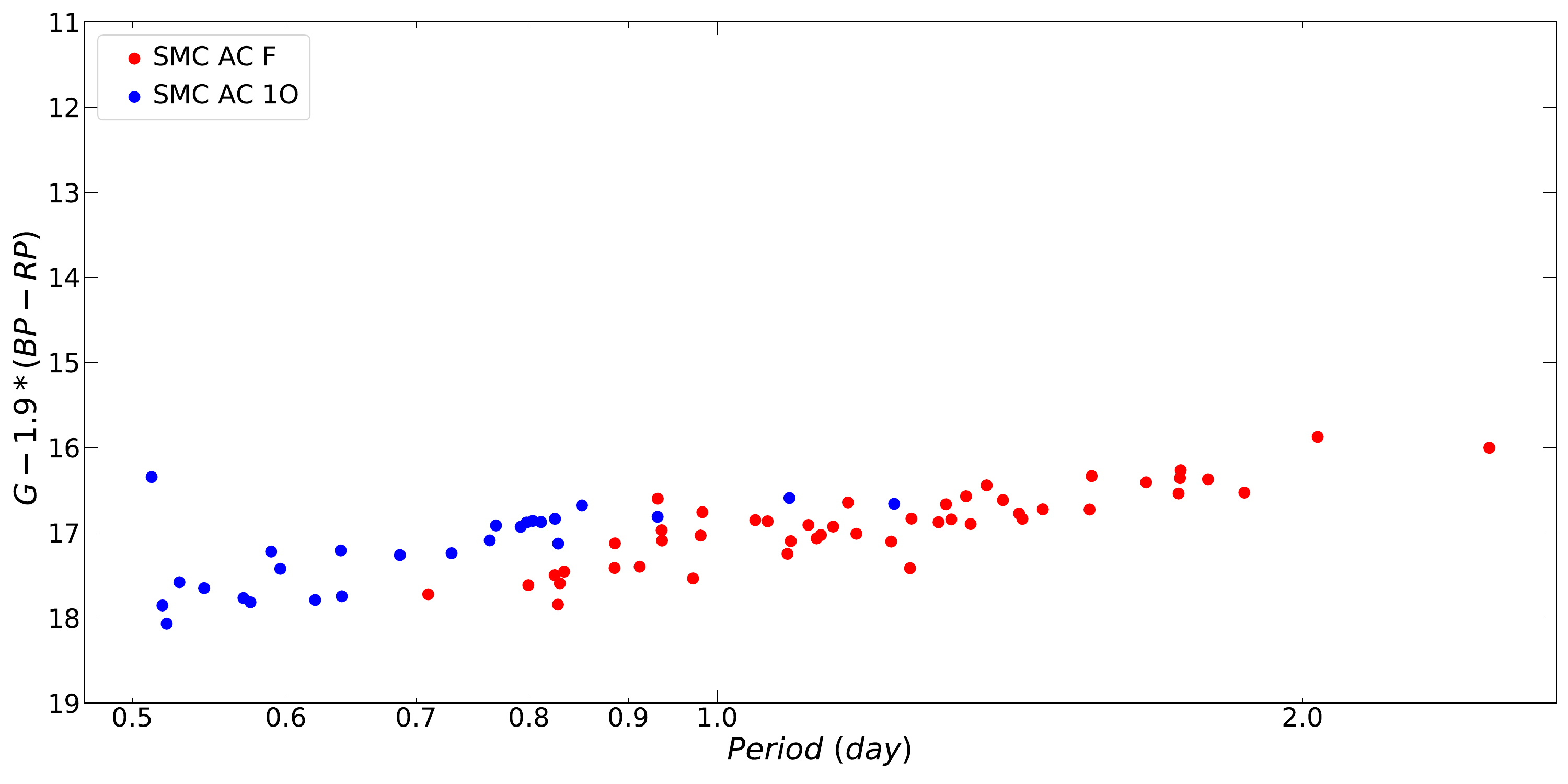}
    }
    \hbox{
    \includegraphics[width=0.34\textwidth]{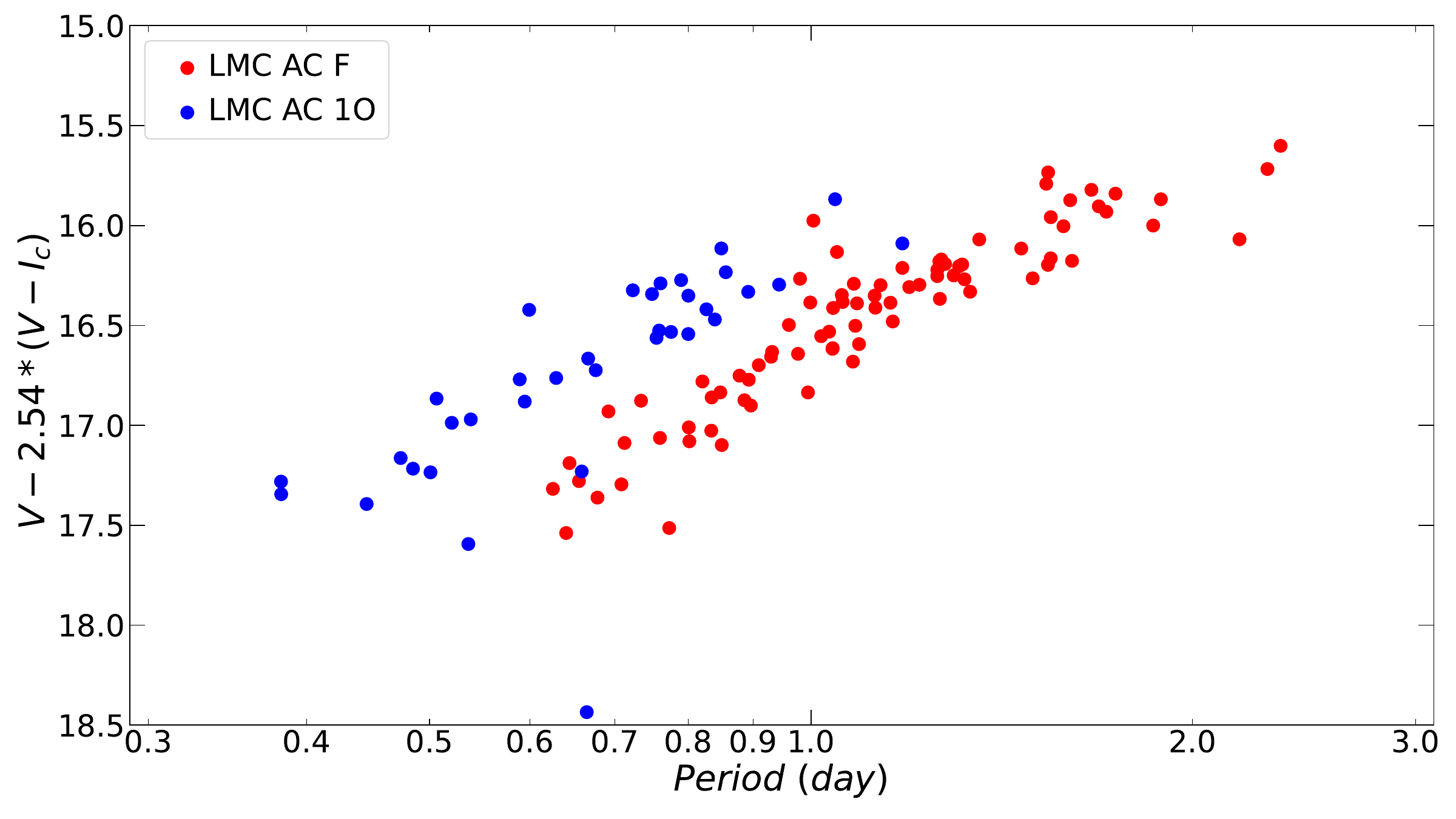}
    \includegraphics[width=0.34\textwidth]{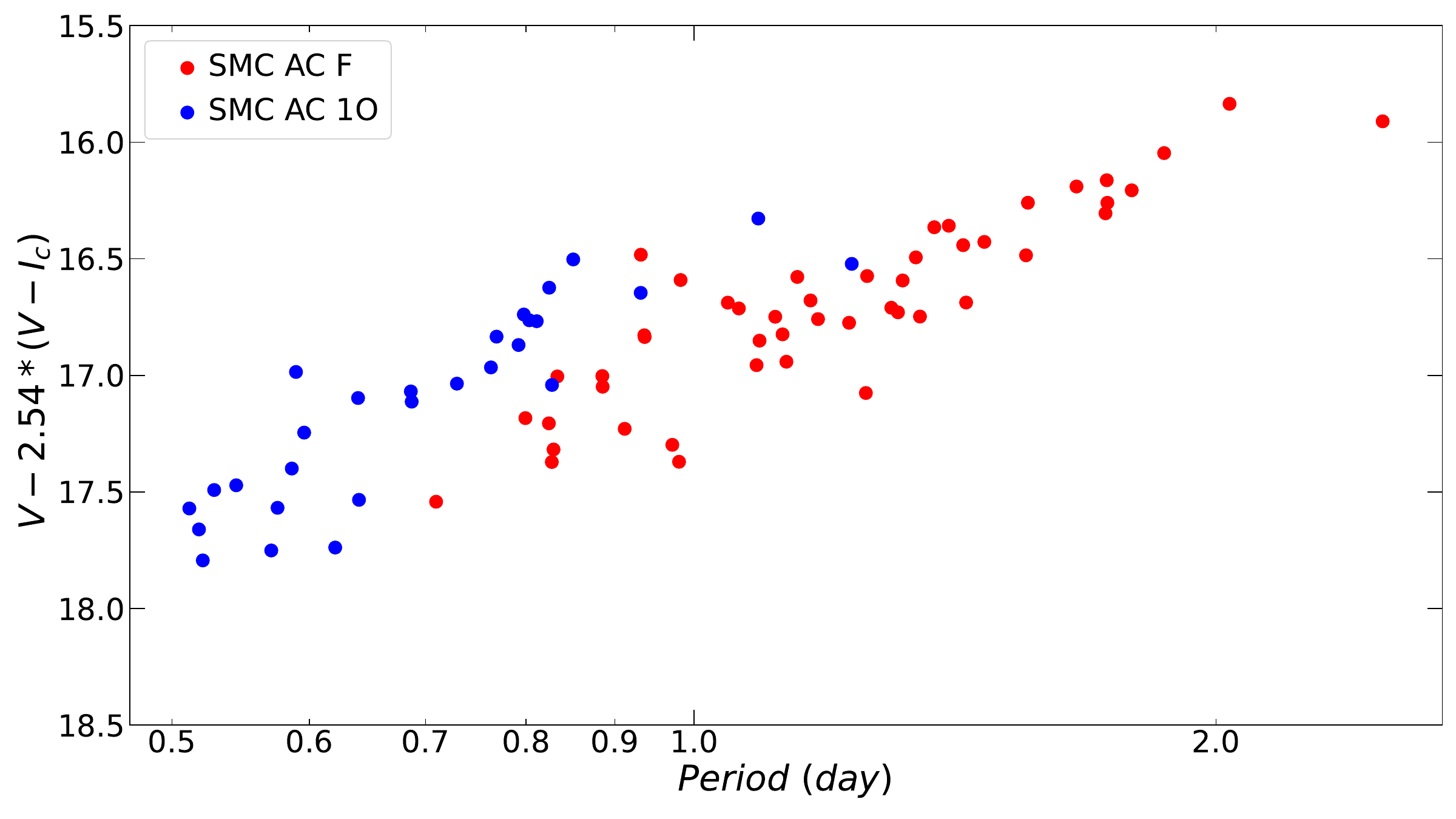}
    \includegraphics[width=0.34\textwidth]{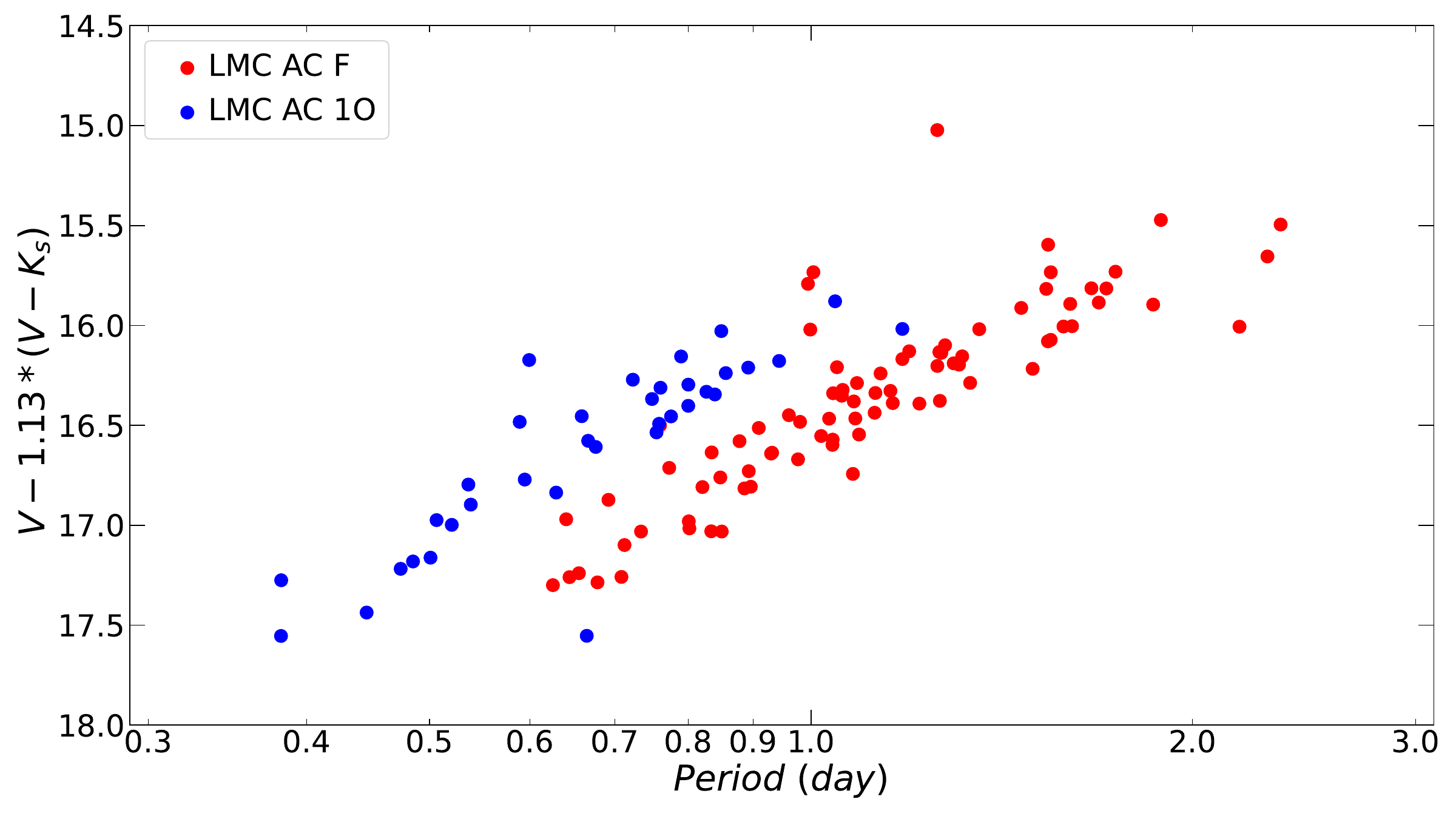}
    }
    \hbox{
    \includegraphics[width=0.34\textwidth]{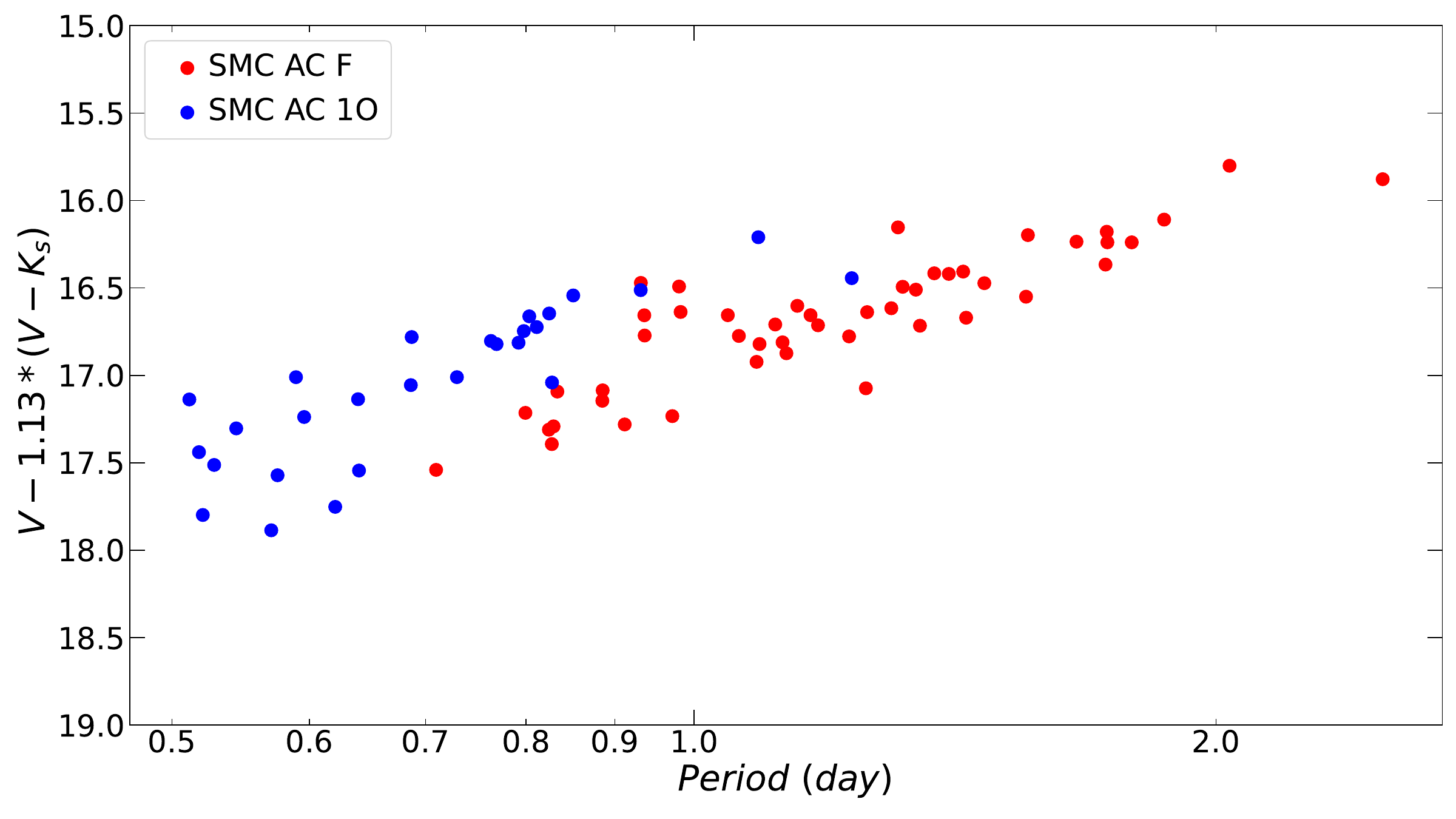}
    \includegraphics[width=0.34\textwidth]{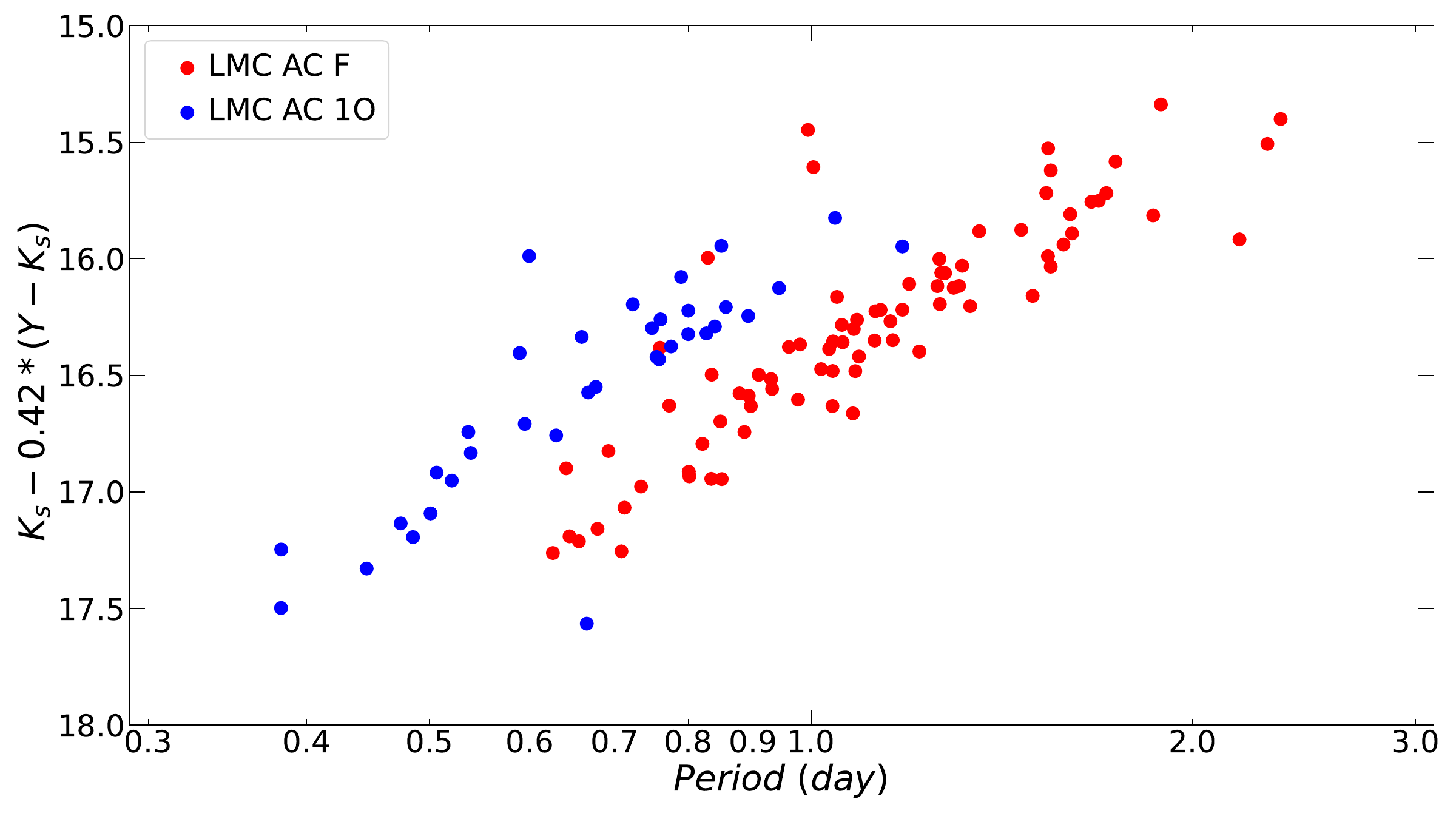}
    \includegraphics[width=0.34\textwidth]{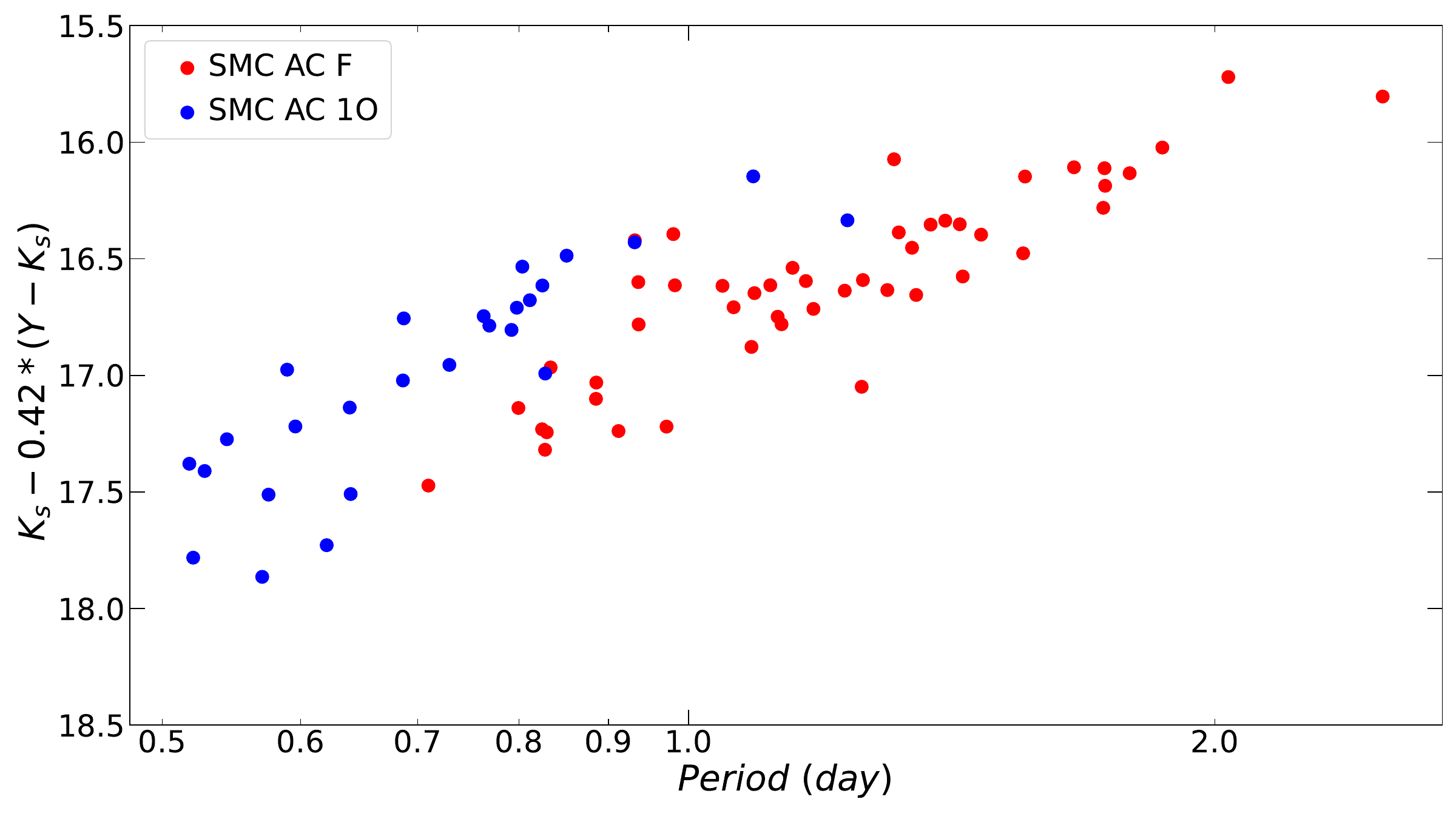}
    }
    }
    \caption{ Same as Fig. \ref{pllmc}, but for the remaining LMC and SMC PL and PW relations.}	
	\label{plsmc}
    \end{figure*}

\begin{figure*}
    %\begin{adjustwidth}{-2cm} 
    \hbox{
    \includegraphics[width=0.34\textwidth]{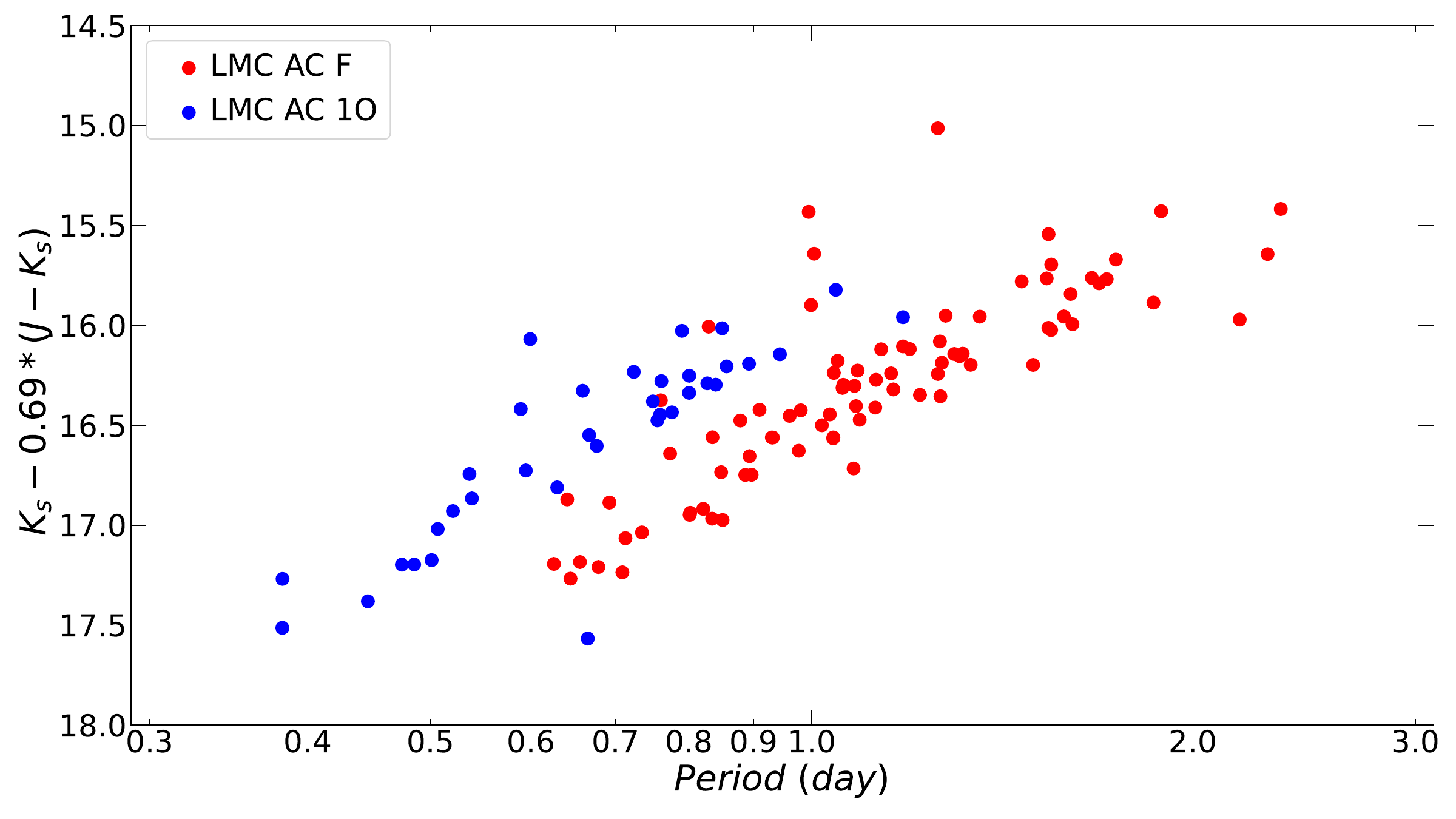}
    \includegraphics[width=0.34\textwidth]{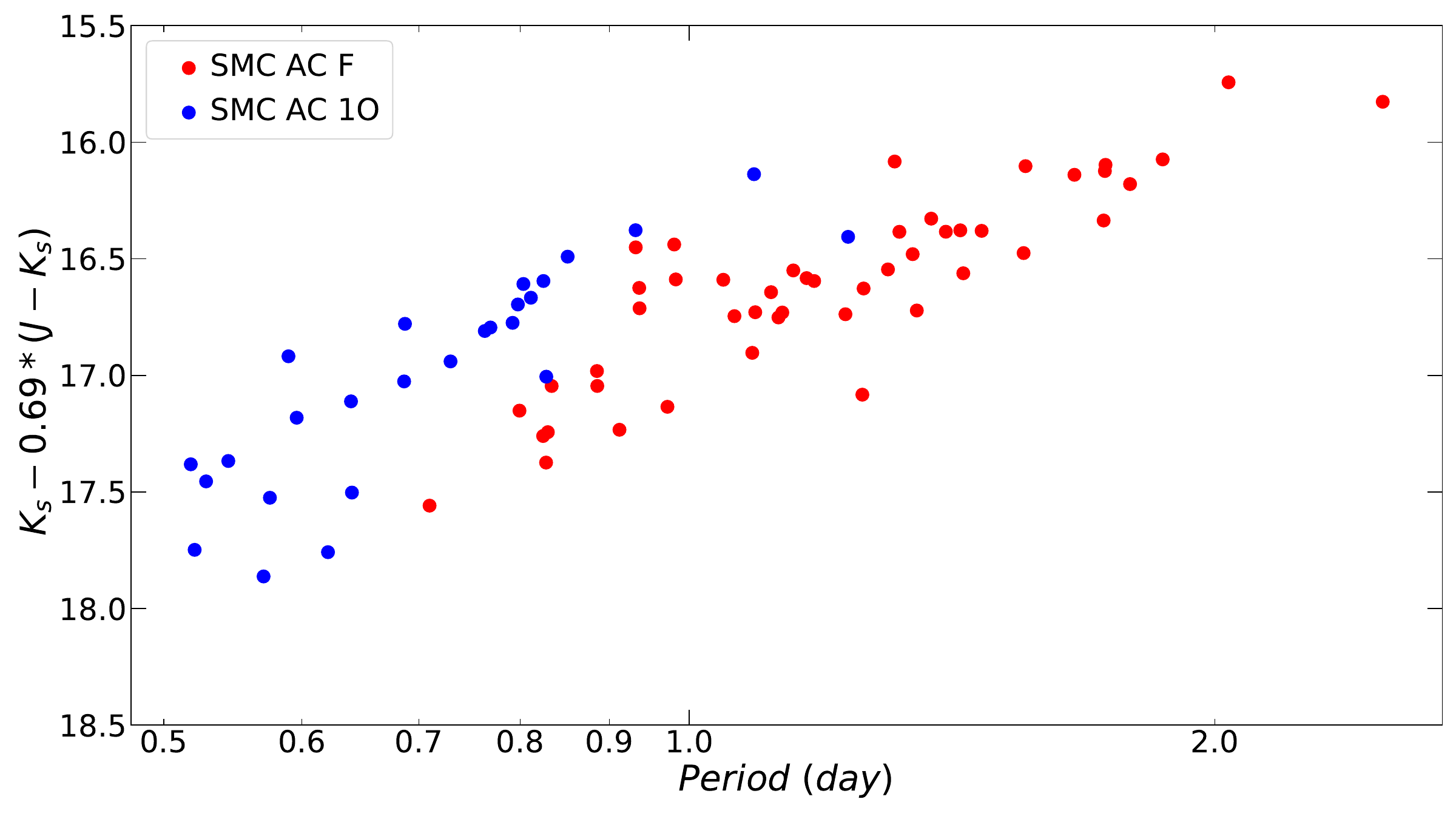}
    }
    \ContinuedFloat
    \caption{continued. }  
	\end{figure*}

\begin{figure*}
%    \begin{adjustwidth}{-10 cm}{}	
    \vbox{
    \hbox{
    \includegraphics[width=0.34\textwidth]{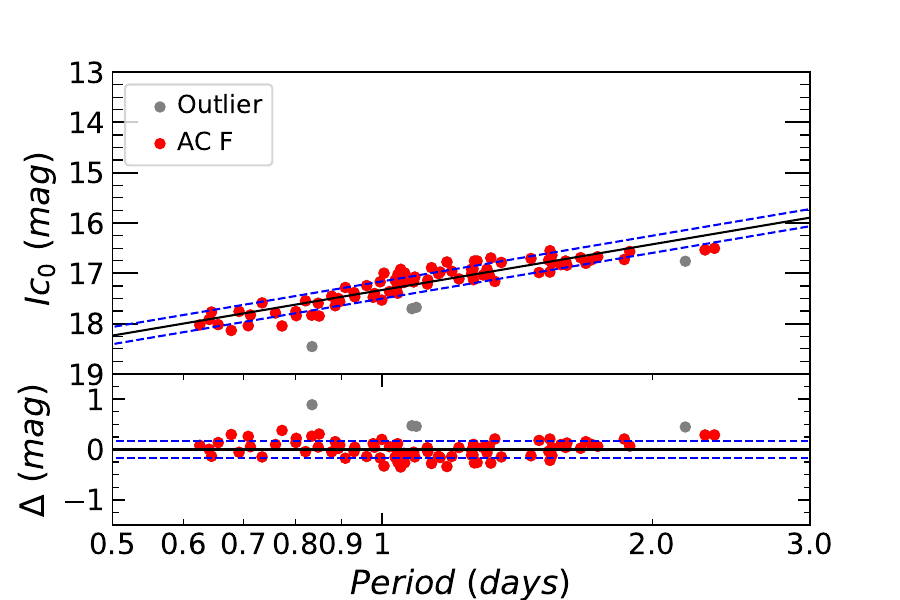}
    \includegraphics[width=0.34\textwidth]{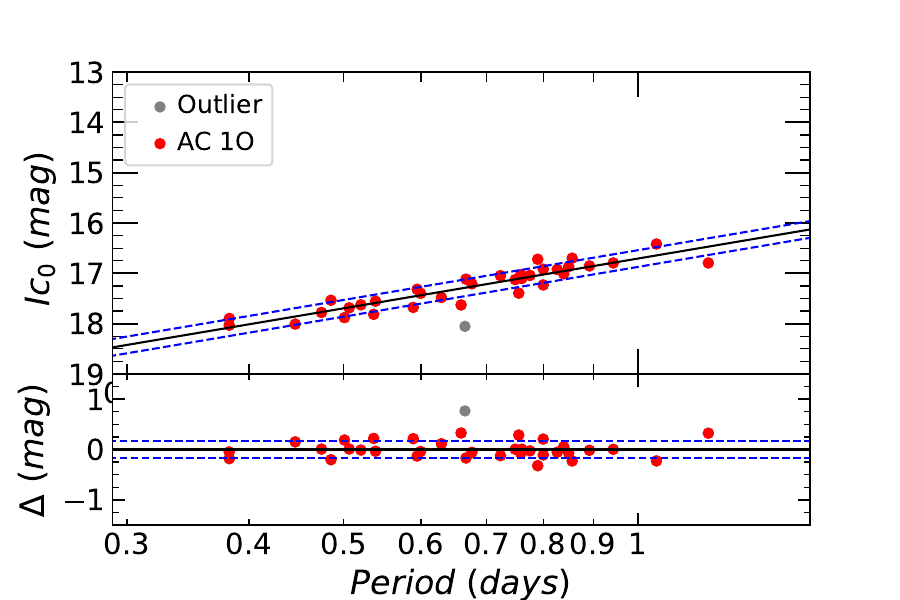}
    \includegraphics[width=0.34\textwidth]{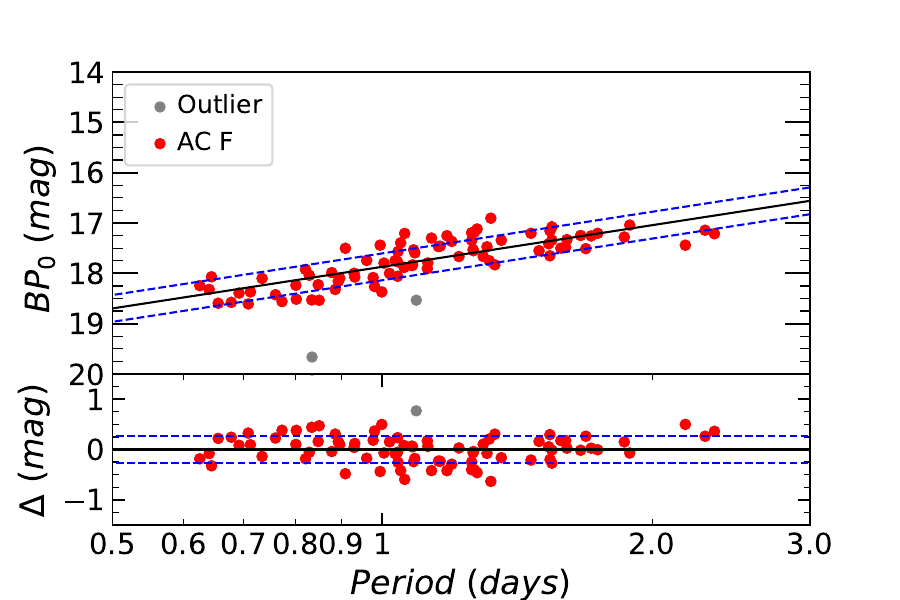}
    }
    \hbox{
    \includegraphics[width=0.34\textwidth]{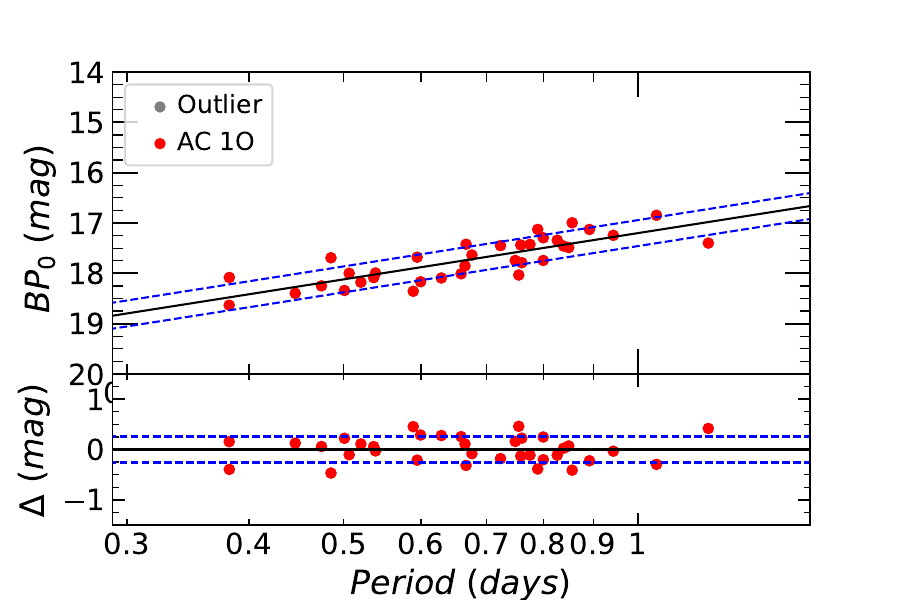}
    \includegraphics[width=0.34\textwidth]{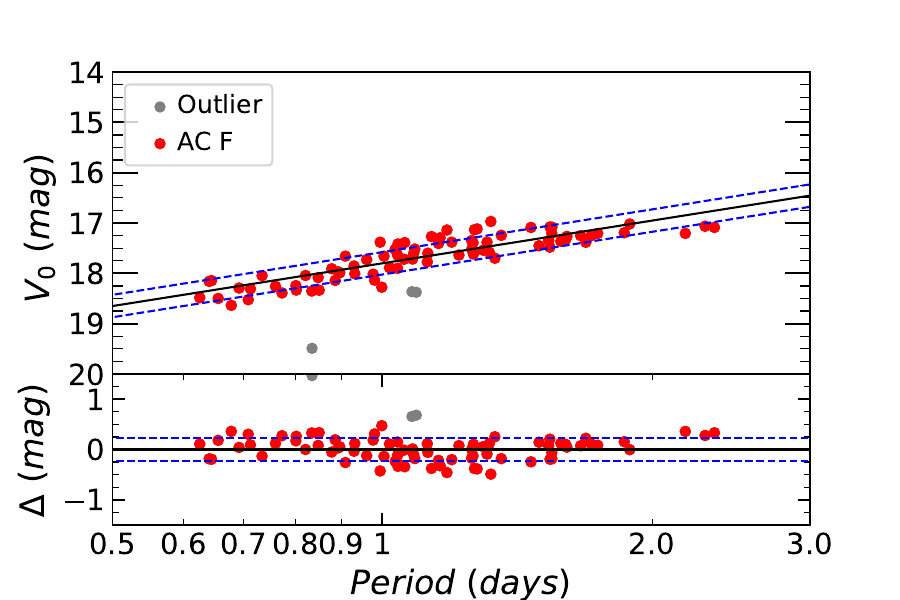}
    \includegraphics[width=0.34\textwidth]{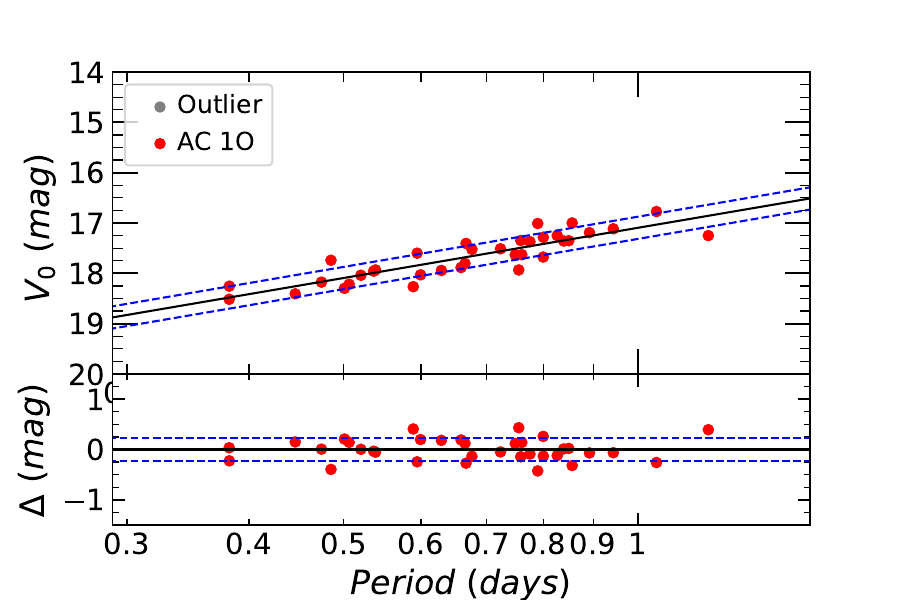}    }
    \hbox{
    \includegraphics[width=0.34\textwidth]{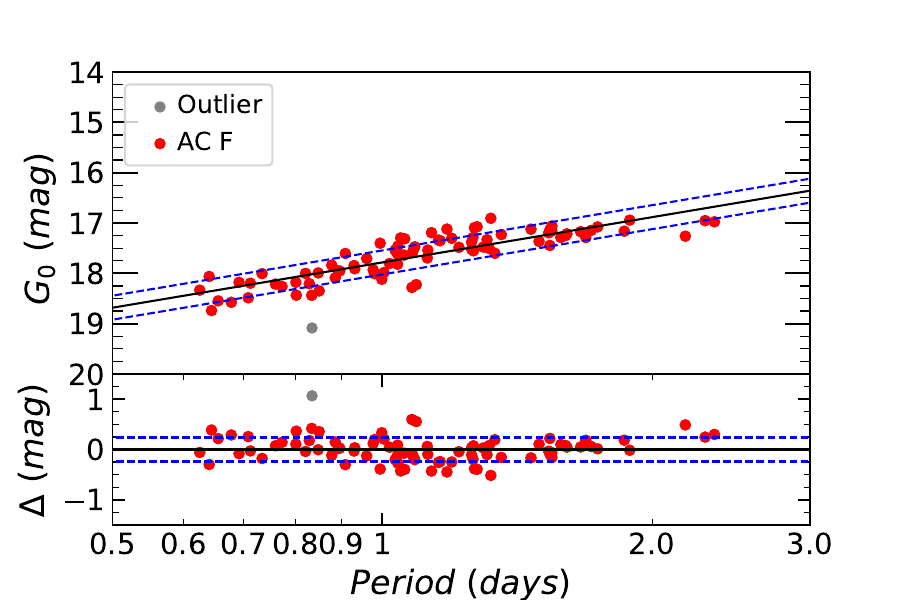}
    \includegraphics[width=0.34\textwidth]{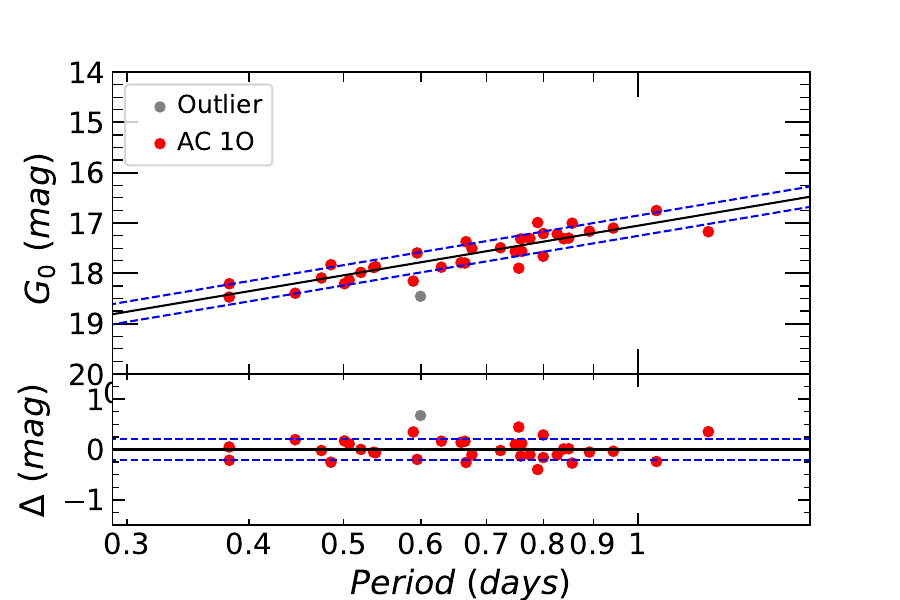}
    \includegraphics[width=0.34\textwidth]{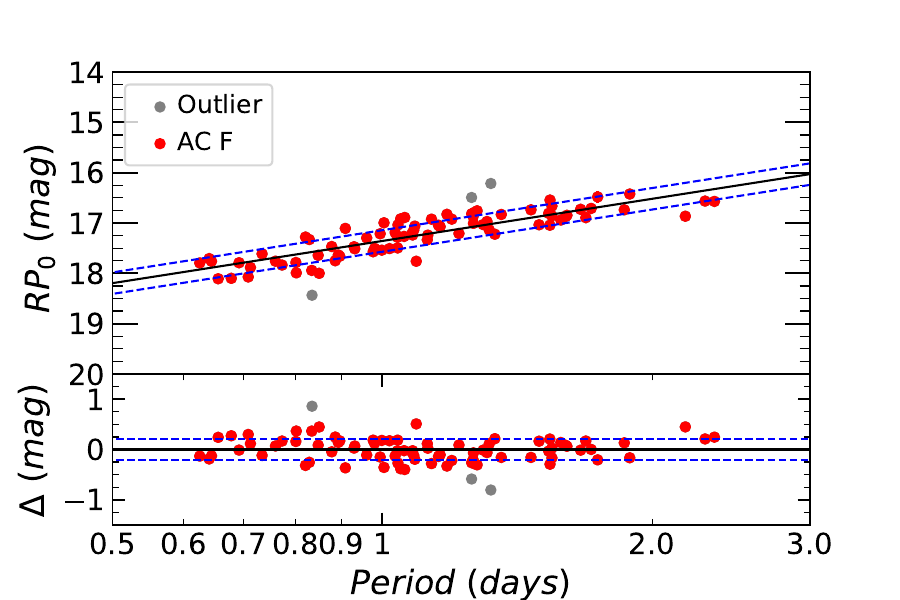}
    }
    \hbox{
    \includegraphics[width=0.34\textwidth]{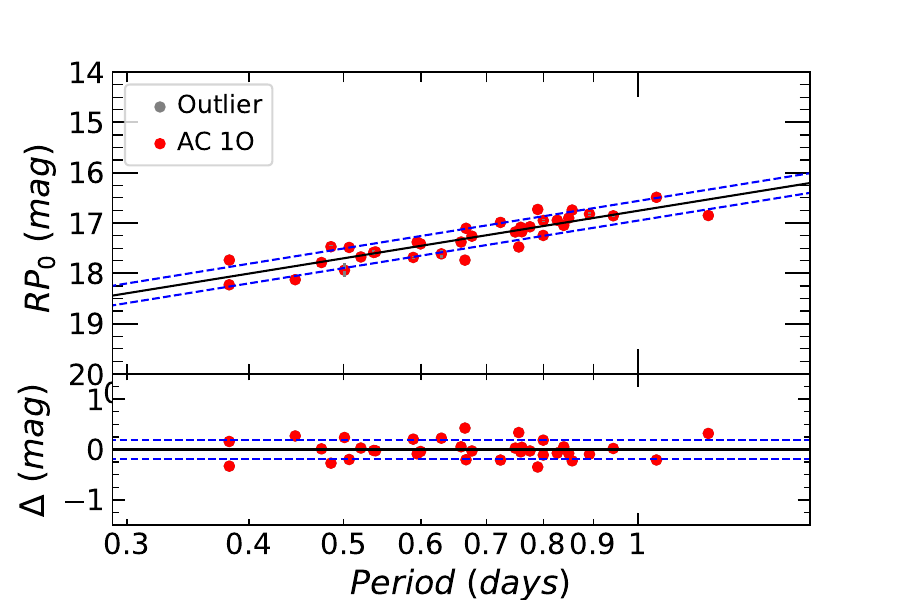}
    \includegraphics[width=0.34\textwidth]{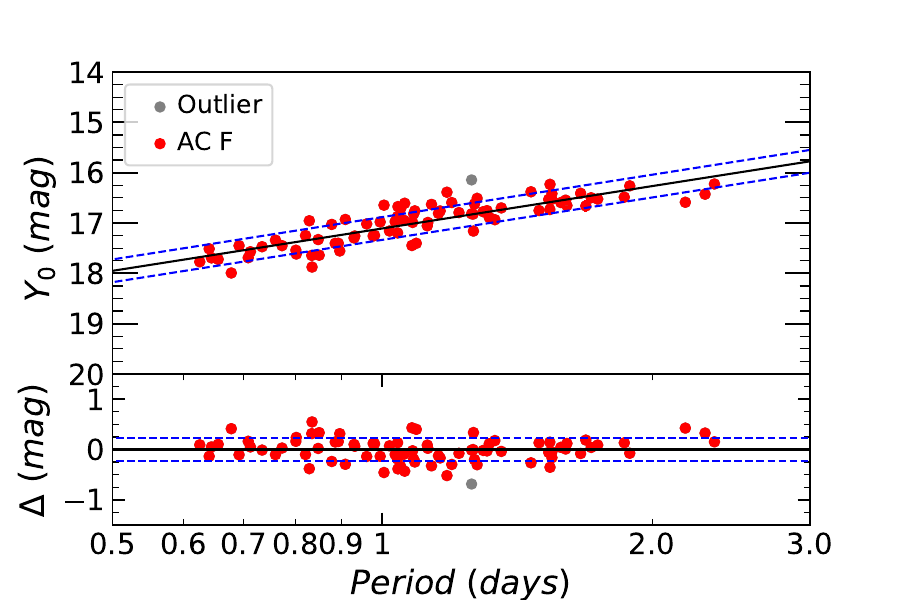}
    \includegraphics[width=0.34\textwidth]{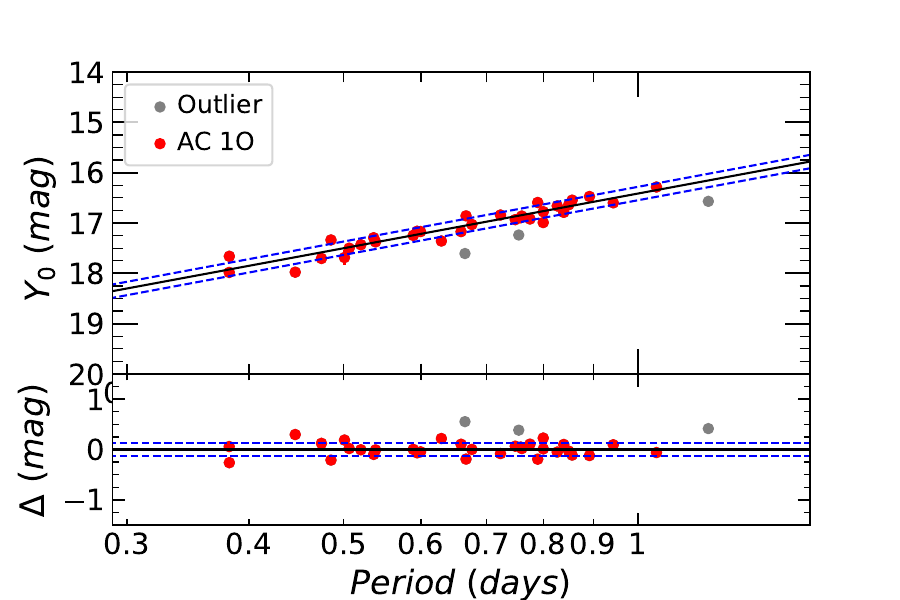}
    }
    }
    \caption{ \label{fit1lmc} Relation fitting in different bands for ACs pulsators in the LMC.}
    \end{figure*}
 
\begin{figure*}
    %\begin{adjustwidth}{-2cm} 
    \vbox{
    \hbox{
    \includegraphics[width=0.34\textwidth]{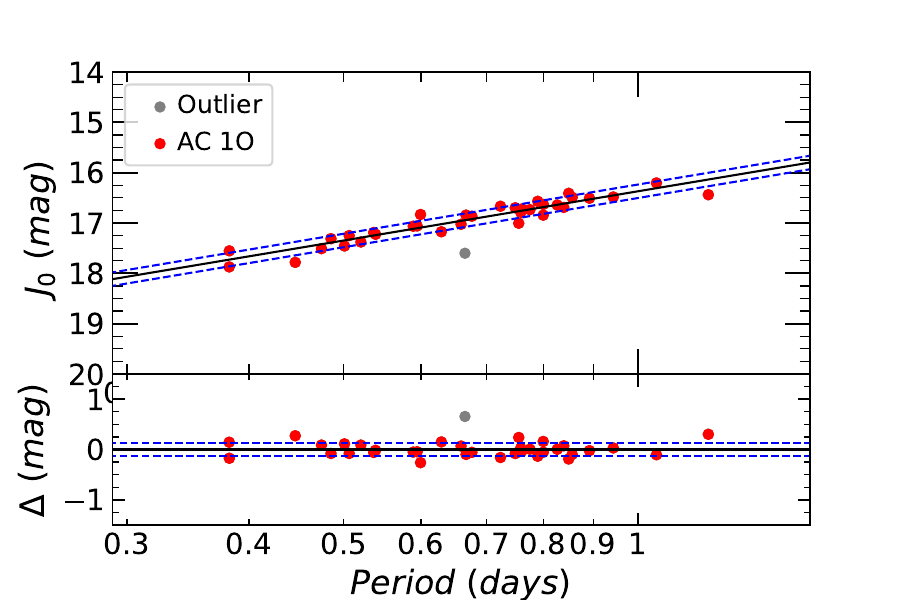}
    \includegraphics[width=0.34\textwidth]{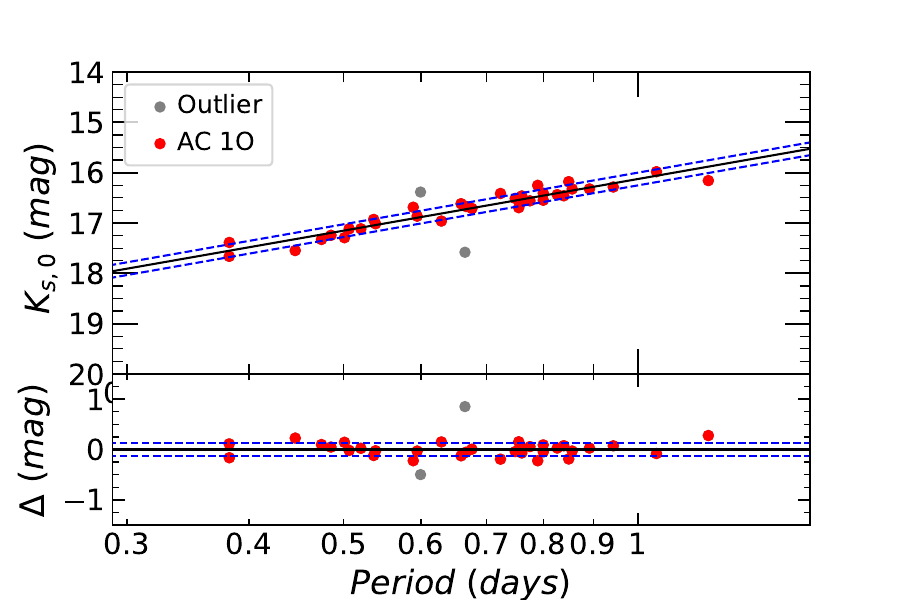}
    \includegraphics[width=0.34\textwidth]{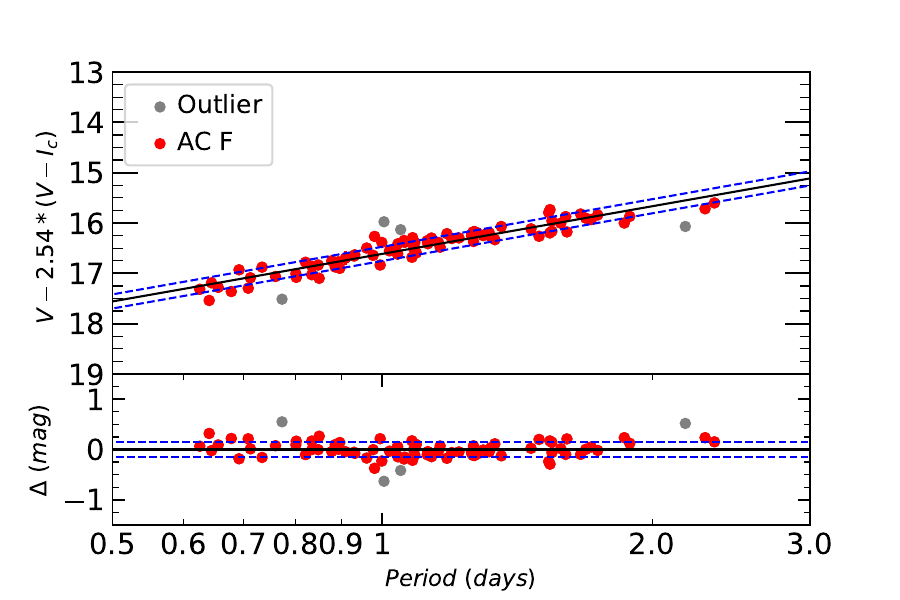}
    }
    \hbox{
    \includegraphics[width=0.34\textwidth]{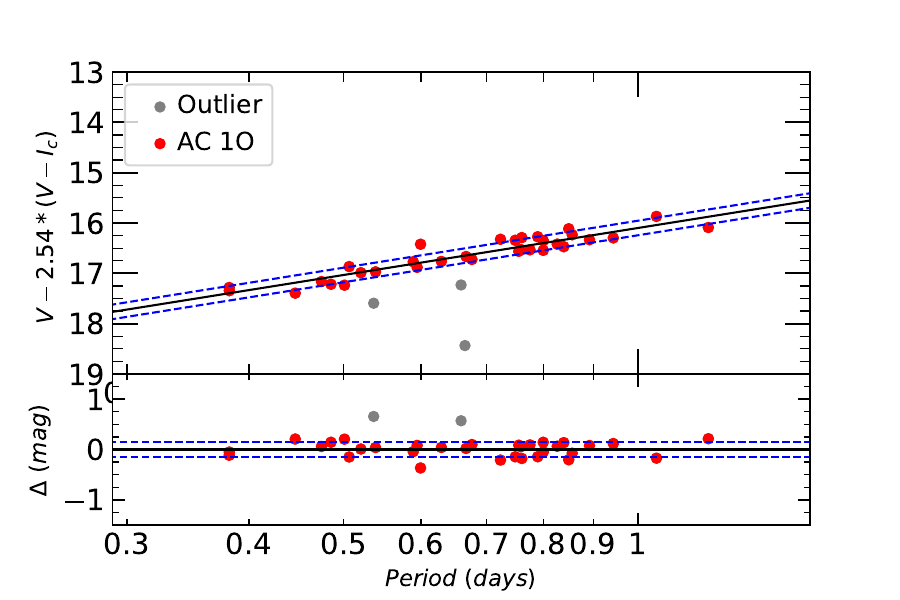}
    \includegraphics[width=0.34\textwidth]{PWVI_1O_lmc_2.pdf}
    \includegraphics[width=0.34\textwidth]{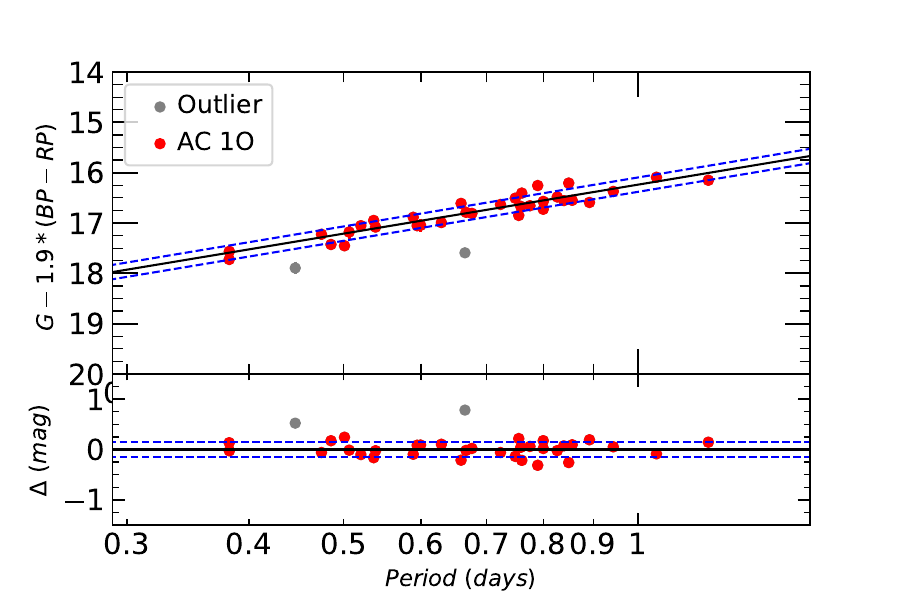}
    }
    \hbox{
    \includegraphics[width=0.34\textwidth]{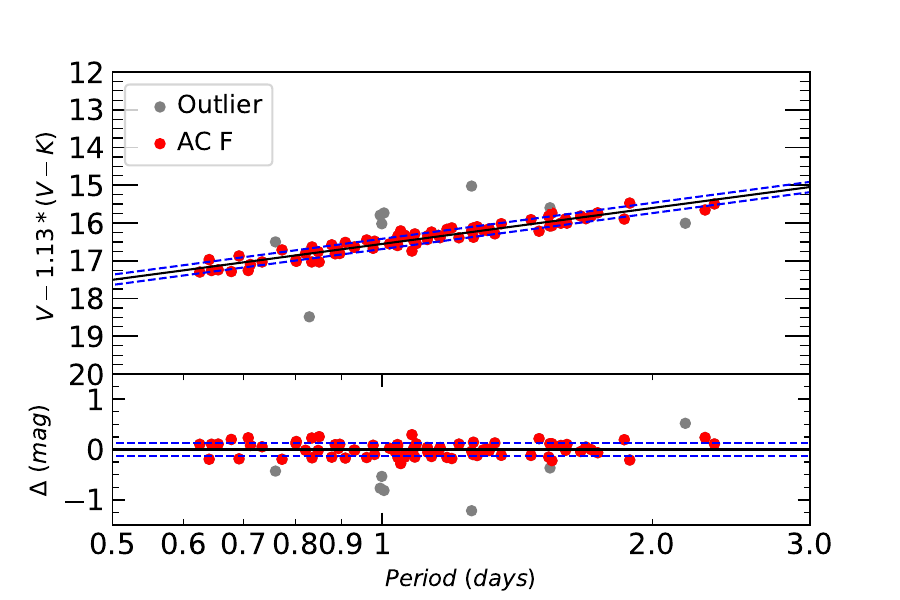}
    \includegraphics[width=0.34\textwidth]{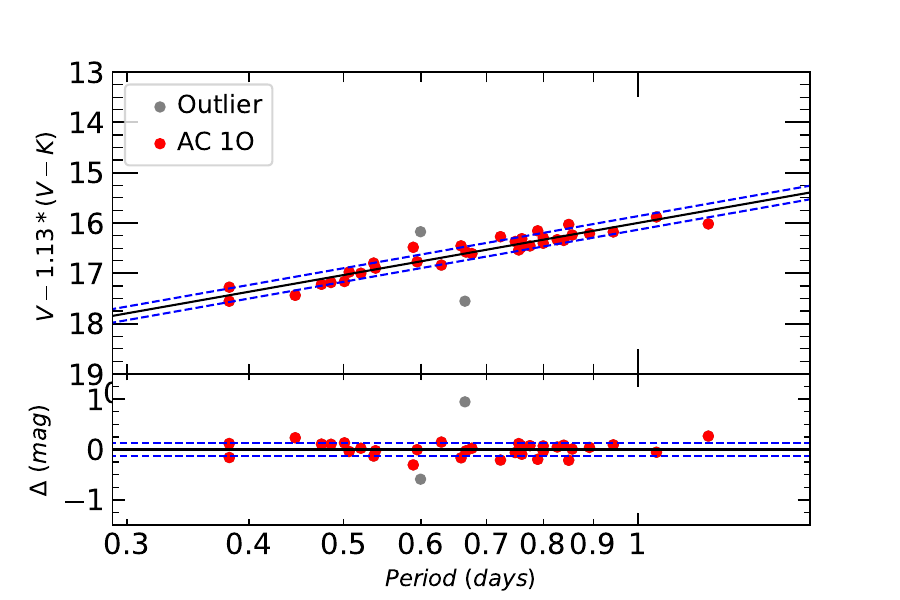}
    \includegraphics[width=0.34\textwidth]{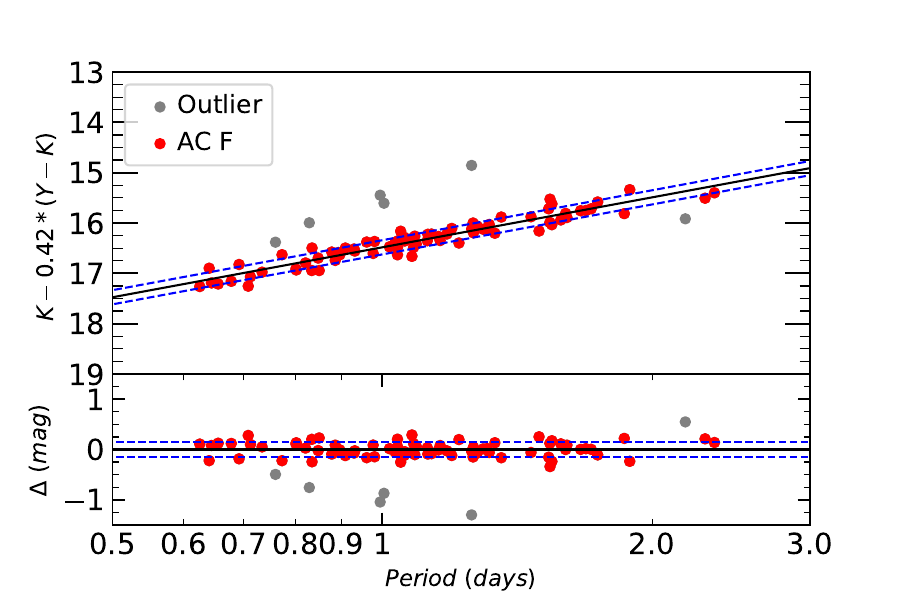}
    }
    \hbox{
    \includegraphics[width=0.34\textwidth]{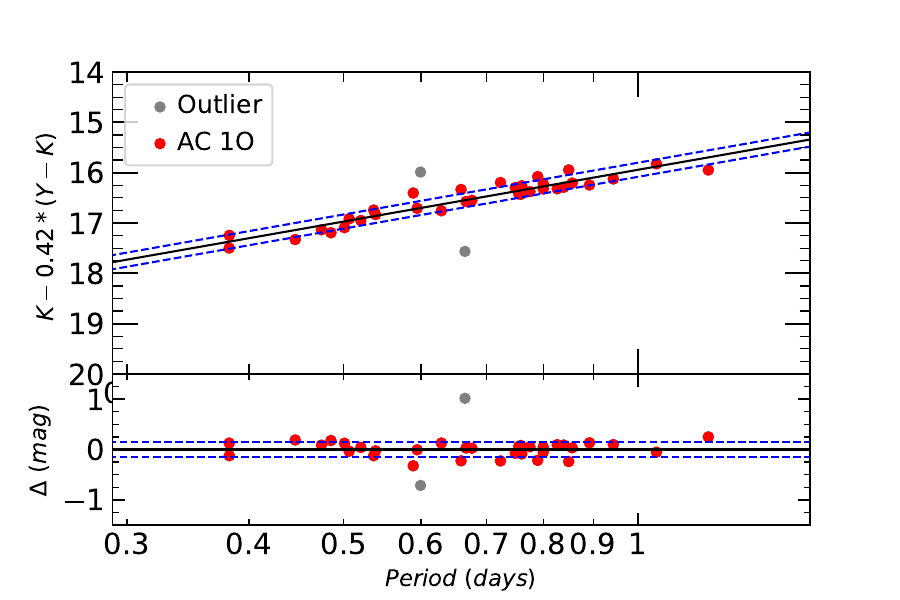}
    \includegraphics[width=0.34\textwidth]{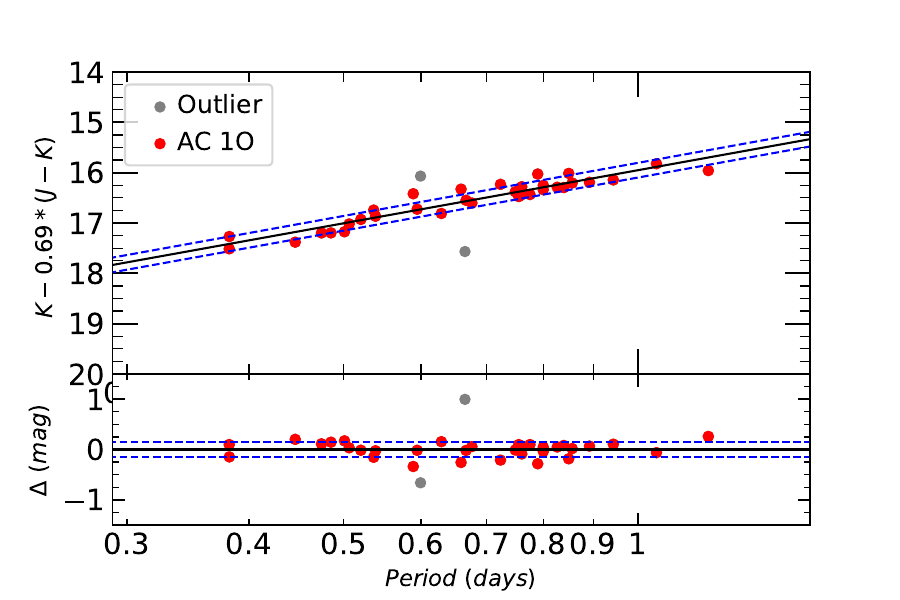}
    }
    }
    \ContinuedFloat
    \caption{continued. }  
	\end{figure*}
\begin{figure*}[h]
    %\begin{adjustwidth}{-2cm} 	
    \vbox{
    \hbox{
    \includegraphics[width=0.34\textwidth]{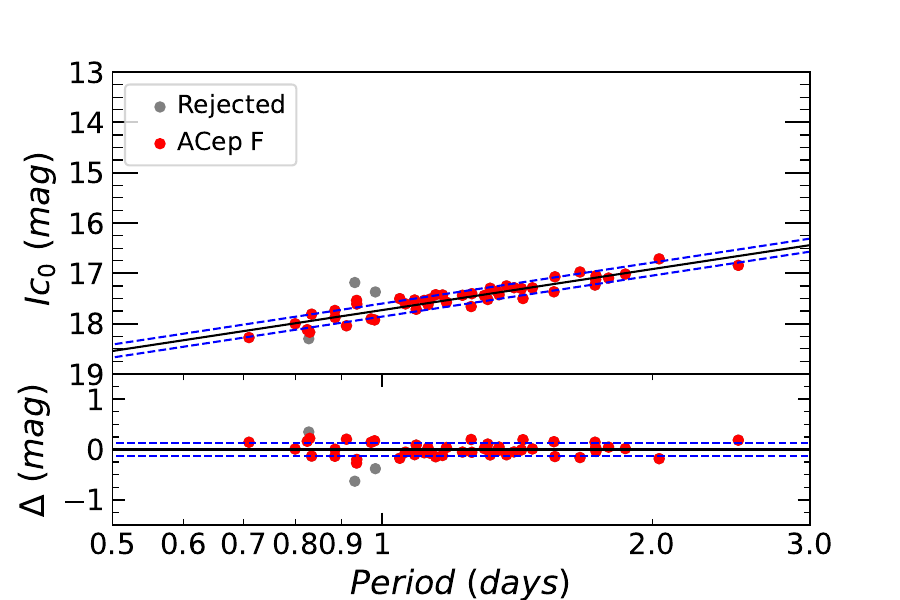}
    \includegraphics[width=0.34\textwidth]{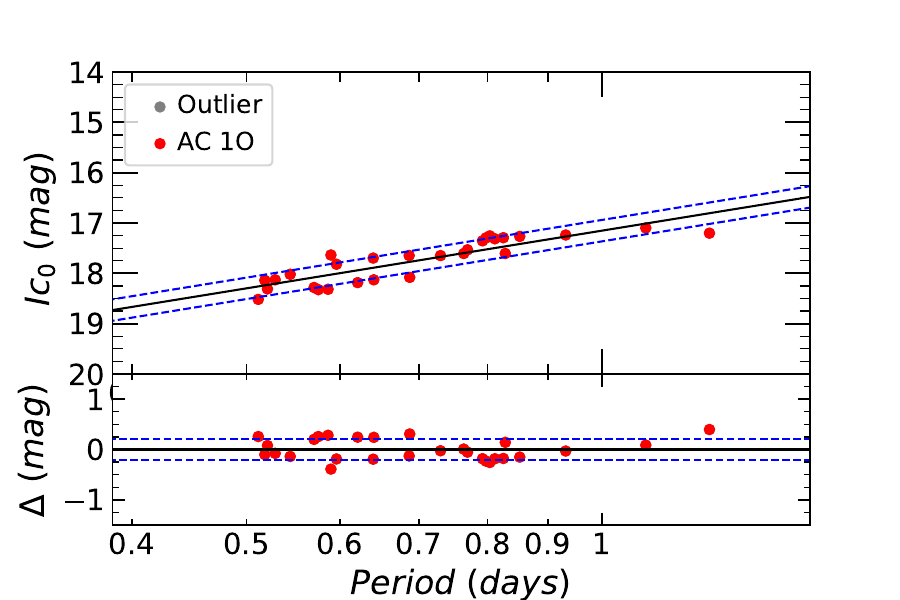}
    \includegraphics[width=0.34\textwidth]{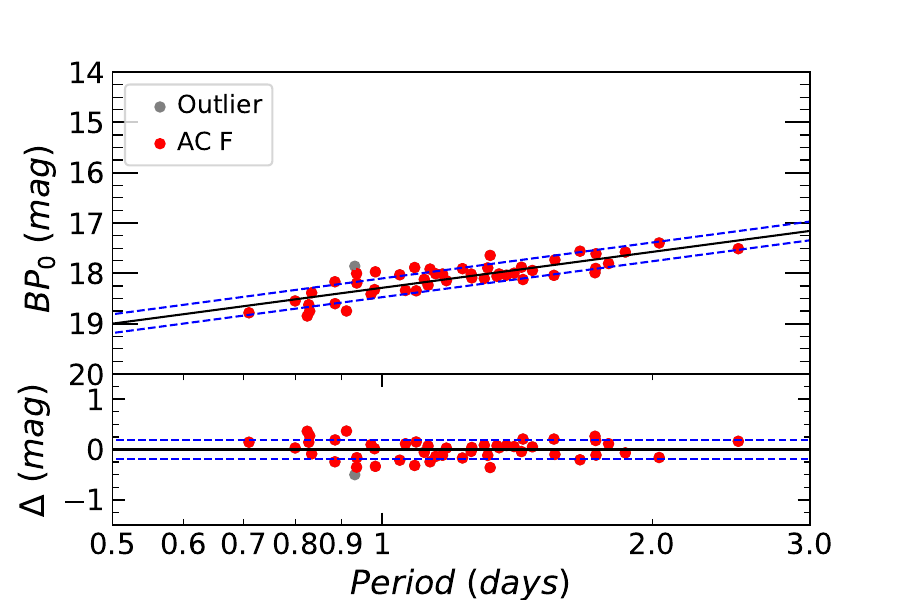}
    }
    }
    \caption{\label{fit1smc} Same as Fig.~\ref{fit1lmc}, but for the SMC. }
    %\end{adjustwidth}
	\end{figure*}

\begin{figure*}[h]
    %\begin{adjustwidth}{-2cm} 	
    \vbox{
    \hbox{
    \includegraphics[width=0.34\textwidth]{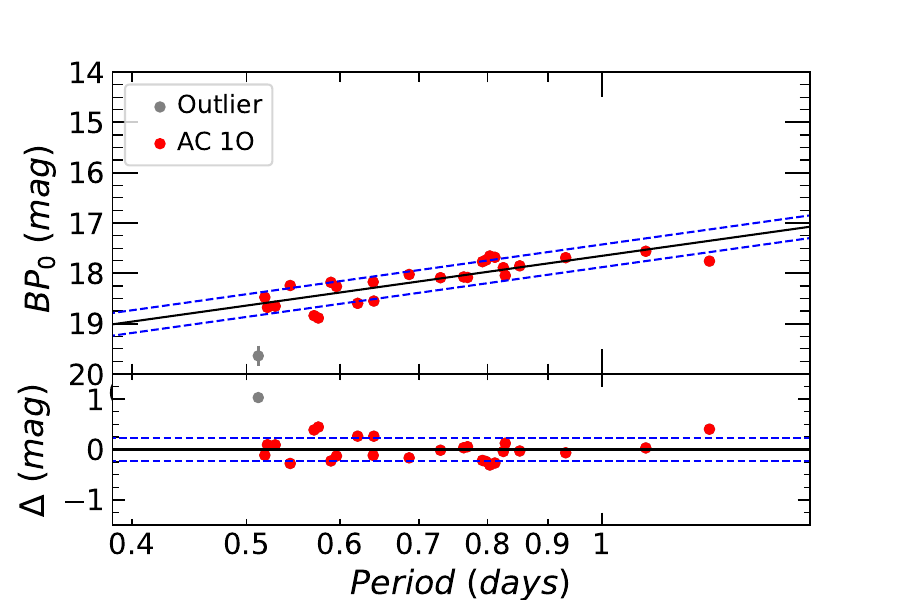}
    \includegraphics[width=0.34\textwidth]{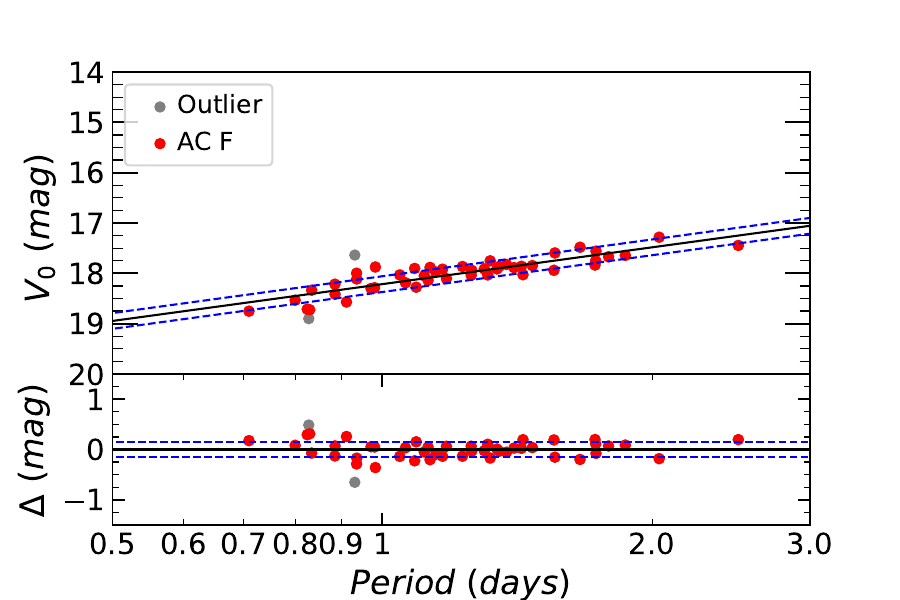}
    \includegraphics[width=0.34\textwidth]{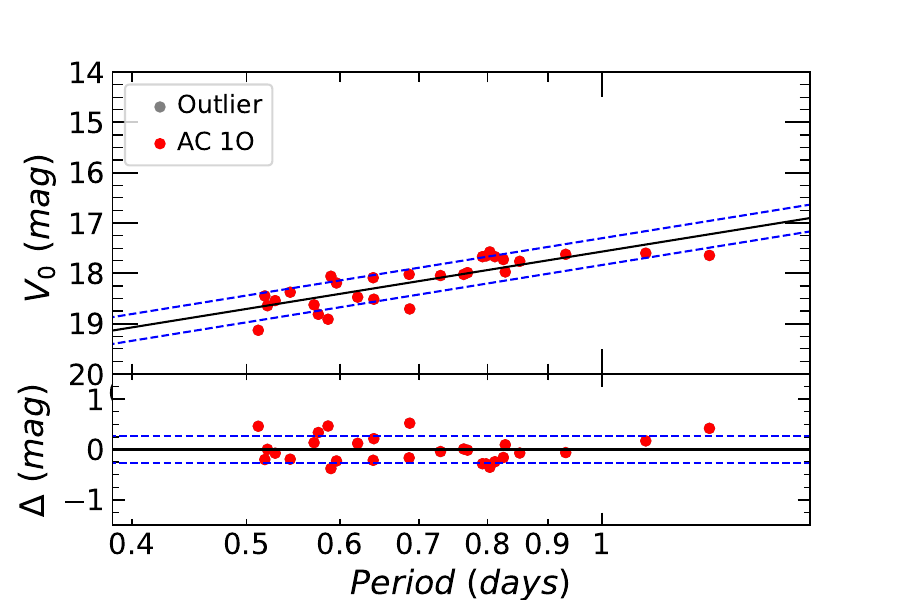}
    }
    \hbox{
    \includegraphics[width=0.34\textwidth]{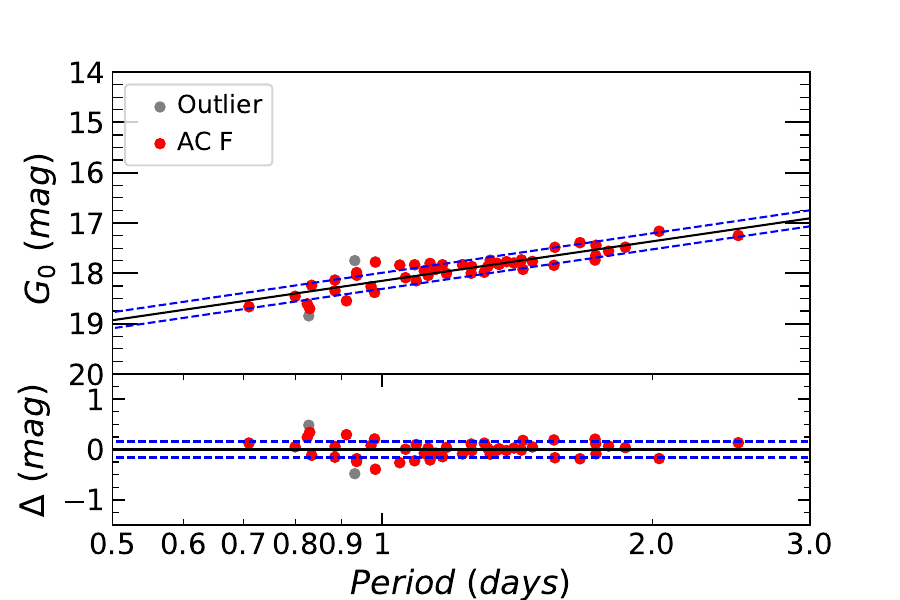}
    \includegraphics[width=0.34\textwidth]{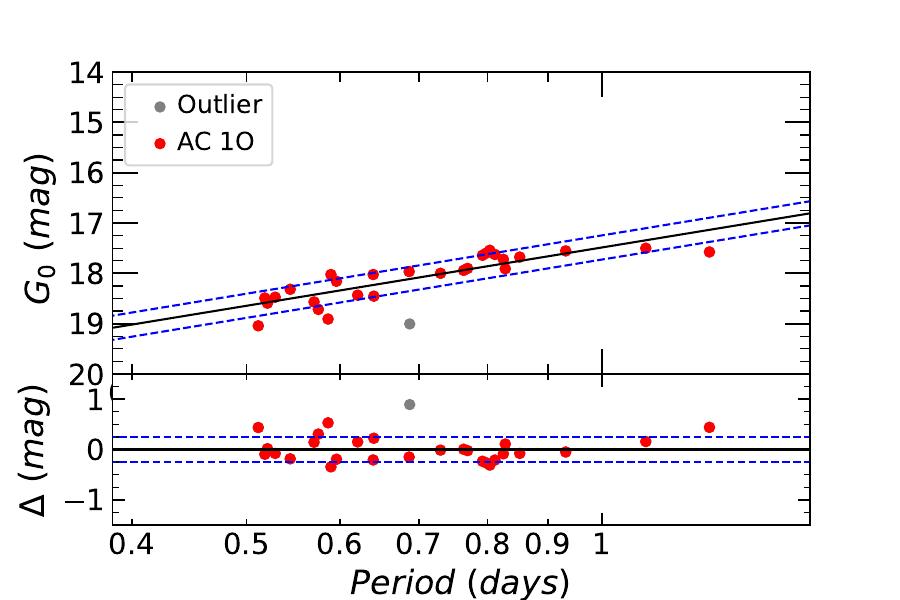}
    \includegraphics[width=0.34\textwidth]{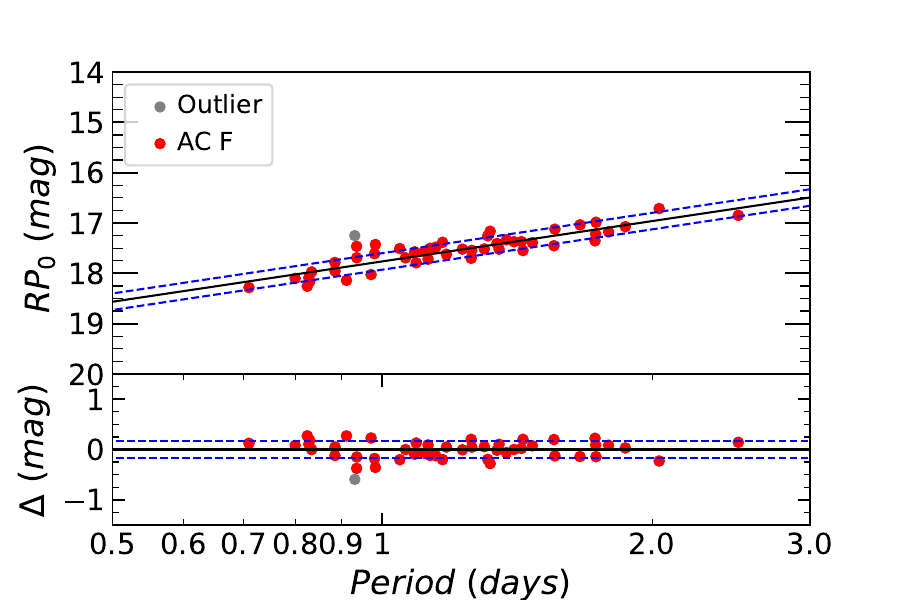}
    }
    \hbox{
    \includegraphics[width=0.34\textwidth]{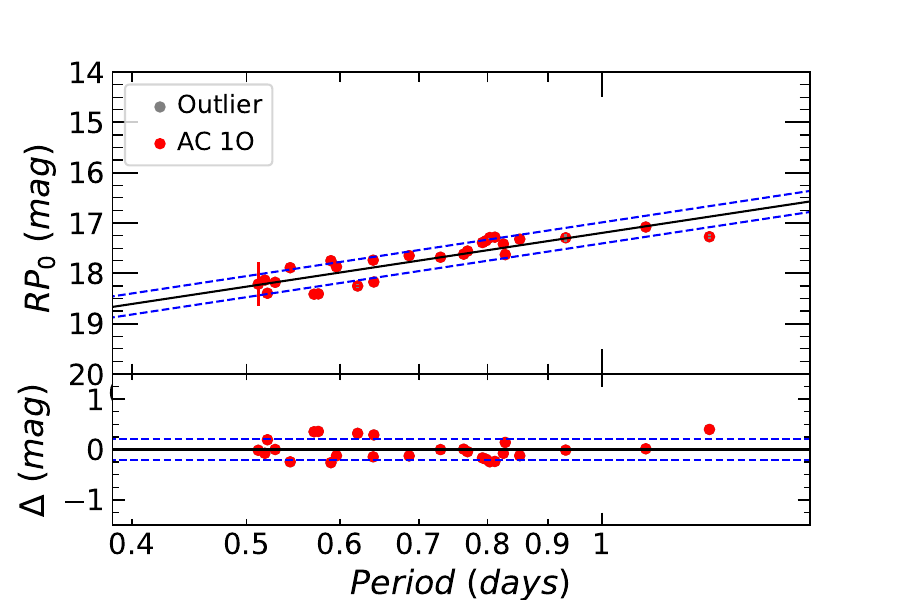}
    \includegraphics[width=0.34\textwidth]{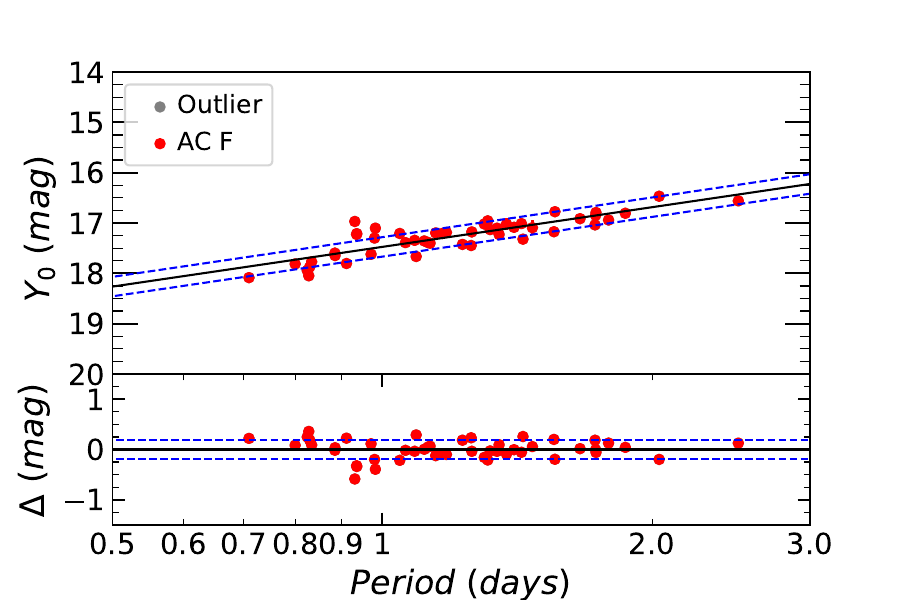}
    \includegraphics[width=0.34\textwidth]{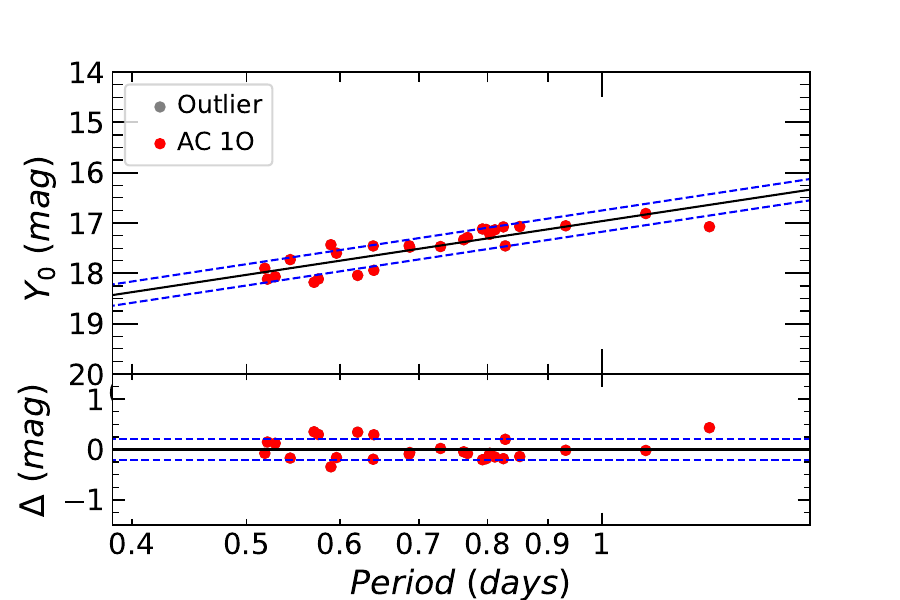}
    }
    \hbox{
    \includegraphics[width=0.34\textwidth]{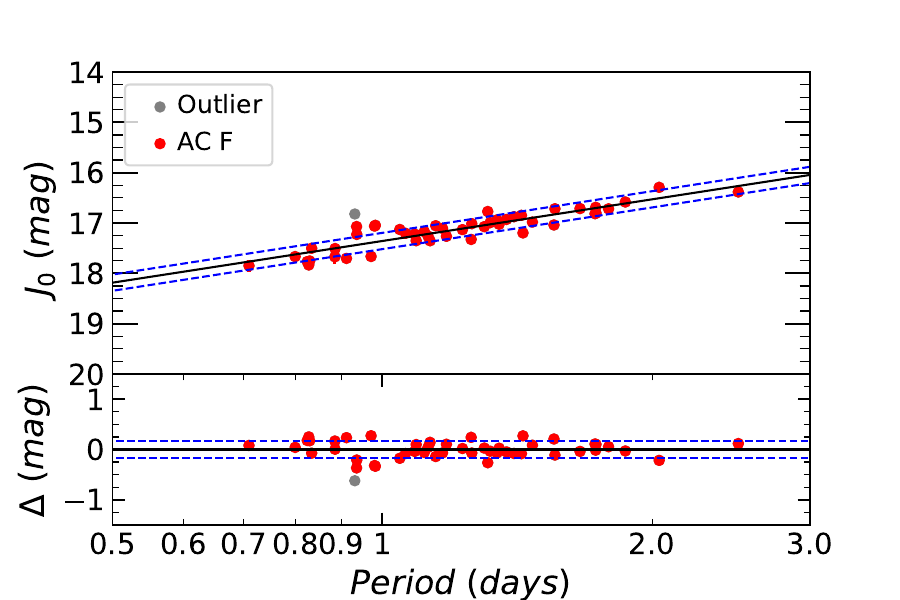}
    \includegraphics[width=0.34\textwidth]{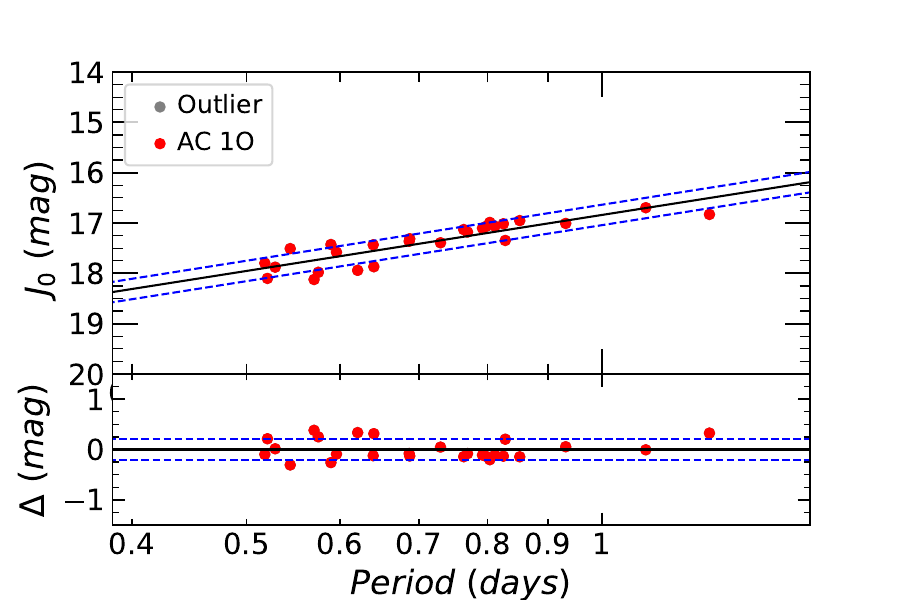}
    \includegraphics[width=0.34\textwidth]{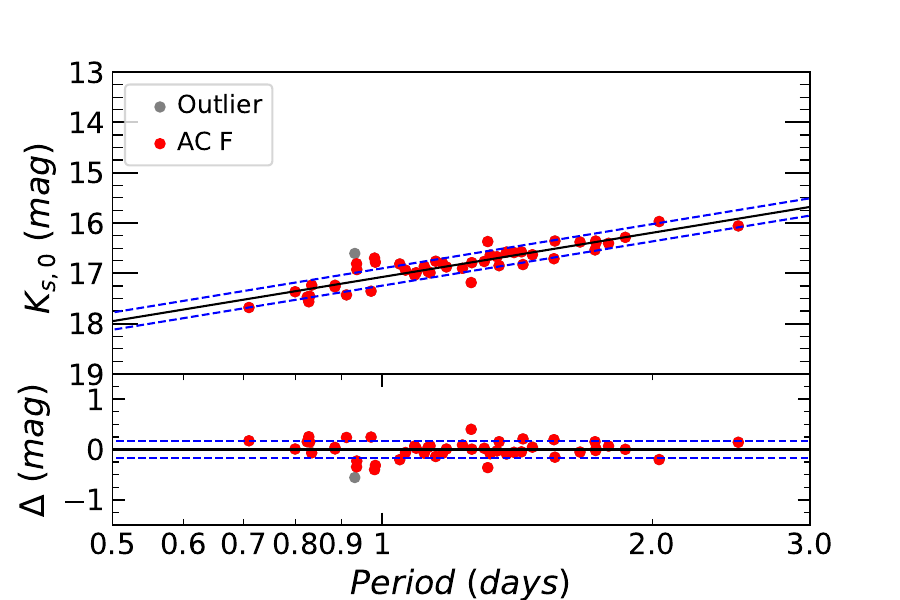}
    }
    \hbox{
    \includegraphics[width=0.34\textwidth]{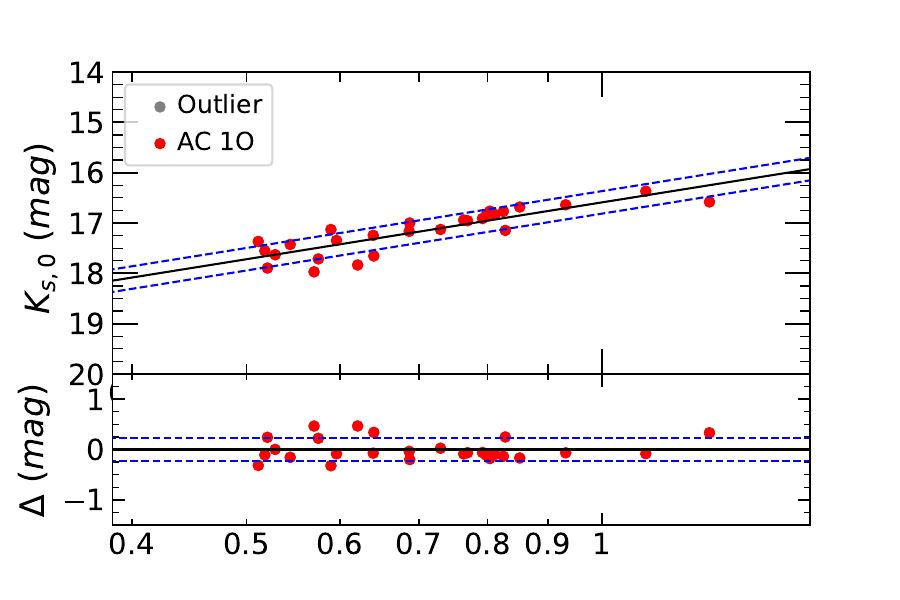}
    \includegraphics[width=0.34\textwidth]{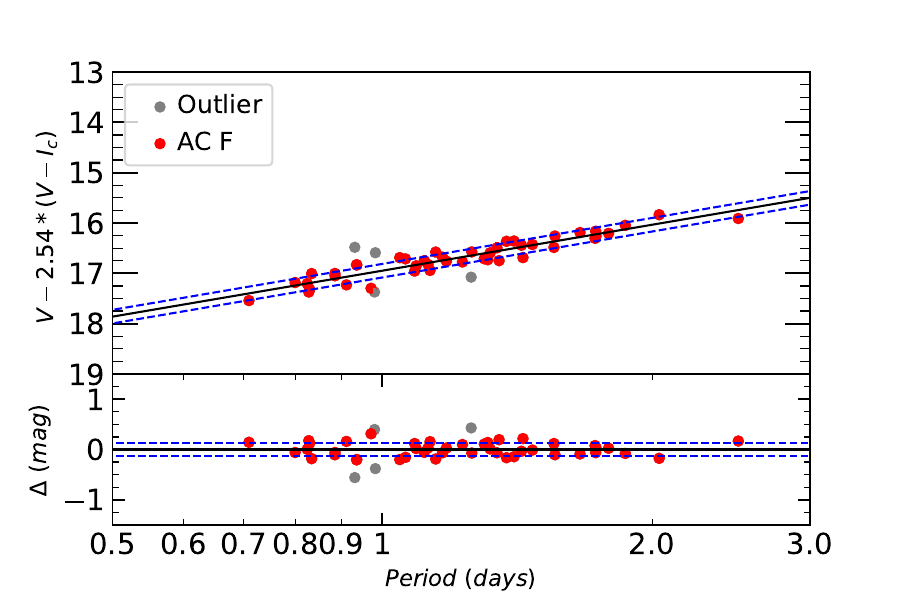}
    \includegraphics[width=0.34\textwidth]{PWVI_1O_lmc_2.pdf}
    }
    \hbox{
    \includegraphics[width=0.34\textwidth]{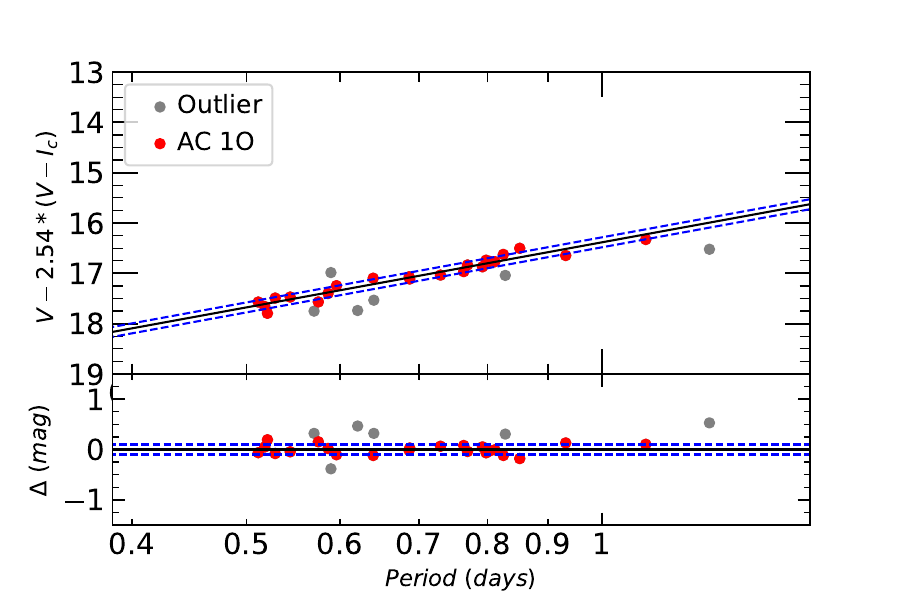}
    \includegraphics[width=0.34\textwidth]{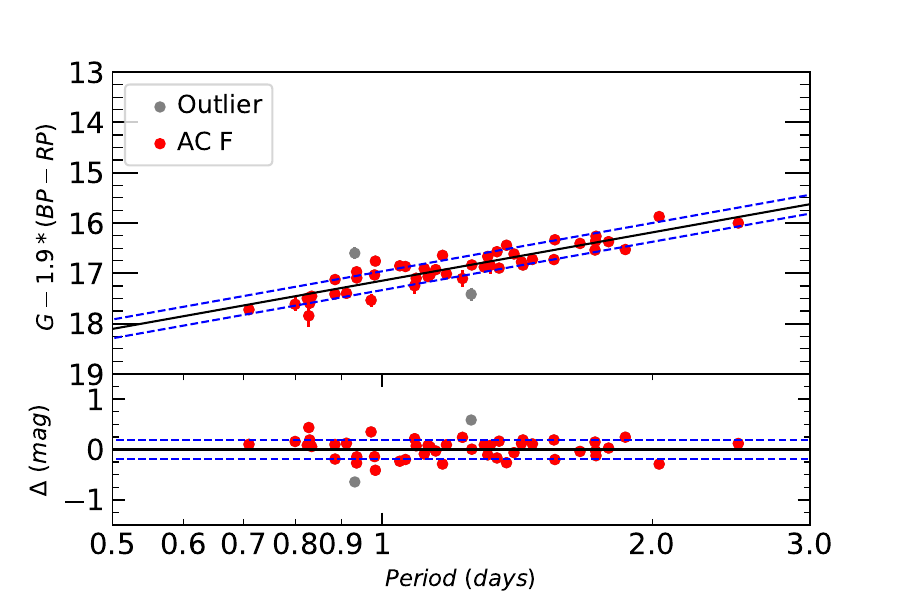}
    \includegraphics[width=0.34\textwidth]{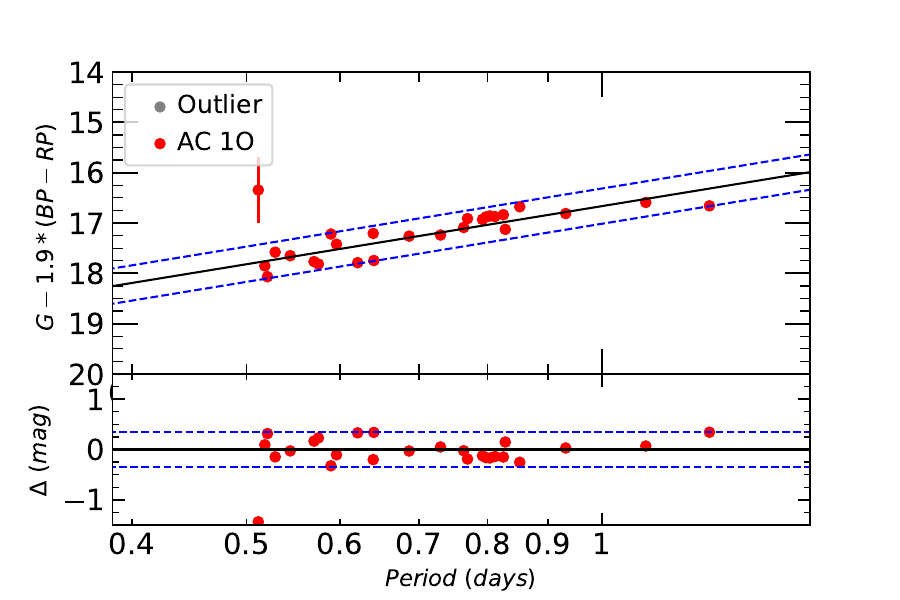}
    }
    }
    \ContinuedFloat
    \caption{continued. }
	\end{figure*}

\begin{figure*}[h]
    %\begin{adjustwidth}{-2cm} 	
    \vbox{
    \hbox{
    \includegraphics[width=0.34\textwidth]{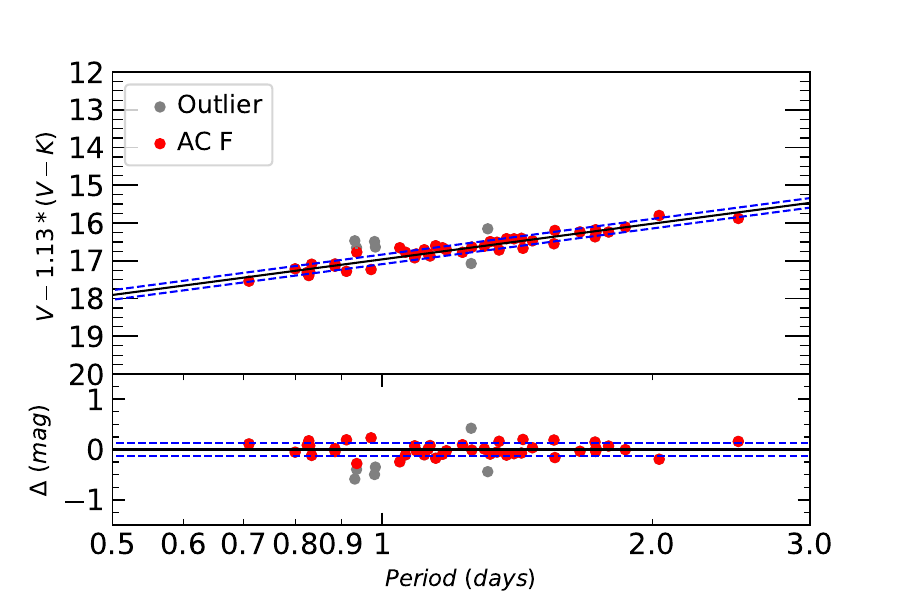}
    \includegraphics[width=0.34\textwidth]{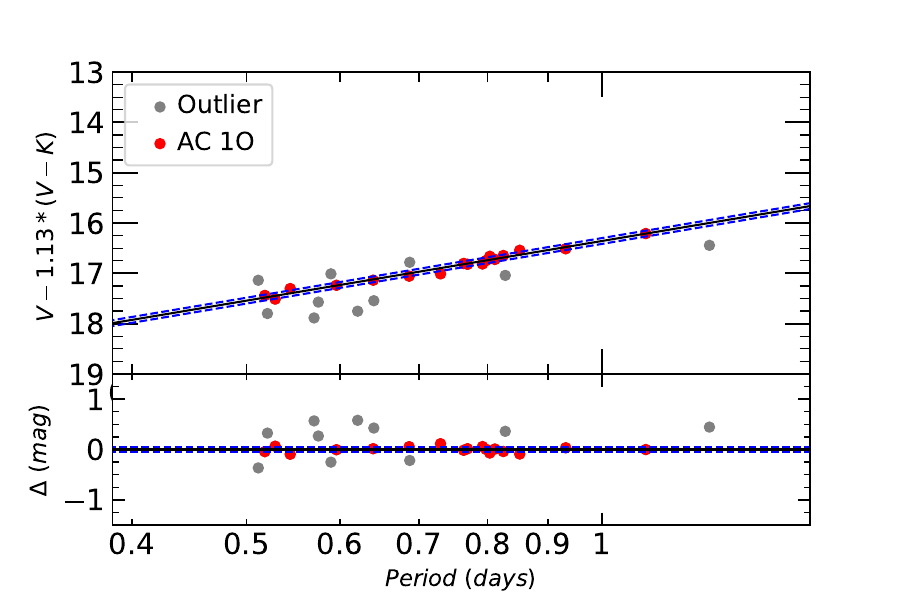}
    \includegraphics[width=0.34\textwidth]{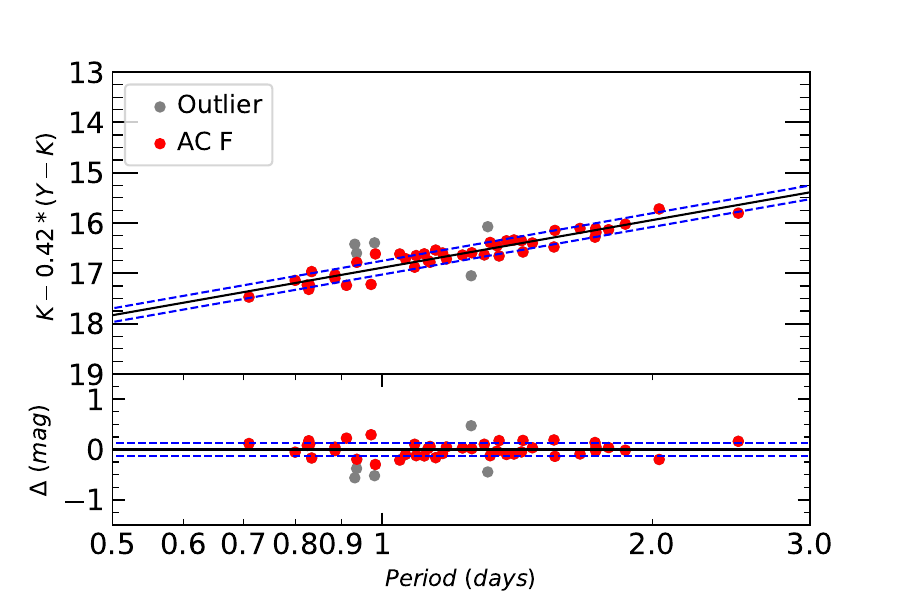}
    }
    \hbox{
    \includegraphics[width=0.34\textwidth]{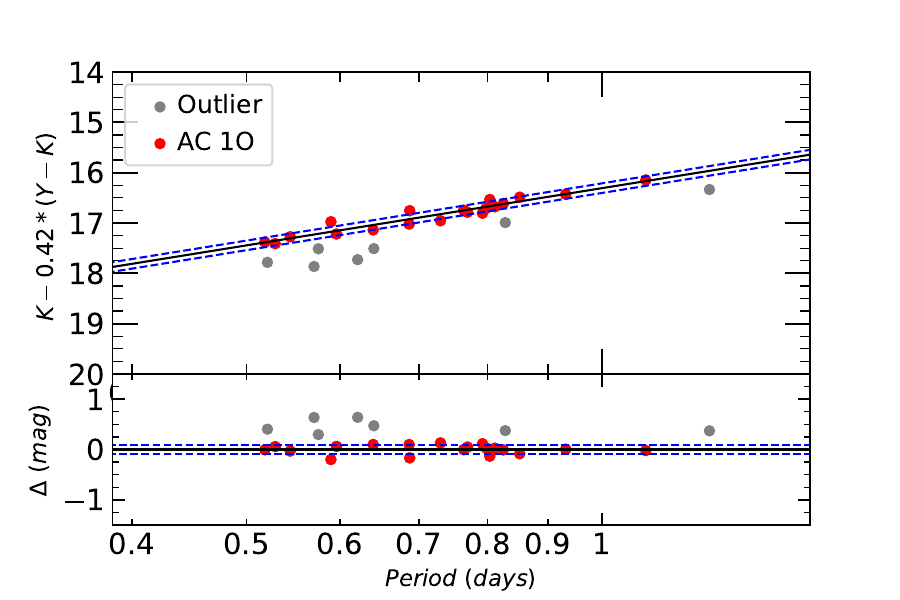}
    \includegraphics[width=0.34\textwidth]{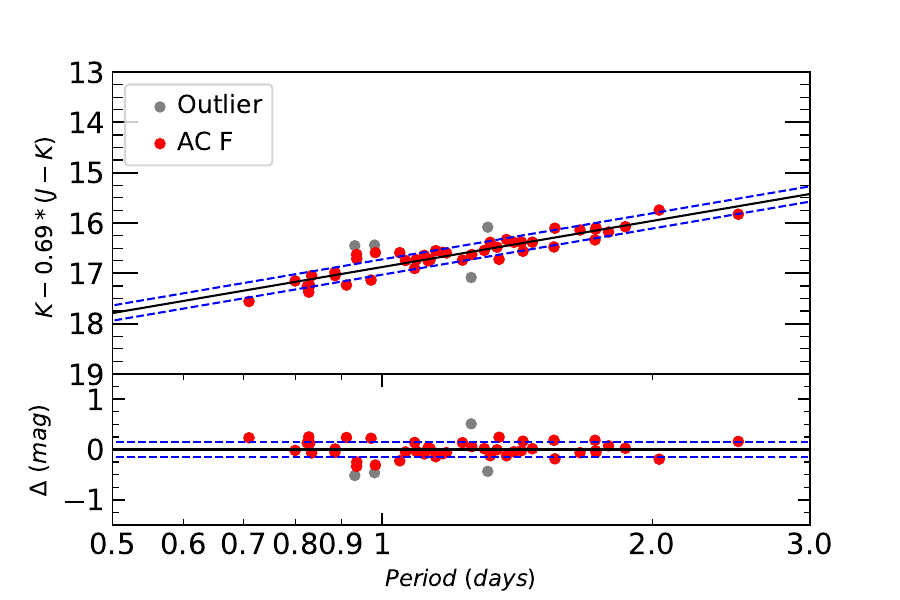}
    \includegraphics[width=0.34\textwidth]{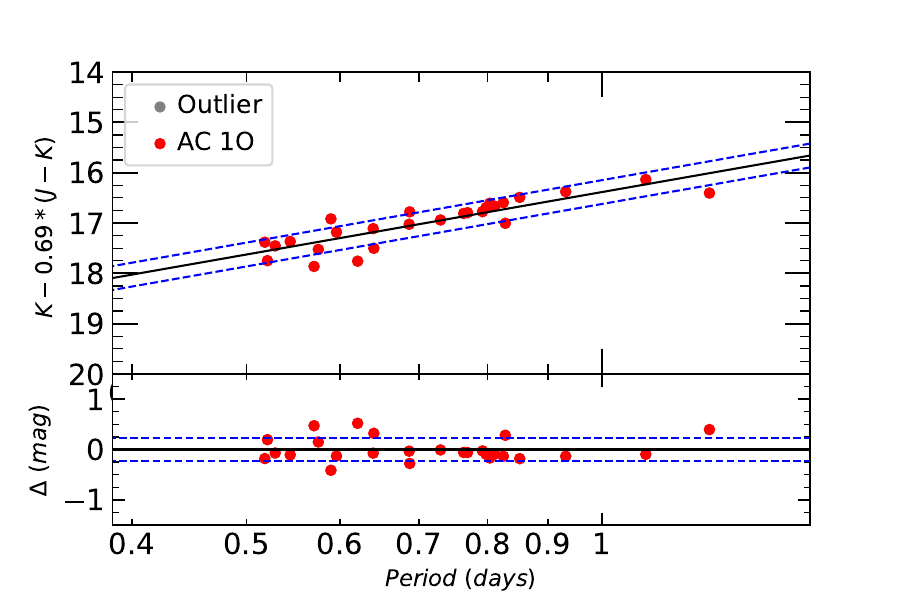}
    }
    }
    \ContinuedFloat
    \caption{continued. }  
	\end{figure*}

 \begin{figure*}
\vbox{\includegraphics[width=0.5\textwidth]{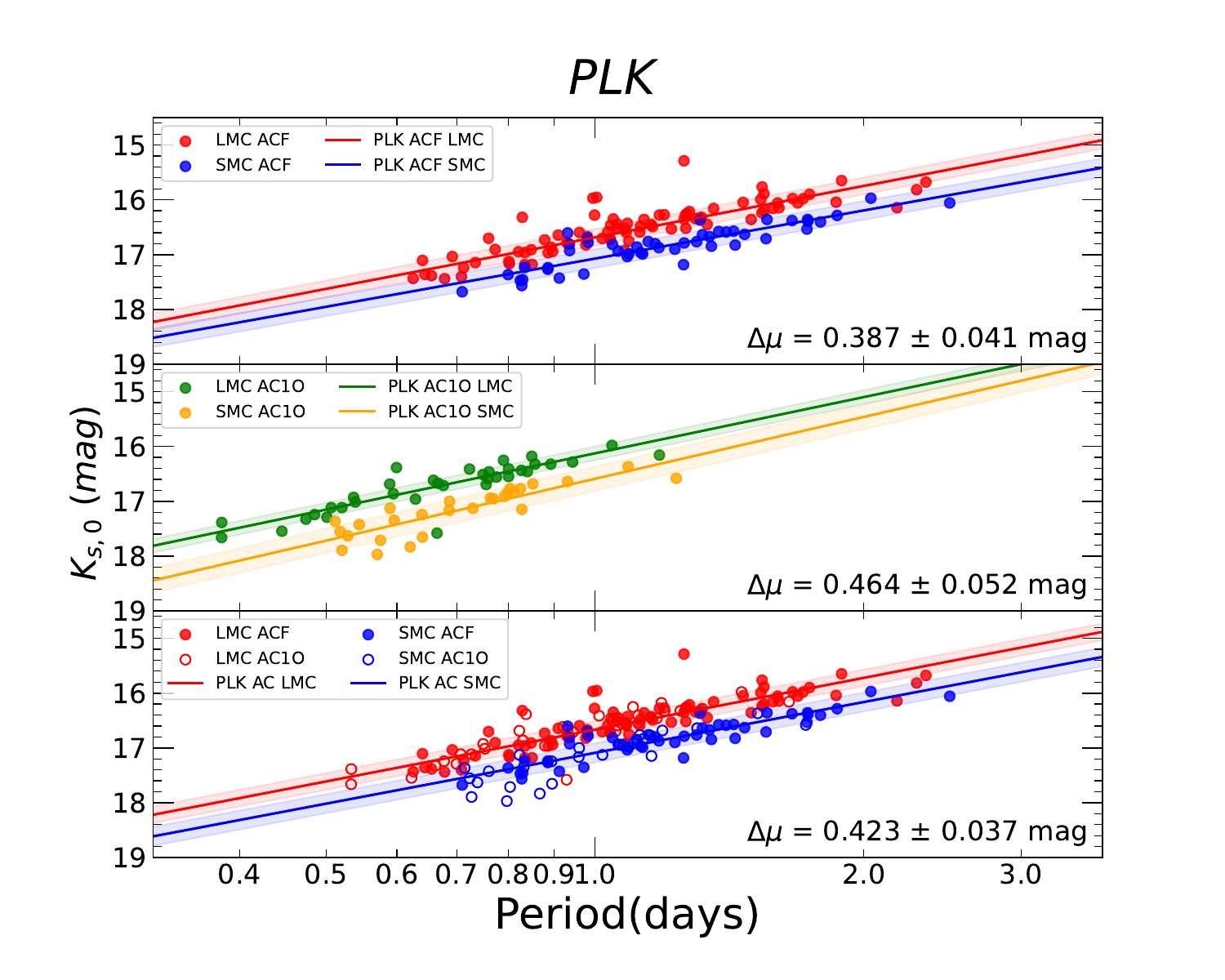}
\includegraphics[width=0.445\textwidth]{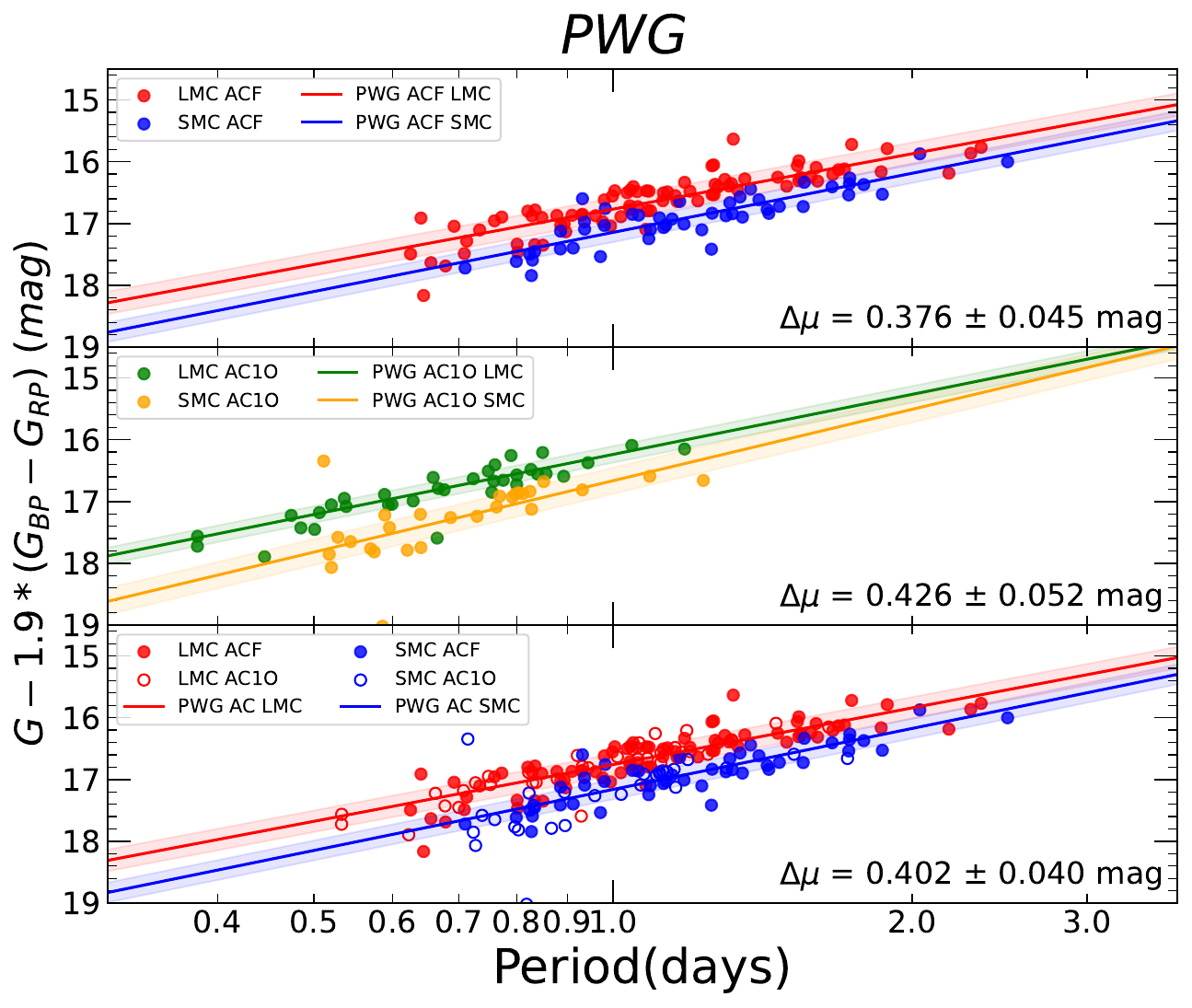}
}
    
    \caption{As in Fig.\ref{lmc_smc} but for the PLK and PWG relations. }
\end{figure*}

\end{appendix}

\end{document}